\documentclass[10pt]{article}
\pdfoutput=1

\usepackage[T2A]{fontenc}
\usepackage[utf8]{inputenc}
\usepackage[english]{babel}
\usepackage{amsmath, amsfonts, amssymb}
\usepackage{nccmath}
\usepackage{mathrsfs}
\usepackage[mathscr]{euscript}
\usepackage{color}
\usepackage{amsthm}
\usepackage{psfrag}
\usepackage{graphicx}
\usepackage{url}
\usepackage[hyperfootnotes=false]{hyperref}
\usepackage{multirow}
\usepackage{makecell}

\renewcommand{\theequation}{\arabic{section}.\arabic{equation}}
\renewcommand{\thesection}{\arabic{section}}
\renewcommand{\thesubsection}{\arabic{section}.\arabic{subsection}}
\catcode`@=11
\@addtoreset{equation}{section}
\catcode`@=12

\newcommand{\be}{\begin{equation}}
\newcommand{\ee}{\end{equation}}
\newcommand{\bl}{\begin{align}}
\newcommand{\el}{\end{align}}
\newcommand{\ba}{\begin{aligned}}
\newcommand{\ea}{\end{aligned}}
\newcommand{\beqa}{\begin{eqnarray}}
\newcommand{\eeqa}{\end{eqnarray}}

\renewcommand\l{\lambda}
\newcommand\m{\mu}
\newcommand\D{\Delta}

\newcommand\s{\sigma}

\newcommand{\al}{\alpha}
\newcommand{\bt}{\beta}
\newcommand{\p}{\partial}
\newcommand{\n}{\nabla}
\newcommand{\di}{\mathrm d}
\newcommand{\g}{\gamma}

\def\e{{\rm e}}
\def\d{\partial}

\newcommand{\bseq}{\begin{subequations}}
\newcommand{\eseq}{\end{subequations}}

\renewcommand{\ln}{\mathop{\rm ln}\nolimits}

\newcommand{\Tr}{{\rm Tr}}

\allowdisplaybreaks[4]

\begin{document}
\title{\vspace{-2cm}
\begin{flushright}
{\normalsize
INR-TH-2021-020}
\end{flushright}
\vspace{0.5cm}
{\bf  Beta functions of $\bf (3+1)$-dimensional projectable\\ 
Ho\v rava gravity}}
\author{Andrei O. Barvinsky$^a$~, Alexander V. Kurov$^a$~, Sergey
  M. Sibiryakov$^{b,c,d}$\\[2mm] 
{\small\it $^a$Theory Department, Lebedev Physics Institute,
Leninsky Prospect 53, Moscow 117924, Russia}\\
{\small\it $^b$Department of Physics \& Astronomy, McMaster
University, Hamilton, Ontario, L8S 4M1, Canada}\\
{\small\it $^c$Perimeter Institute for Theoretical Physics, Waterloo,
 Ontario, N2L 2Y5, Canada}\\
{\small\it $^d$Institute for Nuclear Research of the Russian Academy
  of Sciences,}\\
{\small\it 60th October Anniversary Prospect, 7a,
  117312 Moscow, Russia}
}
\date{}
\maketitle

\begin{abstract}
We derive the full set of beta functions for the marginal essential
couplings of projectable Ho\v rava gravity in $(3+1)$-dimensional
spacetime. To this end we compute the divergent part of the one-loop
effective action in static background with arbitrary spatial
metric. The computation is done in several steps: reduction of the
problem to three dimensions, extraction of an operator square root from
the spatial part of the fluctuation operator, and evaluation of its
trace using the method of universal functional traces. This provides
us with the renormalization of couplings in the potential part of the
action which we combine with the results for the
kinetic part obtained previously. The calculation uses
symbolic computer algebra and is performed in four different gauges
yielding identical results for the essential beta functions. We
additionally check the calculation by evaluating the effective action
on a special background with spherical spatial slices using an
alternative method of spectral summation. We conclude with a
preliminary discussion of the properties of the beta functions and the
resulting renormalization group flow, identifying several candidate
asymptotically free fixed points. 

\end{abstract}

\tableofcontents

\section{Introduction and summary}

\subsection{State of the art}

Ho\v rava gravity (HG) has been suggested \cite{Horava:2009uw} as an
approach to quantum gravity within the framework of renormalizable
quantum field theory (see
\cite{Mukohyama:2010xz,Sotiriou:2010wn,Blas:2010hb,Wang:2017brl,Blas:2018flu}
for reviews). The key idea is borrowed from condensed matter physics
and uses the notion of anisotropic (Lifshitz) scaling of time and
space coordinates,\footnote{Throughout the paper we use Latin indices
  to denote the spatial directions, $i=1,\ldots,d$.}
\be
\label{scaling}
t\mapsto b^{-d} t~,~~~~~x^i\mapsto b^{-1} x^i\;,
\ee
where $d$ is the number of spatial dimensions and $b$ is a positive
scaling parameter. A theory whose bare (tree-level) action does not
contain any irrelevant operators under this scaling is power-counting
renormalizable, implying that it has fair chances to be perturbatively
renormalizable in the strict sense, i.e. all divergences generated
within perturbation theory can be absorbed into redefinition of the
couplings in the action.

For any space dimension $d$ bigger than $1$, time and space in
(\ref{scaling}) scale differently. Clearly, a theory of gravity
formulated in this setting cannot be Lorentz invariant nor completely
diffeomorphism invariant. The special role of time restricts possible
symmetries to foliation preserving diffeomorphisms (FDiff) which leave
the constant-time slices invariant,
\be
\label{FDiff}
t\mapsto t'(t)~,~~~~~x^i\mapsto x'^i(t,{\bf x})\;,
\ee
where $t'(t)$ is a monotonic function. In other words, we are left
with time reparameterizations and time-dependent spatial
diffeomorphisms. 

To formulate the action of HG, the metric is expanded into time and
space components like in the Arnowitt--Deser--Misner (ADM)  
decomposition,
\be
\di s^2 = N^2 \di t^2 - \gamma_{ij} (\di x^i + N^i \di t) (\di x^j +
N^j \di t)\,.
\ee
The lapse $N$, the shift $N^i$ and the spatial metric $\gamma_{ij}$
transform in the usual way under FDiff,
\be
\label{FDiff1}
N\mapsto N\frac{\di t}{\di t'}~,~~~~~
N^i\mapsto \Big(N^j\frac{\d x'^i}{\d x^j}-\frac{\d x'^i}{\d
  t}\Big)\frac{\di t}{\di t'}~,~~~~~
\gamma_{ij}\mapsto \gamma_{kl}\frac{\d x^k}{\d x'^i}\frac{\d x^l}{\d
  x'^j}\;. 
\ee
They are assigned the following dimensions under the scaling
(\ref{scaling}):\footnote{We say that a field $\Phi$ has dimension
  $[\Phi]$ if it transforms under (\ref{scaling}) as $\Phi\mapsto
  b^{[\Phi]}\Phi$.}
\be
\label{dims}
[N]=[\gamma_{ij}]=0~,~~~~~[N^i]=d-1\;.
\ee
The Lagrangian is then built out of all local FDiff-invariant
operators that can be constructed from these fields and have dimension
less or equal $2d$. The latter bound comes from the scaling dimension
of the spacetime integration measure $[\di t\di^d x]=-2d$ and ensures
that the terms in the action corresponding to relevant ($[{\cal
  O}]$<2d) and marginal  ($[{\cal O}]$=2d) operators have non-positive
dimensions. This is a hallmark of power-counting renormalizability and
is necessary, but in general not sufficient, prerequisite of the true
perturbative renormalizability. 

The precise form of the action and its physical content depends on the
assumption about the form of the lapse function $N$. In the {\it
  non-projectable} version of HG $N$ is assumed to be a full-fledged
dynamical field that has dependence both on space and time
coordinates. In this case the Lagrangian can depend on the spatial
gradients of $N$ through the FDiff-covariant vector $a_i=\d_i \ln N$
\cite{Blas:2009qj}  and the number of independent operators in the
Lagrangian is very large (order $O(100)$) \cite{Kimpton:2013zb}. The
action simplifies at low energies where it reduces to General
Relativity (GR) plus a sector describing the dynamics of the preferred
foliation \cite{Blas:2009yd,Blas:2010hb}. The latter sector is stable
and, by appropriately choosing the couplings, its interactions
with gravity and visible matter can be suppressed. In other words,
despite the absence of Lorentz invariance or general covariance as
fundamental principles in HG, it can reproduce the known phenomenology
of GR at the scales at which it has been tested
\cite{Blas:2014aca}. While the parameter space of the theory has been
strongly constrained by the exquisite tests of Lorentz invariance in
the matter sector \cite{Liberati:2013xla} and in gravity
\cite{Monitor:2017mdv}, it still remains phenomenologically viable
\cite{Gumrukcuoglu:2017ijh}. 

However, the question about renormalizability of the non-projectable
HG remains open. Although its Lagrangian is power-counting
renormalizable, the dynamical lapse component of the metric induces an
instantaneous interaction \cite{Blas:2010hb,Blas:2011ni} 
which can potentially lead to
non-local divergences \cite{Barvinsky:2015kil}. Whether such
divergences actually occur 
or not is presently unknown.    
 
In this paper we focus on a different version of HG called {\it
  projectable} whose quantum properties are better understood. In
projectable HG the lapse $N$ is restricted to be a function of time
only, $N=N(t)$. This assumption is compatible with the FDiff
transformations (\ref{FDiff1}). Further, by using the
time-reparameterization, $N$ can be set to any given constant value,
say $N=1$. In this way lapse is completely eliminated from the
model. The action reads \cite{Horava:2009uw},  
\be
\label{Sgen}
S=\frac{1}{2G}\int \di t\di^d x \sqrt{\g}\big(K_{ij}K^{ij}-\l
K^2-{\cal V}\big)\;,
\ee
where
    \begin{align}
    K_{ij}=\frac{1}{2}\left(\dot{\gamma}_{ij}
    -\nabla_i N_j -\nabla_j N_i\right)\;,       \label{K}
    \end{align}
is the extrinsic curvature of the foliation and $K\equiv
K_{ij}\gamma^{ij}$ is its trace. Here dot denotes the time derivative
and $\nabla_i$ is the covariant derivative associated to the spatial
metric $\g_{ij}$; $G$ and $\l$ are dimensionless couplings. 
The potential part ${\cal V}$ does not involve time derivatives and is
constructed out of the curvature tensor corresponding to 
$\g_{ij}$. Its form depends on the number of spatial dimensions. In
$d=3$ it reads \cite{Sotiriou:2009gy},
\be
\label{pot31}
\begin{split}
{\cal V}=&2\Lambda-\eta R+\mu_1 R^2+\mu_2 R_{ij}R^{ij}\\
&+\nu_1 R^3+\nu_2 RR_{ij}R^{ij}+\nu_3R^i_jR^j_kR^k_i +\nu_4 \nabla_i
R\nabla^i R+\nu_5 \nabla_iR_{jk}\nabla^i R^{jk}\;,
\end{split}
\ee
where we have used that in 3 dimensions the Riemann tensor is
expressed in terms of Ricci $R_{ij}$. This expression includes all
relevant and marginal terms that cannot be reduced to each other upon
integration by parts and use of Bianchi identities. It contains 
9 couplings $\Lambda, \eta, \mu_1,\mu_2$ and $\nu_a,
a=1,\ldots,5$.   

The spectrum of perturbations propagated by this action contains a
transverse-traceless (tt) graviton and an additional scalar mode. Both
modes have positive kinetic terms as long as $G$ is positive and $\l$
is either less than $1/3$ or bigger than $1$,
\be
\label{lambdaunitary}
\lambda<1/3~~\text{or}~~\lambda>1\;,
\ee
implying that the theory
admits unitary quantization. Their dispersion relations around a flat
background\footnote{By this we mean the configuration $N^i=0$,
  $\g_{ij}=\delta_{ij}$. It is a solution of equations following from
  (\ref{Sgen}), (\ref{pot31}) provided the cosmological constant
  $\Lambda$ is set to zero.} are \cite{Blas:2010hb,Barvinsky:2015kil} 
\bseq
\label{disp}
\begin{align}
\label{disp1}
&\omega_{tt}^2=\eta k^2+\mu_2 k^4+\nu_5k^6\;,\\
\label{disp2}
&\omega_s^2=\frac{1-\l}{1-3\l} \big(-\eta k^2+(8\mu_1+3\mu_2)k^4\big)+\nu_sk^6\;,
\end{align}
\eseq
where $k$ is the spatial momentum and we have defined
\be
\label{nus}
\nu_s\equiv\frac{(1-\l)(8\nu_4+3\nu_5)}{1-3\l}\;.
\ee
These dispersion relations are problematic at low energies where they
are dominated by the $k^2$-terms. Due to the negative sign in front of
this term, either the scalar mode or the graviton 
behaves as a tachyon at low energies,
implying that flat space is not a stable vacuum of the model. Attempts
to suppress the instability by choosing $\l$ close to $1$ lead to the
loss of perturbative control
\cite{Blas:2009yd,Koyama:2009hc,Blas:2010hb} (see, however,
\cite{Mukohyama:2010xz,Izumi:2011eh,Gumrukcuoglu:2011ef}
for suggestions to restore the control by rearranging the perturbation
theory). Alternatively, the instability can be eliminated by tuning
$\eta=0$ or by expanding around a curved vacuum. In both cases the
theory does not appear to reproduce GR in the low-energy limit, as
there is no regime where the dispersion relation of the $tt$-graviton
would have the relativistic form $\omega_{tt}^2\propto k^2$. This
low-energy problem does not affect the ultraviolet (UV) behavior of
the model: At high momenta both dispersion relations (\ref{disp}) are
perfectly regular, as long as $\nu_5,\nu_s>0$.

Projectable HG has been proven to be perturbatively renormalizable
\cite{Barvinsky:2015kil,Barvinsky:2017zlx} in any number of spacetime
dimensions. Furthermore, for $d=2$ its renormalization group (RG) flow
was computed and shown to possess an asymptotically free UV fixed
point, implying that the theory is UV complete
\cite{Barvinsky:2017kob}. Also in $d=2$ Ref.~\cite{Griffin:2017wvh} 
computed the
renormalization of the cosmological constant. 

Partial results about the RG flow of projectable HG in $d=3$ were
obtained in Ref.~\cite{towards}. The UV behavior of the
theory is parameterized by seven couplings in front of the marginal
operators in (\ref{Sgen}), (\ref{pot31}) $G$, $\l$, $\nu_a$,
$a=1,\ldots,5$. However, not all these couplings have physical
meaning, which is reflected in the dependence of their
$\beta$-functions on the choice of gauge used in the loop
calculations. There are in total six {\it essential} couplings which
enter into the on-shell effective action and whose $\beta$-functions are
thus gauge invariant. These can be chosen as follows
\cite{towards},\footnote{The coupling $u_s$ is related to
  the coupling $\alpha$ used in \cite{towards} as
  $\alpha=u_s^2$.} 
\be
{\cal G} = \frac{G}{\sqrt{\nu_5}},\qquad \lambda,
\qquad u_s=\sqrt{\frac{\nu_s}{\nu_5}},
\qquad v_a = \frac{\nu_a}{\nu_5}, \quad   a=1,2,3, \label{new_couplings}
\ee
where $\nu_s$ is defined in (\ref{nus}). The one-loop $\beta$-function
of $\l$ depends only on the first three of these couplings and reads
\cite{towards}, 
 \begin{eqnarray}
    &&\beta_\lambda=
    \frac{\cal G}{120\pi^2(1-\lambda)(1+u_s)u_s}\,\big[27(1-\lambda)^2
    +3u_s(11-3\lambda)(1-\lambda)
    -2u_s^2(1-3\lambda)^2\big]\;.              \label{betalambda}
\end{eqnarray}
The gauge-dependent $\beta$-function of $G$ (not ${\cal G}$) was also
computed for several gauge choices, the results are summarized in
Appendix~\ref{app:betaG}.   

The purpose of this work is to provide $\beta$-functions of the
remaining couplings in the list (\ref{new_couplings}).

\subsection{Outline of the method and main results }

The calculation in \cite{towards} was done using a
combination of the background field approach followed by diagrammatic
expansion around a special background. The latter was chosen to have
flat spatial metric, $\g_{ij}=\delta_{ij}$, and arbitrary shift vector
$N^i$. This allowed us to avoid the complicated vertices originating
from the potential part of the action and profit from the relatively
simple structure of the kinetic term, which for this background choice
reduces to a linear combination of two quadratic invariants   
$(\partial_i N_j)^2$ and $(\partial_i N_i)^2$. Computing the one-loop
effective action in this background we were able to extract the
$\beta$-functions of $G$ and~$\l$.

However, application of this method to the renormalization of the
couplings $\nu_a$, $a=1,\ldots,5$ is infeasible. It would require
evaluation of a huge number of Feynman diagrams with two and tree
background 
metric insertions in the external legs. Instead, we use the powerful
tools based on Schwinger--DeWitt 
\cite{Schwinger,DeWitt,DeWitt:2003pm,PhysRep,twoloop,Scholarpedia} 
or Gilkey--Seeley \cite{Gilkey-Seeley,Avramidi,Vassilevich} heat kernel
method. They provide an effective resummation of the field
perturbation series and allow one to obtain the UV divergences not as
expansion in powers of field perturbations, but as full nonlinear
counterterms --- local nonlinear functionals of the generic background
field. 
Pioneering application of this method in quantum Einstein
theory \cite{tHooft-Veltman} proved to be very efficient and now
underlies the majority of results on renormalization of
(super)gravitational models. The basic tool of this method is
the heat equation kernel whose proper time expansion coefficients ---
the so-called HAMIDEW \cite{Gibbons} or Gilkey--Seeley coefficients ---
carry a full information about UV divergences and can be
systematically calculated. 

Despite powerful calculational advantages of the heat kernel method,
its application to HG encounters two major difficulties.
This method is most efficient for covariant
operators in which all spacetime derivatives form covariant
d'Alembertians or spatial Laplacians and are treated on equal
footing. Existence of preferred time foliation obviously violates this
property. Several approaches have been put forward to
circumvent this
problem and extend the heat kernel method to 
Lifshitz-type
theories 
\cite{Nesterov-Solodukhin,DOdorico:2014tyh,DOdorico:2015pil,HKLT,
Grosvenor:2021zvq}. 
However, applications to HG models are marred
by an extra difficulty --- {\em non-minimal} operators arising in these
models have higher order derivative terms which are not exhausted by
powers of the spacetime d'Alembertian $\Box\equiv
g^{\mu\nu}\nabla_\mu\nabla_\nu$ or spatial Laplacian 
$\Delta\equiv\gamma^{ij}\nabla_i\nabla_j$. The principal symbol term of
these operators is non-diagonal in derivatives whose indices are
contracted with the tensor field indices. To circumvent this
difficulty we use the technique of the so-called {\em universal
  functional traces} (UFT) applicable to this class of higher-derivative
non-minimal operators. 

Originally this method was developed for spacetime covariant operators
in \cite{PhysRep,twoloop} (see also \cite{JackOsborn}). Universal functional
traces are the coincidence limits of the kernels of nonlocal
(pseudo-differential) operators of the form 
    \begin{align}
    \nabla_{\mu_1}...\nabla_{\mu_m}
    \frac{\hat 1}{\Box^n}\delta(x,y)\,\Big|_{\,y=x}, \label{UFT0}
    \end{align}
which are defined in curved spacetime with a generic metric
$g_{\mu\nu}$ and with the covariant d'Alembertian operator
$\Box$ acting on generic set of fields
$\varphi=\varphi^A(x)$ (the hat symbol denoting the matrix structure
of the operator kernel in the vector space of $\varphi$, $\hat
1\varphi=\delta^A_B\varphi^B$, etc.). For appropriate values of
parameters $m$ and $n$ these coincidence limits are UV divergent and
contain all the information about one-loop UV divergences of the
theory. 

The latter property can easily be demonstrated on the example, say, of
the higher derivative theory with the inverse propagator
$\Box^N+P(\nabla)$, where $P(\nabla)$ is its lower derivative part. By
expanding the one-loop functional determinant ${\rm
  Tr}\,\ln\big(\Box^N+P(\nabla)\big)=N\,{\rm Tr}\,\ln\Box+{\rm
  Tr}\,\ln\big(1+P(\nabla)/\Box^N\big)$ in powers of the nonlocal term
$P(\nabla)/\Box^N$ and commuting to the left the powers of $P(\nabla)$
and to the right -- the inverse powers of $\Box$, one finds that the
result will be given by the infinite series of terms (\ref{UFT0})
multiplied by tensors of ever growing dimensionality. Only a finite
number of these terms will be UV divergent, so that the overall
one-loop divergent part will be known, provided one can calculate
divergent parts of separate universal functional traces
(\ref{UFT0}). One can easily and systematically do this by the
Schwinger-DeWitt heat kernel technique, just like it can be done for
the first term $N\,{\rm Tr}\,\ln\Box$ of the expansion. This
calculation is equally possible for any spacetime dimension and,
moreover, can be extended to noninteger values of $n$ in
(\ref{UFT0}). As shown in \cite{PhysRep}, similar technique applies to
non-minimal operators. 

This method admits the generalization to Lorentz violating Lifshitz
theories with regular propagators (the class of HG models
in which the proof of their perturbative renormalizability holds
\cite{Barvinsky:2015kil}) --- see, for instance, \cite{Saueressig}.
Fortunately,
however, this generalization is not really needed in our case.
Due to the properties of the projectable HG, the
renormalization of its potential term can be done via a special
3-dimensional reduction, upon which the one-loop effective action is
represented as the trace of the square root of a certain 6th order
operator fully covariant in the 3-dimensional space. The operator is
non-minimal and bringing its square root to the form suitable for the
application of the UFT technique presents a major computational
challenge. We overcome it by the use of symbolic computer algebra. As
a result, we obtain the one-loop effective action as a sum of UFTs
(\ref{UFT0}) with half-integer $n$ that are fully covariant in the
3-dimensional space and, therefore, fully manageable by the technique of
\cite{PhysRep}. 

The divergent part of the one-loop background effective action
provides us with the renormalized coupling constants $\nu_a$,
$a=1,\ldots,5$, and allows us to determine the $\beta$-functions of
${\cal G}$ and the other four essential couplings 
collectively denoted by $\chi = (u_s,v_1,v_2,v_3,)$. The corresponding
expressions have the form,
\bseq
\label{betas_new}
 \begin{align}
    &\beta_{\cal G}  = \frac{ {\cal G}^2}{26880\pi^2(1-\lambda)^2(1-3\lambda)^2
    (1+u_s)^3 u_s^3} \sum_{n=0}^7 u_s^n\, {\cal P}^{\cal
    G}_n[\l,v_1,v_2,v_3],                    \label{beta_calG} \\
    &\beta_\chi  = A_\chi \frac{ {\cal G}}{26880\pi^2(1-\lambda)^3(1-3\lambda)^3
    (1+u_s)^3 u_s^5} \sum_{n=0}^9 u_s^n\, {\cal
    P}^{\chi}_n[\l,v_1,v_2,v_3],             \label{beta_chi}
\end{align}
\eseq
where the prefactor coefficients $A_\chi=(A_{u_s},A_{v_1},A_{v_2},A_{v_3})$ equal
    \be
   A_{u_s} = u_s (1-\lambda),\quad A_{v_1} = 1,\quad A_{v_2} =
   A_{v_3} = 2. 
   \label{Achi}
    \ee
Note that the coupling ${\cal G}$ factorizes and its powers enter the
$\beta$-functions only as overall coefficients. The functions 
${\cal
  P}^{\cal G}_n[\l,v_1,v_2,v_3,]$, 
${\cal
  P}^{\chi}_n[\l,v_1,v_2,v_3,]$ are polynomials in $\l$ and $v_a$,
$a=1,2,3$, with integer coefficients.
${\cal P}^{\cal G}_n$, ${\cal P}^{u_s}_n$ and ${\cal
  P}^{v_a}_n$ are respectively of fourth, fifth and sixth order in
$\lambda$. The maximum overall power of the couplings $v_a$ is two
for ${\cal P}^{\cal G}_n$, ${\cal P}^{u_s}_n$ and three for  ${\cal
  P}^{v_a}_n$.
Explicit expressions for these polynomials
 are lengthy and
are collected in Appendix~\ref{betaexpl}. 

Equations (\ref{betas_new}), (\ref{Achi}) and
(\ref{Pol_calG})--(\ref{Pol_v3}) represent the main results of this
paper. Together with 
the $\beta$-function of $\lambda$, Eq.~(\ref{betalambda}), they
comprise the full set of $\beta$-functions for essential coupling
constants of projectable HG in $d=3$. The files containing the
$\beta$-functions in the Mathematica \cite{Mathematica} format are
available as Supplemental Material. \\

In the rest of the paper we describe in detail our calculation and
various checks, to which we subject our results. In Sec.~\ref{sec:2} we
discuss the gauge-fixing procedure the choice of the background and
get the general expression for the one-loop effective action. In
Sec.~\ref{Sect4} we perform the reduction to 3 dimensions, reformulate the
problem as extraction of an operator square root, and describe
an algorithm to perform this extraction. In Sec.~\ref{UFT} we review
the UFT technique and classify the UFTs required for our calculation. In
Sec.~\ref{btfuns} we extract the $\beta$-functions of the essential
couplings and discuss their gauge invariance. In Sec.~\ref{spherecheck}
we perform and independent check of 
our results by computing the one-loop effective action on a
sphere with a different method based on spectral
decomposition. In Sec.~\ref{concl} we discuss our results and 
make several preliminary observations about the structure of the
$\beta$-functions and the corresponding RG flow.
Some lengthy formulas and technical details are relegated to Appendices.

\section{Gauge fixing and one-loop effective action}
\label{sec:2}

\subsection{The choice of the background and background covariant
  gauge fixing}\label{gaugefix}

We focus on the part of the action consisting of the marginal
operators with respect to the scaling (\ref{scaling}). They form a
closed set under renormalization and determine the UV behavior of the
theory. From now on we switch to the imaginary ``Euclidean'' time
$\tau=it$. In this ``signature'' the tree-level action reads,
\begin{equation}
\begin{aligned}
S= \frac{1}{2G} \int d\tau\, d^3x \,\sqrt{\gamma}\,(K_{ij}K^{ij}-\lambda K^2 + \nu_1 R^3+\nu_2 R R_{ij}R^{ij}
+\nu_3 R^i_j R^j_k R^k_i + \nu_4 \nabla_i R \nabla^i R
+ \nu_5 \nabla_i R_{jk} \nabla^i R^{jk} ).           \label{action31}
\end{aligned}
\end{equation}
Renormalization of the theory implies the calculation of the UV
divergent part of the effective action, which has the same covariant
structure as the classical tree-level action (\ref{action31}) provided
one works in the class of the so-called background covariant gauges
\cite{DeWitt} which were discussed in context of HG in
\cite{Barvinsky:2015kil,Barvinsky:2017zlx,towards}. For the
renormalization of the potential part of the action it is, therefore,
sufficient to consider the metric background on which all its five
tensor structures are nonvanishing and can be distinctly
separated. This is the spacetime metric with a generic static
3-dimensional part $g_{ij}({\bf x})$, and vanishing shift functions
$N^i=0$. Static nature of $g_{ij}$ and zero shift functions lead to 
zero kinetic term of (\ref{action31}) whose contribution is not needed
for the renormalization of couplings $\nu_1,...\nu_5$.\footnote{This
  is a strategy opposite to that of \cite{towards}, where the
  renormalization of the kinetic term with its couplings $G$ and
  $\lambda$ was performed on a 3-dimensional flat space background
  with nontrivial $N^i({\bf x})$ treated by perturbations in external
  lines of Feynman diagrams.} 

Thus we perform the split of the full set of fields into this
background and quantum fluctuations $h_{ij}(\tau,{\bf x})$ and
$n^i(\tau,{\bf x})$, 
\be
\gamma_{ij}(\tau,{\bf x}) = g_{ij}({\bf x}) + h_{ij}(\tau,{\bf x}),
\quad N^i(\tau,{\bf x}) = 0 + n^i(\tau,{\bf x})\,,           \label{split}
\ee
retain the relevant quadratic part of the full action on this
background and take the resulting Gaussian path integral over the
fluctuations. We
begin this procedure by considering first a special gauge breaking
part of the action which preserves gauge invariance of the
counterterms and is compatible with the anisotropic scaling
(\ref{scaling}) 
\cite{Barvinsky:2015kil,Barvinsky:2017zlx}. 

The gauge-fixing action is chosen as
\begin{equation}\label{LgfG}
S_{\rm gf} =\frac{\sigma}{2G}\int d\tau\, d^3x\,\sqrt{g}\, F^i {\cal O}_{ij} F^j,
\end{equation} 
which is the quadratic form in the gauge-condition functions $F^i$
with the kernel ${\cal O}_{ij}$, both parametrically depending on
the background fields in the way that this action is invariant under
the simultaneous diffeomorphisms of both the full field (\ref{split})
and the metric background --- the so-called {\em background gauge
  transformations}. Notice that under these background diffeomorphisms
the quantum fields $n^i$ and $h_{ij}$ transform as a vector and a
rank-two tensor, respectively. 
For a static $g_{ij}$ and vanishing background
shifts the gauge conditions and the gauge-fixing
matrix take the form
\bseq\label{OFgf}
\begin{align}
F^i &= \dot{n}^i + \frac1{2\sigma} {\cal O}^{-1}_{ij}\big(\nabla_k h_{jk}
- \lambda\nabla_j h\big),\\
{\cal O}_{ij} &=
\left( g^{ij}\Delta^2 + \xi \nabla^i \Delta\nabla^j \right)^{-1}.
\end{align}
\eseq
Here and below the covariant derivatives are defined using the
background metric $g_{ij}$.
The gauge functions $F^i$ are local linear combinations of quantum
fields $h_{ij}$ and $n^i$ with operator coefficients. 
The gauge-fixing matrix ${\cal
  O}_{ij}$ is the nonlocal Green's function of the covariant
fourth-order differential operator ${\cal O}^{-1}_{ij}=g^{ij}\Delta^2
+ \xi \nabla^i \Delta\nabla^j$.  
This non-locality can be resolved by integrating in an auxiliary field
and does not spoil the locality of
counterterms~\cite{Barvinsky:2015kil}. 
Under background gauge transformations $F^i$ and ${\cal O}_{ij}$
transform respectively as the vector and the second rank tensor, so
that the gauge-breaking action (\ref{LgfG}) is indeed invariant and
provides explicit gauge invariance of quantum counterterms
\cite{Barvinsky:2017zlx}.\footnote{Strictly speaking, for $F^i$ to be
  a vector under background transformations $\dot n^i$ should be
  replaced by the covariant time derivative $D_\tau n^i=\dot
  n^i-\bar N^k\partial_k n^i+\partial_k\bar N^i n^k$, where $\bar N^i$
  is the background shift function \cite{Barvinsky:2015kil}, but on
  our background $\bar N^i=0$.} Further, this gauge fixing leads to a
homogeneous fall-off of all field propagators in UV and ensures that
all counterterms are compatible with the naive power-counting
arguments~\cite{Barvinsky:2015kil}. 

The gauge conditions (\ref{OFgf}) are parameterized by two constants
$\xi$ and $\sigma$. The gauge-breaking action (\ref{LgfG}) reads
explicitly, 
\begin{equation}\label{gaugebr}
\begin{aligned}
S_{\rm gf} =
\frac{1}{2G}\int d\tau\, d^3x\sqrt{g}
\bigg(& \sigma\dot{n}^i {\cal O}_{ij} \dot{n}^j 
+ \dot{n}^i \big(\nabla^j h_{ij} - \lambda \nabla_i h\big)\\
&+ \frac{1}{4\sigma}\nabla^k h_{ik}\, {\cal O}^{-1}_{ij}\nabla^l h_{jl}
-\frac{\lambda}{2\sigma} \nabla_i h \, {\cal O}^{-1}_{ij}\nabla^k h_{jk}
+\frac{\lambda^2}{4\sigma}\nabla_i h\, {\cal O}^{-1}_{ij}\nabla_j h
\, \bigg)\;.
\end{aligned}
\end{equation}
The distinguished nature of this two-parameter family
is that the cross-terms between $n^i$ and $h_{ij}$ in $S_{\rm gf}$
completely cancel the analogous terms in the kinetic part of the
classical action at the quadratic order,
\be
\begin{split}
\label{Lkin}
S_{\rm kin}=\frac{1}{2G}\int d\tau\,d^3x\,\sqrt{\gamma} \big(&K_{ij}K^{ij}-\l K^2)
=
\frac{1}{2G}\int d\tau\,d^3x\,\sqrt{g} \bigg[-\frac{1}{4}h^{ij}\ddot
h_{ij}+\frac{\lambda}{4}h\ddot h \\
&-\dot n^i\,\big(\nabla^j h_{ij}-\l
\nabla_i h\big)-\frac{1}{2} n_i\D n^i-\bigg(\frac{1}{2}-\l\bigg)n^i\nabla_i\nabla_j
n^j
-\frac{1}{2}n^i R_{ij}n^j\bigg]+...\;,
\end{split}
\ee
where dots mean higher order terms of the expansion. In this
expression we have integrated by parts both in space and time
and used the staticity of the background 3-metric,
$\dot g_{ij}=0$. 
The cancellation of the cross-terms between $n^i$ and $h_{ij}$ implies
that the shift and metric sectors can be treated separately.

\subsection{Shift and ghost parts of the action}

From the sum of the kinetic action (\ref{Lkin}) and the gauge-breaking
term (\ref{gaugebr}) we obtain the 
quadratic in $n^i$ part of the gauge-fixed action,
\begin{equation}
\label{shift_part}
\begin{aligned}
S_n&=\frac{1}{2G}\int d\tau d^3x \sqrt{g}\,n^i \left[-\sigma {\cal
    O}_{ij} \partial_\tau^2 +  \lambda
  \nabla_i\nabla_j-\frac12\nabla_j\nabla_i -\frac12 g_{ij}\Delta
\right] n^j\\ 
&=\frac{\sigma}{2G}\int d\tau d^3x \sqrt{g}\,n^i {\cal O}_{ij}\left[ -
  \delta^j_k\partial_\tau^2 +  \mathbb{B}^j_{~k}(\nabla)\right] n^k, 
\end{aligned}
\end{equation}
where the differential operator $\mathbb{B}^i_{~j}(\nabla)$ in spatial derivatives reads
\be
\begin{aligned}
\mathbb{B}^i_{~j}(\nabla) =-\frac1{2\sigma}\delta^i_j\Delta^3
-\frac1{2\sigma}\Delta^2\nabla_j\nabla^i
-\frac{\xi}{2\sigma}\nabla^i\Delta\nabla^k\nabla_j\nabla_k
-\frac{\xi}{2\sigma}\nabla^i\Delta\nabla_j\Delta
 +\frac{\lambda}{\sigma}\Delta^2\nabla^i\nabla_j
 + \frac{\lambda\xi}{\sigma}\nabla^i\Delta^2\nabla_j.
\end{aligned}                                     \label{B}
\ee

It is quite remarkable that the chosen two-parameter family of gauge
conditions on a static background provides another very useful
property --- modulo the multiplication by the gauge fixing matrix the
corresponding Faddeev-Popov ghost operator coincides with the operator
in the shift action (\ref{shift_part}). Indeed, the action of
the ghost fields $c^i$ and $\bar c_j$ reads 
\begin{equation}\label{ghact}
S_{\rm gh} = -\frac1G \int d\tau d^3x\sqrt{g}\, \bar{c}_i({\bf s}F^i),
\end{equation}
where ${\bf s}F^i$ is the BRST
transform of the gauge conditions. This is computed using
the BRST transformations of the quantum fields $h_{ij}$ and
$n^i$ which coincide with the infinitesimal diffeomorphisms of the full
fields $\gamma_{ij}$ and $N^i$ with the gauge parameter replaced by
the Grassmann ghost $c^i$, 
\bseq\label{metrictr}
\begin{align}
{\bf s}h_{ij} &= \nabla_i c_j + \nabla_j c_i + h_{ik}\nabla_j c^k
+ h_{jk}\nabla_i c^k + c^k \nabla_k h_{ij}, \quad c_i=g_{ij}c^j,\\
{\bf s}n^i &= \dot{c}^i - n^j \nabla_j c^i + c^j \nabla_j n^i. \label{shifttr}
\end{align}
\eseq
After the substitution of (\ref{metrictr}) into (\ref{ghact}), the
ghost action in the quadratic order of all quantum fields takes the
following form (bearing in mind zero background values of ghosts) 
\begin{equation}
\begin{aligned}
S_{\rm gh} &=
\frac1G \int d\tau d^3x\sqrt{g}\,
\bar{c}_i \left(-\delta^i_j\p_\tau^2 + \mathbb{B}_{~j}^i (\nabla)\right)c^j,
\end{aligned}
\end{equation}
where the operator $\mathbb{B}_{~j}^i$ exactly coincides with that of
(\ref{shift_part}). This property is an artifact of the special choice
of the gauge fixing action and the static nature of the metric
background, and it significantly simplifies further calculations,
because the contributions of ghosts and shift functions are expressed
through the functional determinant of one and the same operator.

\subsection{Metric part of the action}\label{expansion}

The kinetic part for the metric perturbations has the form
\begin{eqnarray}
-\frac{\sqrt{g}}{2G}\,h^A  \mathbb{G}_{AB}\d_\tau^2 h^B\;,
\end{eqnarray}
where we introduced a collective notation for the indices of an
symmetric rank-2 tensor, $h^A\equiv h_{ij}$. The DeWitt metric
in the space of such tensors and its inverse read,
\be
\label{DeWittG}
\mathbb{G}^{ij,kl}=\frac{1}{8}(g^{ik}g^{jl}+g^{il}g^{jk})
-\frac{\l}{4}g^{ij}g^{kl}\;,~~~~~~
\mathbb{G}^{-1}_{ij,kl}=2(g_{ik}g_{jl}+g_{il}g_{jk})
+\frac{4\l}{1-3\l}g_{ij}g_{kl}\;.
\ee

The part of the quadratic action with space derivatives of the metric
is too lengthy
(contains hundred of terms) to be written
explicitly. We have obtained it using the tensor computer albegra 
package {\it xAct}
\cite{xAct,xPerm,Brizuela:2008ra,Nutma:2013zea} 
for Mathematica \cite{Mathematica}. 
Schematically, it has the form,
\be
\label{Lpotgen}
{\cal L}_{\rm pot,\,hh}+{\cal L}_{{\rm gf},\,hh}=\frac{\sqrt{g}}{2G}\;
h^A \mathbb{D}_{AB} h^B, 
\ee
where $\mathbb{D}_{AB}$ is a purely 3-dimensional differential
operator of 6th order. Note that in the indices $A,B$ it is
not an operator, but rather a quadratic form. In flat background (\ref{Lpotgen})
reduces to terms with exactly 6 derivatives \cite{towards},
\be
\label{Lpotflat}
\begin{split}
{\cal L}_{{\rm pot},\,hh}+{\cal L}_{gf,\,hh}=\frac{1}{2G}
\bigg[&-\frac{\nu_5}{4}h^{ij}\D^3h_{ij}
+\bigg(\frac{\nu_5}{2}-\frac{1}{4\sigma}\bigg)
h^{ik}\D^2\d_i\d_j h^{jk}
+\bigg(-\nu_4-\frac{\nu_5}{2}-\frac{\xi}{4\sigma}\bigg)
h^{ij}\D\d_i\d_j\d_k\d_l h^{kl}\\
&+\bigg(2\nu_4+\frac{\nu_5}{2}
+\frac{\l(1+\xi)}{2\s}\bigg)
h\D^2\d_k\d_l h^{kl}
+\bigg(-\nu_4-\frac{\nu_5}{4}-\frac{\l^2(1+\xi)}{4\s}\bigg)h\D^3h\bigg]\;.
\end{split}
\ee

\subsection{Total one-loop action}

The one-loop effective action is given by the Gaussian path integral
\begin{equation}
\exp\left(-\varGamma^{\rm 1-loop}\right)=\sqrt{{\rm Det}\, {\cal O}_{ij}}\int \left[\,dh^A\,dn^i\, dc^i\, d\bar{c}_j \right]\, \exp\big(-S^{(2)}[\,h^A,n^i,c^i,c_j\,]\,\big),       \label{PI}
\end{equation}
where the quadratic part of the full action consists of three
contributions --- metric, shift vector and ghost ones,
\be
\begin{aligned}
S^{(2)}[\,h^A,n^i,c^i,\bar c_j\,] = \frac{1}{G}\int d\tau d^3x \sqrt{g}\bigg[&\,
\frac12 h^A\left(-\mathbb{G}_{AB}\partial_\tau^2 + \mathbb{D}_{AB}
\right)h^B \\ 
&+\frac12\sigma n^i {\cal
    O}_{ik}\left(-\delta^k_j\partial_\tau^2 + \mathbb{B}^k_{~j}\right)n^j 
+ \bar{c}_i \left(-\delta^i_j\partial_\tau^2 + \mathbb{B}^i_{~j} 
\right)c^j\bigg].
\end{aligned}
\ee
The normalization factor $\sqrt{{\rm Det}\, {\cal O}_{ij}}$ comes from
the smearing of the gauge-fixing conditions with a Gaussian weight
which leads to the gauge-breaking term (\ref{LgfG})
\cite{Barvinsky:2015kil}. 
The result of the integration
\begin{equation}
\exp\left(-\varGamma^{\rm 1-loop}\right) = \sqrt{{\rm Det}\,{\cal O}_{ij}}
\frac{{\rm Det} \left(-\delta^i_j\partial_\tau^2 + \mathbb{B}^i_{~j}
  \right)}{\sqrt{{\rm Det}\left(-\mathbb{G}_{AB}\partial_\tau^2 +
      \mathbb{D}_{AB} \right)} 
\sqrt{{\rm Det}\left[{\cal O}_{ik}\left(-\delta^k_j\partial_\tau^2 +
      \mathbb{B}^k_{~j}\right)\right]}}. 
\end{equation}
immediately shows that the contribution of the operator ${\cal
  O}_{ij}$ cancels out, while the shift and ghost parts reduce to the
contribution of a single functional determinant. Factoring out and
disregarding the ultralocal determinant of the DeWitt
metric,\footnote{Which is actually cancelled by
  the local measure $\sqrt{{\rm Det}\,\mathbb{G}_{AB}\delta(x,x')}$
  arising in the Lagrangian path integral after the transition from the
  canonical one \cite{Henneaux}.} 
we write the effective action as the sum of two parts,  
\be
\label{Slog}
\varGamma^{\rm 1-loop} = \frac{1}{2}\Tr
\ln(-\delta^{A}_{B}\d_\tau^2+{\mathbb{D}}^A_{\;B})- 
\frac12{\rm Tr}\ln \left(-\delta^i_j\partial_\tau^2 + \mathbb{B}^i_{~j}\right)\;,
\ee
where 
\be
\label{Dtilde}
{\mathbb{D}}_{~B}^{A}=(\mathbb{G}^{-1})^{AC} \mathbb{D}_{CB}
\ee
is now an {\em operator} in the condensed indices $A$ and $B$. 
We presently turn to the computation of the functional traces entering
in (\ref{Slog}).

\section{3D reduction --- one-loop effective action as the trace of an
  operator square root }
\label{Sect4}

We begin by using the proper-time representation for the 
trace of the logarithm of an operator 
\be
{\rm Tr}\,\ln \mathbb{F}=
-\int_0^\infty\frac{ds_6}{s_6}\,
\Tr\, \e^{-s_6(-\d^2_\tau+\mathbb{F})},        \label{tracelog}
\ee
where $\mathbb{F}$ is either $\mathbb{D}^A_{~B}$ or
$\mathbb{B}^i_{~j}$. 
The subscript of the parameter $s_6$ emphasizes 
its scaling dimension, $[\,s_6]=-6$, which provides that the exponent
is dimensionless (recall that the dimension of the operator
$-\d_\tau^2+\mathbb{F}$ is $6$). 
Thus, we obtain the expression for the metric tensor part of the
effective action, 
\be
\label{Ssqrt0}
\begin{split}
\varGamma^{\rm 1-loop}_{\rm metric} 
=&-\frac{1}{2}\int_0^\infty\frac{ds_6}{s_6}\,\Tr\,
\e^{-s_6(-\delta^A_{B}\d_\tau^2+ 
{\mathbb{D}}^A_{\;B})}\\
=&-\frac{1}{2}\int d\tau\, d^3x\,\int
\frac{ds_6}{s_6}\,{\rm tr}\,\e^{-s_6(-\delta^A_{B}\d_\tau^2+
{\mathbb{D}}^A_{\;B})}
\delta(\tau-\tau')\,\delta({\bf x}-{\bf
  x}')\,\Big|_{\,\tau=\tau',\,{\bf x}={\bf x}'}\,,
\end{split}
\ee
where the operator acts on the first arguments of the
$\delta$-functions before taking the coincidence limit 
and ``tr'' stands 
for the simple trace over matrix indices $A=(ij)$.
For a static ($\tau$-independent) background it can be transformed by
the following chain of relations 
\begin{align}
\label{Ssqrt}
\varGamma^{\rm 1-loop}_{\rm metric}
=&-\frac{1}{2}\int d\tau\, d^3x\int \frac{ds_6}{s_6}\,{\rm
  tr}\,\e^{-s_6(-\delta^A_{B}\d_\tau^2+
{\mathbb{D}}^A_{\;B})}\int\frac{d\omega}{2\pi}\e^{i\omega(\tau-\tau')}
\delta({\bf x}-{\bf x}')\,\Big|_{\,\tau-\tau',\,{\bf x}={\bf x}'}\notag\\
=&-\frac{1}{2}\int d\tau\, d^3x\int \frac{ds_6}{s_6}\,\int\frac{d\omega}{2\pi}
\,\e^{-s_6\omega^2}{\rm tr}\,\e^{-s_6{\mathbb{D}}^A_{\;B}}
\delta({\bf x}-{\bf x}')\,\Big|_{\,{\bf x}={\bf x}'}\notag\\
=&-\frac{1}{4\sqrt{\pi}}\int d\tau\,d^3x\int \frac{ds_6}{s_6^{3/2}}\,
{\rm tr}\,\e^{-s_6{\mathbb{D}}^A_{\;B}}
\delta({\bf x}-{\bf x}')\,\Big|_{\,{\bf x}={\bf x}'}\notag\\
=&-\frac{\Gamma(-1/2)}{4\sqrt{\pi}}\int d\tau\,d^3x\,
{\rm tr}\,\sqrt{{\mathbb{D}}^A_{\;B}}
\delta({\bf x}-{\bf x}')\,\Big|_{\,{\bf x}={\bf x}'}\notag\\
=&\frac{1}{2}\int d\tau\;  \Tr_3\sqrt{{\mathbb{D}}^A_{\;B}}\;.
\end{align}
Note that $\Tr_3$ in the final formula is the
functional trace in the 3-dimensional sense (contrary to the
four-dimensional one $\Tr\equiv\Tr_4$). Thus, we conclude that the
calculation of the 1-loop effective action boils down to the 
calculation of the functional trace of the square root of
${\mathbb{D}}^A_{\;B}$, which we denote by 
$\mathbb{Q_{\,D}}^A_{\;B}\equiv\sqrt{{\mathbb{D}}^A_{\;B}}$. This is a
purely 3-dimensional problem. Analogous procedure can be carried out
for the 
 vector part of the trace. Introducing the notation
 $\mathbb{Q_{\,B}}^i_{~j}\equiv \sqrt{\mathbb{B}^i_{~j}}$ the full
 one-loop action can be expressed as 
 \begin{equation}
\varGamma^{\rm 1-loop} =\frac12\int d\tau
\left[{\rm Tr}_3\,\mathbb{Q_{\,D}}^A_{~B} - {\rm
    Tr}_3\,\mathbb{Q_{\,B}}^i_{~j}   \right].            \label{root} 
\end{equation}

Let us outline the strategy for evaluation of the above operator
traces. By commuting the covariant derivatives contracted with each
other to the right and collapsing them into powers of the Laplacian
the local operators  
$\mathbb{F}=({\mathbb{D}},\mathbb{B})$ 
can be brought into the following schematic form:
    \begin{eqnarray}
    \mathbb{F}=\sum_{a=0}^6 {\cal R}_{(a)}\!
    \sum_{6\geq 2k\geq a}\alpha_{a,k}\nabla_{1}...\nabla_{2k-a}
    (-\Delta)^{3-k}\;.
    \end{eqnarray}
Here ${\cal R}_{(a)}$ are background field tensors built of the
curvature and its derivatives of the following dimensionality in units
of inverse length, 
    \begin{eqnarray}
    {\cal R}_{(a)}=O\Big(\,\frac1{l^a}\,\Big).   \label{dimR}
    \end{eqnarray}
We will refer to them as ``coefficient functions''. 
On the other hand, 
$\alpha_{a,k}$ are dimensionless scalar coefficients depending on
the couplings $\l$, $\nu_1,\ldots,\nu_5$. Overall powers of
derivatives and Laplacians are related to the dimensionality of the
coefficient functions to maintain the total dimensionality of
$\mathbb{F}$ which is six. 

The square root of such operators can be obtained by the perturbation
theory in powers of the background curvature and the derivatives of
this curvature, that is again in powers of $1/l$. However, in contrast
to $\mathbb{F}$ this is not a finite length polynomial, but rather a
nonlocal {\em pseudo-differential} operator given by an infinite
series in ${\cal R}_{(a)}$, 
    \begin{eqnarray}
    \sqrt{\mathbb{F}}
    =\sum\limits_{a=0}^\infty
    {\cal R}_{(a)}\sum\limits_{k\geq a/2}^{K_a}\tilde\alpha_{a,k}
    \nabla_{1}...\nabla_{2k-a}
    \frac1{(-\Delta)^{k-3/2}}\;,     \label{root_structure}
    \end{eqnarray}
with some other coefficients $\tilde\alpha_{a,k}$ obtained from
$\alpha_{a,k}$ above. At each dimensionality $a$ the powers of
derivatives are bounded from above by some finite number $2K_a-a$.
Indeed, the number of free tensor indices $K_{\rm free}$ of the
operator is fixed by the nature of the space it acts on: $K_{\rm
  free}=2$ for $\sqrt\mathbb{B}$ and $K_{\rm
  free}=4$ for $\sqrt\mathbb{D}$. All the indices of the derivatives
that are not contracted with each other, minus the number of free
indices, must be contracted with the indices of ${\cal R}_{(a)}$. The
latter is bounded by $a$. Thus, we have
\be
\label{indexbound}
2k-a\leq K_{\rm free}+a\;.
\ee

Altogether this
means that the UV divergent part of (\ref{root}) follows
from the calculation of UFTs of the form 
    \begin{eqnarray}
\label{neededUFTs}
    \int d^3x\,
    {\cal R}_{(a)}({\bf x})
    \nabla_{1}...\nabla_{2k-a}
    \frac1{(-\Delta)^{k-3/2}}\delta({\bf x},{\bf x}')\,\Big|_{\,{\bf
        x}={\bf x}'}.
    \end{eqnarray}
Since the divergences of HG have at maximum the dimensionality $a=6$,
only finite number of such traces will be needed. This method splits
the problem into two steps --- calculation of the operator square root
(\ref{root_structure}) and the evaluation of UFTs (\ref{neededUFTs}) ---
which makes it computationally efficient. 

The first step is the perturbative calculation of the square root
(\ref{root_structure}). This calculation is based on the fact that in
the lowest order approximation of the expansion in curvature the covariant
derivatives commute. Thus, the procedure reduces to the extraction of
the square root from a c-number matrix --- the principal symbol of the
operator, obtained by replacing the covariant derivatives with c-number
momenta and neglecting all terms proportional to curvature. 
Going back in the resulting matrix from these momenta to
covariant derivatives one gets the operator $\mathbb{Q}^{(0)}$. By
denoting all curvature corrections in $\sqrt{\mathbb{F}}$ as
$\mathbb{X}$, 
\be
\label{Qaeq0}
\sqrt{\mathbb{F}}=\mathbb{Q}^{(0)}+\mathbb{X}
\ee
one obtains the equation for this correction term
    \be
    \mathbb{Q}^{(0)}\mathbb{X}+\mathbb{X}\,\mathbb{Q}^{(0)}
    =\mathbb{F}-\big(\mathbb{Q}^{(0)}\big)^2-
    \mathbb{X}^2.                        \label{Xeq}
    \ee
This nonlinear equation can be solved by iterations because its right
hand side is at least linear in curvature. Indeed, the difference
$\mathbb{F}-\big(\mathbb{Q}^{(0)}\big)^2\propto R$ is nonzero entirely
due to the commutation of covariant derivatives, proportional to the
Riemann tensor $R$. At each stage of this iteration procedure one has to
to go from the operator $\mathbb{X}$ to its c-number symbol. Then one 
finds this symbol from the matrix equation (\ref{Xeq}) in which the
right hand side is known with a needed accuracy from the previous
iteration stages. This is the so called Sylvester equation,  
its solution will be constructed below.
In the meantime we focus on the square root of the principal symbol of
$\mathbb{F}$.

\subsection{Square root of the principal symbol and four gauge
  choices}
\label{sec:3.1}
\subsubsection{Tensor sector}

From Eq.~(\ref{Lpotflat}) we read off the quadratic form
${\mathbb{D}}^{mn,kl}$ in flat space. Replacing the derivatives with the
momenta, $\d_i\mapsto ip_i$, and contracting with the inverse DeWitt
metric $\mathbb{G}^{-1}_{ij,mn}$ we obtain the principal symbol of the
operator $\mathbb{D}$,  
\be
\label{Dtildeflat}
\begin{split}
{\mathbb{D}}({\bf p})_{ij}^{\;\;kl}=p^6\bigg[&
\frac{\nu_5}{2}(\delta_i^{k}\delta_j^{l}+\delta_i^{l}\delta_j^{k})
+\frac{4\nu_4(1-\l)+\nu_5}{1-3\l}\delta_{ij}\delta^{kl}
+\bigg(-\frac{\nu_5}{2}+\frac{1}{4\s}\bigg)
\big(\delta_i^{k}\hat p_j\hat p^l+
\delta_i^{l}\hat p_j\hat p^k+
\delta_j^{k}\hat p_i\hat p^l+
\delta_j^{l}\hat p_i\hat p^k\big)\\
&+\bigg(\frac{-4\nu_4(1-\l)-\nu_5}{1-3\l}\bigg)\delta_{ij}\hat p^k\hat p^l
+\bigg(\!\!-\!4\nu_4\!-\!\nu_5\!-\!\frac{\l(1+\xi)}{\s}\bigg) \hat p_i \hat
p_j\delta^{kl}
+\bigg(\!4\nu_4\!+\!2\nu_5\!+\!\frac{\xi}{\s}\bigg)\hat p_i\hat p_j\hat p^k \hat p^l\bigg]\,,
\end{split}
\ee
where $\hat{\bf p}={\bf p}/p$ is the unit vector along the momentum.
This is a $6\times 6$ matrix acting in the space of symmetric tensors
$h_{kl}$. To extract its square root, we need to find its eigenvalues
and eigenvectors.
We do it by 
decomposing $h_{kl}$ into a transverse-traceless, vector and scalar
parts. Namely , we write,
\be
\label{decomp}
h_{kl}=T_{(r)} e^{(r)}_{kl}
+V_{(r)}\frac{1}{\sqrt{2}}(e_k^{(r)}\hat p_l+\hat p_k e_l^{(r)})
+\phi \frac{1}{\sqrt{2}}(\delta_{kl}-\hat p_k\hat p_l) +\psi \hat p_k
\hat p_l\;,
\ee
where $e^{(r)}_k,\;r=1,2$ form the basis of unit vectors orthogonal to
$\hat{\bf p}$, $e^{(r)}_{kl},\;r=1,2$ are the two transverse traceless
polarization tensors, and $T_{(r)}$, $V_{(r)}$, $\phi$, $\psi$ are
coefficients. It is straightforward to see that $e^{(r)}_{kl}$ are
eigenvectors of (\ref{Dtildeflat}) with the eigenvalue
$\kappa_{T}=\nu_5p^6$, whereas the vector polarizations $(e_k^{(r)}\hat
p_l+\hat p_k e_l^{(r)})/\sqrt{2}$ are eigenvectors with the eigenvalue
$\kappa_V=p^6/2\s$. The projectors on the corresponding subspaces are,
\bseq
\label{ProjTV}
\begin{align}
\mathbb{P}_{ij}^{(T)\;kl}\equiv&\sum_{r=1,2}e^{(r)}_{ij}e^{(r)\;kl}=
\frac{1}{2}(\delta_i^{k}\delta_j^{l}+\delta_i^{l}\delta_j^{k})
-\frac{1}{2}\delta_{ij}\delta^{kl}\notag\\
&-\frac{1}{2}\big(\delta_i^{k}\hat p_j\hat p^l+
\delta_i^{l}\hat p_j\hat p^k+
\delta_j^{k}\hat p_i\hat p^l+
\delta_j^{l}\hat p_i\hat p^k\big)
+\frac{1}{2}\big(\delta_{ij}\hat p^k\hat p^l+ \hat p_i \hat
p_j\delta^{kl}\big)
+\frac{1}{2}\hat p_i\hat p_j\hat p^k \hat p^l\;,
\label{ProjT}\\
\mathbb{P}_{ij}^{(V)\;kl}\equiv&\sum_{r=1,2}\frac{1}{2}
(e_i^{(r)}\hat p_j+\hat p_i e_j^{(r)})(e^{(r)\,k}\hat p^l+\hat p^k
e^{(r)\,l})\notag\\
=&\frac{1}{2}\big(\delta_i^{k}\hat p_j\hat p^l+
\delta_i^{l}\hat p_j\hat p^k+
\delta_j^{k}\hat p_i\hat p^l+
\delta_j^{l}\hat p_i\hat p^k\big)
-2\hat p_i\hat p_j\hat p^k \hat p^l\;.
\label{ProjV}
\end{align}
\eseq

In the scalar sector the situation is more subtle. Here we have two
eigenvalues that in general are not degenerate. To see this we act
with ${\mathbb{D}}({\bf p})$ on the scalar part of (\ref{decomp}) and
find
\be
\label{Dactscalar}
\begin{split}
{\mathbb{D}}({\bf p})_{ij}^{\;\;kl}h_{kl}\Big|_{\rm scalar}=
p^6\bigg[\,\phi\,\frac{\nu_s}{\sqrt{2}}(\delta_{ij}-\hat p_i\hat p_j)
+\bigg(\phi\,\sqrt{2}\l\Big(\frac{\nu_s}{1-\l}-\frac{1+\xi}{\s}\Big)
+\psi\frac{(1-\l)(1+\xi)}{\s}\bigg)\,\hat p_i\hat p_j\bigg]\;,
\end{split}
\ee
where $\nu_s$ is defined by Eq.(\ref{nus}). Thus, in the
two-dimensional subspace of vectors 
$\Upsilon=(\phi,~\psi)^{\rm T}$
the operator ${\mathbb{D}}({\bf p})$ acts as a matrix
\be
p^6\begin{pmatrix}
\nu_s&0\\
\sqrt{2}\l\Big(\frac{\nu_s}{1-\l}-\frac{1+\xi}{\s}\Big)&
\frac{(1-\l)(1+\xi)}{\s}
\end{pmatrix}\;.
\ee
The corresponding eigenvalues and eigenvectors are 
\bseq
\label{eigenscalar}
\begin{align}
\label{eigenscalar1}
&\kappa_{S1}=\nu_sp^6~,&&
\Upsilon_{S1}=\begin{pmatrix} 1\\
\frac{\sqrt{2}\l}{1-\l}\end{pmatrix}\;,\\
\label{eigenscalar2}
&\kappa_{S2}=\frac{(1-\l)(1+\xi)}{\s}p^6~,&&
\Upsilon_{S2}=\begin{pmatrix} 0\\
1\end{pmatrix}\;.
\end{align}
\eseq
It is convenient to construct the operators
$\mathbb{P}^{(S1)}$ and $\mathbb{P}^{(S2)}$
projecting on the vectors $\Upsilon_{S1}$ and $\Upsilon_{S2}$
respectively. This is done using the linear forms conjugate to these vectors
\be
\label{eigendagger}
\Upsilon_1^\dag=\begin{pmatrix}
1&0\end{pmatrix}~,~~~~~~~~
\Upsilon_2^\dag=\begin{pmatrix}
-\frac{\sqrt{2}\l}{1-\l}&1\end{pmatrix}\;,
\ee
that have the property
$\Upsilon_r^\dag\Upsilon_q=\delta_{rq},\; r,q=1,2$.
Then
\be
\label{Projss}
\mathbb{P}^{(S1)}=\Upsilon_{S1}\otimes\Upsilon_{S1}^\dag~,~~~~~~
\mathbb{P}^{(S2)}=\Upsilon_{S2}\otimes\Upsilon_{S2}^\dag\;.
\ee
Restoring the spatial indices we have,
\be
\label{ProjS}
\mathbb{P}_{ij}^{(1)\,kl}
=\frac{1}{2}\bigg(\delta_{ij}-\frac{1-3\l}{1-\l}\hat p_i\hat p_j\bigg)
(\delta^{kl}-\hat p^k\hat p^l)\;,\qquad
\mathbb{P}_{ij}^{(2)\,kl}=\hat p_i\hat
p_j\bigg(-\frac{\l}{1-\l}\delta^{kl}
+\frac{1}{1-\l}\hat p^k\hat p^l\bigg)\;.
\ee

It is now straightforward to verify that the principal symbol
(\ref{Dtildeflat}) decomposes into the sum of projectors,
\be
\label{DProjs}
{\mathbb{D}}({\bf p})_{ij}^{\;\;kl}=\nu_5 p^6\,
\mathbb{P}_{ij}^{(T)\,kl}
+\frac{p^6}{2\s}\, \mathbb{P}_{ij}^{(V)\,kl}
+\nu_s p^6 \,\mathbb{P}_{ij}^{(S1)\,kl}
+\frac{(1-\l)(1+\xi)}{\s}p^6\, \mathbb{P}_{ij}^{(S2)\,kl}\;.
\ee
Then its square root is
obtained by taking the square roots of the coefficients,
\be
\label{Qproj}
\mathbb{Q_{\,D}}({\bf p})
=\sqrt{\nu_5}\,p^3\sum_{\al} u_\al \mathbb{P}^{(\al)}\;,\qquad
\al=T,V,S1,S2\;,
\ee
where
\be
\label{allus}
u_T=1,\quad u_V=\frac1{\sqrt{2\sigma\nu_5}}, \quad
u_{S1}=u_s,\quad u_{S2}=\sqrt{\frac{(1-\l)(1+\xi)}{\s\nu_5}}\;,
\ee
and $u_s$ is defined in (\ref{new_couplings}).
Expanding the projectors we finally arrive at,
\be
\label{Qnot}
\begin{split}
\mathbb{Q_{\,D}}({\bf p})_{ij}^{\;\;kl}&=\sqrt{\nu_5}\,p^3\bigg[\frac{1}{2}
(\delta_i^k\delta_j^l+\delta_i^l\delta_j^k)
+\frac{u_s-1}{2}\delta_{ij}\delta^{kl}
+\frac{u_V-1}{2}
(\delta_i^k\hat p_j\hat p^l+\delta_i^l\hat p_j\hat p^k+
\delta_j^k\hat p_i\hat p^l+\delta_j^l\hat p_i\hat p^k)\\
&-\frac{u_s-1}{2}\delta_{ij}\hat p^k\hat p^l-\bigg(\frac{u_s}{2}\frac{1\!-\!3\l}{1\!-\!\l}-\frac1{2}
+\frac{\l\, u_{S2}}{1\!-\!\l}\bigg)
\hat p_i\hat p_j \delta^{kl}+\bigg(\frac{1}{2}-2u_V
+\frac{u_s}{2}\frac{1\!-\!3\l}{1\!-\!\l}
+\frac{u_{S2}}{1\!-\!\l}\bigg)
\hat p_i\hat p_j\hat p^k\hat p^l
\bigg].
\end{split}
\ee
This principal symbol plays the central role in the perturbative
calculation of the operator square root $\mathbb{Q_{\,D}}$. For a
general choice of the gauge parameters $\s,\xi$ this calculation is
prohibitively complex. Thus we restrict to four gauge choices that
simplify the expression (\ref{Qnot}).

\paragraph{Gauge (a)} First we consider a choice where the
eigenvalues of the gauge modes coincide with those of the physical
modes. Namely, we take 
\be
\label{cparam}
u_V=1,~~~u_{S2}=u_s
~~~\Longleftrightarrow ~~~
\s=\frac{1}{2\nu_5},~~~\xi=\frac{\nu_s}{2\nu_5(1-\l)}-1\;.
\ee
This yields,
\be
\label{Qnotc}
\begin{split}
\mathbb{Q_{\,D}}({\bf p})_{ij}^{\;\;kl}=\sqrt{\nu_5}\,p^3\bigg[\frac{1}{2}
(\delta_i^k\delta_j^l+\delta_i^l\delta_j^k)
+\frac{u_s-1}{2}\,\delta_{ij}\delta^{kl}
-\frac{u_s-1}{2}\,\delta_{ij}\hat p^k\hat p^l
-\frac{u_s-1}{2}
\hat p_i\hat p_j \delta^{kl}
+\frac{3}{2}(u_s-1)\,
\hat p_i\hat p_j\hat p^k\hat p^l
\bigg].
\end{split}
\ee
Importantly, this choice overlaps with the gauges considered in
Ref.~\cite{towards} (see also Appendix~\ref{app:betaG}) which allows
us to use the results of this paper for the (gauge-dependent)
$\beta$-function of the coupling $G$ in this gauge.

\paragraph{Gauge (b)} The second choice is similar, but now
\be
\label{bparam}
u_V=u_{S2}=1
~~~\Longleftrightarrow ~~~
\s=\frac{1}{2\nu_5},~~~\xi=-\frac{1-2\l}{2(1-\l)}\;,
\ee
and we obtain
\be
\label{Qnotb}
\begin{split}
\mathbb{Q_{\,D}}({\bf p})_{ij}^{\;\;kl}=\sqrt{\nu_5}\,p^3\bigg[&\frac{1}{2}
(\delta_i^k\delta_j^l+\delta_i^l\delta_j^k)
+\frac{u_s-1}{2}\delta_{ij}\delta^{kl}
-\frac{u_s-1}{2}\delta_{ij}\hat p^k\hat p^l\\
&-\frac{1-3\l}{2(1-\l)}(u_s-1)\,
\hat p_i\hat p_j \delta^{kl}
+\frac{1-3\l}{2(1-\l)}(u_s-1)\,
\hat p_i\hat p_j\hat p^k\hat p^l
\bigg].
\end{split}
\ee
This also overlaps with the choices considered in \cite{towards}.

\paragraph{Gauge (c)} Two other choices are adjusted to remove the
term with four vectors $\hat{\bf p}$ in (\ref{Qnot}) which is
challenging from the computational viewpoint.\footnote{When
  transformed back to configuration space, the four momenta become
  four covariant derivatives that must be commuted through the other
  operators in the course of the perturbative procedure, see below.} 
The
most simplifying choice is
\be
\label{aparam}
u_V=1,~~~u_{S2}=\frac{3(1-\l)}{2}-\frac{(1-3\l)u_s}{2}
~~~\Longleftrightarrow ~~~
\s=\frac{1}{2\nu_5},~~~
\xi=\frac{\big(3(1-\l)-(1-3\l)u_s\big)^2}{8(1-\l)}-1\;.
\ee
The principal symbol becomes,
\be
\label{Qnota}
\begin{split}
\mathbb{Q_{\,D}}({\bf p})_{ij}^{\;\;kl}=\sqrt{\nu_5}\,p^3\bigg[\frac{1}{2}
(\delta_i^k\delta_j^l+\delta_i^l\delta_j^k)
+\frac{u_s-1}{2}\delta_{ij}\delta^{kl}
-\frac{u_s-1}{2}\delta_{ij}\hat p^k\hat p^l
-\frac{1-3\l}{2}(u_s-1)
\hat p_i\hat p_j \delta^{kl}
\bigg].
\end{split}
\ee
A drawback of this choice is that it differs from the gauges studied
in \cite{towards}. Therefore, in this gauge we cannot compute the
running of the essential coupling ${\cal G}$ (see
Eq.~(\ref{new_couplings})) which requires the
knowledge of the $\beta$-function for $G$. Nevertheless, we can still
compute the running of $u_s$ and $v_a$, $a=1,2,3$.

\paragraph{Gauge (d)} To remedy the above drawback of gauge {\bf (c)}
we also consider 
\be
\label{dparam}
u_V=u_{S2}=\frac{1-\l+(1-3\l)u_s}{2(1-2\l)}
~~~\Longleftrightarrow ~~~
\s=\frac{2(1-2\l)^2}{\nu_5\big(1-\l+(1-3\l)u_s\big)^2},
~~~\xi=-\frac{1-2\l}{2(1-\l)}\;.
\ee
Here the principal symbol takes the form,
\be
\label{Qnotd}
\begin{split}
\mathbb{Q_{\,D}}({\bf p})_{ij}^{\;\;kl}=\sqrt{\nu_5}\,p^3\bigg[&\frac{1}{2}
(\delta_i^k\delta_j^l+\delta_i^l\delta_j^k)
+\frac{u_s-1}{2}\delta_{ij}\delta^{kl}
-\frac{u_s-1}{2}\delta_{ij}\hat p^k\hat p^l
-\frac{(1-3\l)(u_s-1)}{2(1-2\l)}
\hat p_i\hat p_j \delta^{kl}
\\
&
+\frac{(1-3\l)(u_s-1)}{4(1-2\l)}
(\delta_i^k\hat p_j\hat p^l+\delta_i^l\hat p_j\hat p^k+
\delta_j^k\hat p_i\hat p^l+\delta_j^l\hat p_i\hat p^k)
\bigg].
\end{split}
\ee
This gauge again overlaps with the gauges used in \cite{towards}.

\paragraph{} Comparison of the results obtained in four different
gauges {\bf (a)}---{\bf (d)} provides a strong check of our
calculation.

\subsubsection{Vector sector}
We now repeat the analysis for the vector operator $\mathbb{B}$ given
by the expression (\ref{B}). Its
principal symbol reads
\begin{equation}
\mathbb{B}^i_{~j}({\bf p}) = p^6\left(\frac1{2\sigma}\delta^i_j +
\frac{1-2\lambda+2\xi(1-\lambda)}{2\sigma} \hat{p}^i \hat{p}_j \right).
\end{equation}
This can be easily written in terms of the transverse and longitudinal
projectors, 
\begin{equation}
\mathbb{B}^i_{~j}({\bf p}) =
\frac{p^6}{2\sigma}{\mathbb{P}^{(VT)\:i}}_j +\frac{(1-\l)(1+\xi)}{\s}p^6
 \,{\mathbb{P}^{(VL)\:i}}_j \;, 
\end{equation}
where 
\begin{equation}\label{ghproj}
{\mathbb{P}^{(VT)\:i}}_j = \delta^i_j - \hat{p}^i \hat{p}_j, \qquad
{\mathbb{P}^{(VL)\:i}}_j = \hat{p}^i \hat{p}_j\;.
\end{equation}
Then the square root reads
\begin{equation}
\mathbb{Q_{\,B}}({\bf p})= \sqrt{\nu_5}\, p^3 \sum_\alpha u_\alpha
\mathbb{P}^{(\alpha)}\;, 
\quad \alpha=VT,VL,\quad u_{VT} =u_V,\:u_{VL} =u_{S2}\;.
\end{equation}

\subsection{Canonical form of the pseudo-differential operators}
\label{sec:canonical}

The next step in the procedure outlined at the beginning of this
section (see Eqs.(\ref{Qaeq0})-(\ref{Xeq})) consists in the recovery
of the pseudo-differential operator $\mathbb{Q}^{(0)}$ from its symbol
$\mathbb{Q}({\bf p})$. The result of this procedure is the canonical
form which we formulate as follows: 
\begin{enumerate}
\item
All (fractional) powers of $p^2$ are replaced by covariant Laplacians
$-\Delta$ and put to the right. 

\item
Other occurrences of momenta are replaced by covariant
derivatives, $p_i\mapsto -i\nabla_i$. 
The covariant derivatives whose indices are contracted
with the tensor indices of the metric fluctuations or the shift vector
are placed to the
right. 
\end{enumerate}
As an example consider $\mathbb{Q}^{(0)}_{\,\mathbb{D}}$ in the gauge {\bf (a)}. The above prescription gives
\be
\label{Qnotccurved}
\begin{split}
\mathbb{Q}_{\,\mathbb{D}\;ij}^{(0)\;kl}=&\sqrt{\nu_5}\left[\bigg(\frac{1}{2}
(\delta_i^k\delta_j^l+\delta_i^l\delta_j^k)
+\frac{u_s-1}{2}g_{ij}g^{kl}\bigg)(-\D)^{3/2}
+\frac{u_s-1}{2}g_{ij}\nabla^k\nabla^l(-\D)^{1/2}\right.\\
&\left.+\frac{u_s-1}{2} g^{kl}
\nabla_i\nabla_j (-\D)^{1/2}
+\frac{3(u_s-1)}{2}\nabla_i\nabla_j
\nabla^k\nabla^l(-\D)^{-1/2}\right].
\end{split}
\ee
Note that the result of the action of this operator on a symmetric
metric fluctuation $\mathbb{Q}_{\,\mathbb{D}\;ij}^{(0)\;kl}h_{kl}$ is automatically
symmetric in the indices $(i,j)$. In other words, this operator acts
in the space of symmetric tensors, as it should.
Similarly for the vector operator, its zeroth-order part in  
a generic $(\sigma,\xi)$-gauge takes the form
\begin{equation}
\label{Qnotvect}
\mathbb{Q}^{(0)\:i}_{\,\mathbb{B}\;\;\;\;j} = \sqrt{\nu_5} 
\left[u_{V} \delta^i_j (-\Delta)^{3/2} + (u_{V}- u_{S2})\nabla^i\nabla_j (-\Delta)^{1/2}\right]\;.
\end{equation}

In a more general case of the curvature dependent part $\mathbb{X}$ of
the square root operator 
the number of derivatives is higher and more
ordering ambiguities arise. Thus, we supplement this prescription by
one more rule: 
\begin{itemize}
\item[3.]
The derivatives not covered by rules 1. and 2. are ordered by using the
``SortCovDs'' command of the {\it xAct} package
\cite{xAct}. 
\end{itemize}

\subsection{Solution of the Sylvester equation}

Perturbation theory for the square root operator $\mathbb{Q}$ implies
solving the equation (\ref{Xeq}) for its curvature part
$\mathbb{X}$. At each stage of the corresponding iteration procedure
we will encounter the matrix equation of the 
following form, 
\be
\label{Syl}
\mathbb{Q}({\bf p})\mathbb{X}({\bf p})
+\mathbb{X}({\bf p})\mathbb{Q}({\bf p})=\mathbb{Y}({\bf p})\;.
\ee
Here $\mathbb{Q}({\bf p})$ is the c-number symbol of
$\mathbb{Q}^{(0)}$, the matrix $\mathbb{Y}({\bf p})$, which is the
symbol of the operator
$\mathbb{Y}\equiv\mathbb{F}-\big(\mathbb{Q}^{(0)}\big)^2- 
\mathbb{X}^2$ in the right hand side of Eq.(\ref{Xeq}), is assumed to
be known, and we need to find $\mathbb{X}({\bf p})$. All matrices depend on
the
3-dimensional wavenumber ${\bf p}$. Equation (\ref{Syl})
is a special case of the Sylvester matrix equation and its
solution can be found with the general method of
Ref.~\cite{Sylvester}.
In the case at hand, however, it is easier to obtain the
solution using the representation of
$\mathbb{Q}({\bf p})$ in terms of the projectors ({\ref{Qproj}),
  (\ref{ghproj}). The solution reads,
\be
\label{Sylsol}
\mathbb{X}({\bf p})=\frac{1}{\sqrt{\nu_5}\,p^3}\sum_{\al,\beta}\frac{1}{u_\al+u_\beta}
\mathbb{P}^{(\al)}\,\mathbb{Y}({\bf p})\,\mathbb{P}^{(\beta)}\;,
\ee
where the sum is taken over $\alpha,\beta=T,V,S1,S2$ (tensor sector) or
$\alpha,\beta=VT,VL$ (vector sector).
The proof goes by a direct substitution:
\be
\mathbb{Q}({\bf p})\mathbb{X}({\bf p})
=\sum_{\alpha,\beta}\frac{u_\al}{u_\al+u_\beta}
\mathbb{P}^{(\al)}\,\mathbb{Y}({\bf p})\,\mathbb{P}^{(\beta)}\;,\quad
\mathbb{X}({\bf p})\mathbb{Q}({\bf
  p})=\sum_{\alpha,\beta}\frac{u_\beta}{u_\al+u_\beta} 
\mathbb{P}^{(\al)}\,\mathbb{Y}({\bf p})\,\mathbb{P}^{(\beta)}\;.
\ee
Here we have used the orthogonality of the projectors
$\mathbb{P}^{(\al)}\mathbb{P}^{(\beta)}=\delta^{\al\beta}\mathbb{P}^{(\al)}$.
Summing up the two expression we obtain
\be
\mathbb{Q}({\bf p})\mathbb{X}({\bf p})+\mathbb{X}({\bf
  p})\mathbb{Q}({\bf p})= 
\sum_{\al,\beta}\mathbb{P}^{(\al)}\,\mathbb{Y}({\bf p})\,\mathbb{P}^{(\beta)}
=\Big(\sum_\al\mathbb{P}^{(\al)}\Big) \,\mathbb{Y}({\bf p})\,
\Big(\sum_\beta\mathbb{P}^{(\beta)}\Big)=\mathbb{Y}({\bf p})\;,
\ee
where in the last equality we used the completeness of the projector basis.
In what follows we will denote the linear map from the r.h.s. of the
Sylvester equation to its solution by ``${\rm Syl}$'', so that
we will write,
\be
\label{Sylop}
\mathbb{X}({\bf p})={\rm Syl}[\mathbb{Y}({\bf p})]\;.
\ee

\subsection{Perturbative scheme}
\label{sec:pert}
As discussed in the beginning of this section, to find the one-loop
renormalization of the action we need to construct an operator
$\mathbb{Q}$ whose square coincides with the operator
$\mathbb{F}=({\mathbb{D}},\mathbb{B})$ entering the quadratic action for
the
fluctuations. We perform this construction perturbatively in the
powers of the background curvature and its derivatives. Namely, we write,
\be
\label{Qdecomp}
\mathbb{Q}=\mathbb{Q}^{(0)}+\mathbb{Q}^{(2)}
+\mathbb{Q}^{(3)}+\mathbb{Q}^{(4)}+\mathbb{Q}^{(5)}+\mathbb{Q}^{(6)}
+\ldots\;,
\ee
where the index in the brackets represents the order of 
the
operator in powers of the inverse length scale characterizing the
background curvature.
Here the operator $\mathbb{Q}^{(0)}$
is given by (\ref{Qnotccurved}) or (\ref{Qnotvect}), 
it does not contain any background
curvature.
The operator
$\mathbb{Q}^{(2)}$ is linear in curvature, this is counted as second
order, since the curvature contains two derivatives of the metric.
The operator $\mathbb{Q}^{(3)}$ contains
first derivatives of the curvature (3 derivatives of the metric), and
so on. Dots stand for higher order terms that do not
contribute into the divergent part of the action.

Substitution of this expansion into the defining relation
\be
\label{basic}
\mathbb{Q}^2={\mathbb{F}}
\ee
produces at each order an equation of the form,
\be
\label{Qaeq}
\mathbb{Q}^{(0)}\mathbb{Q}^{(a)}+\mathbb{Q}^{(a)}\mathbb{Q}^{(0)}=
\mathbb{Y}^{(a)}\;,
\ee
with the r.h.s.
\be
\label{newY}
\mathbb{Y}^{(a)}=\mathbb{F}-\sum_{\begin{smallmatrix}b,c<a\\
b+c\leq a\end{smallmatrix}}\mathbb{Q}^{(b)} \mathbb{Q}^{(c)}\;.
\ee
The operator $\mathbb{Y}^{(a)}$ contain terms of order $a$ and higher. Then
$\mathbb{Q}^{(a)}$ is constructed by the following algorithm:
\begin{enumerate}
\item
Pick up the part of $\mathbb{Y}^{(a)}$ which is exactly of order $a$;
let us denote it by $\mathbb{Y}_a^{(a)}$.
\item
Replace the covariant derivatives acting on the metric fluctuations
in $\mathbb{Y}_a^{(a)}$ by the wavevector,
$\nabla_i\mapsto ip_i$. This gives the c-matrix symbol
$\mathbb{Y}_a^{(a)}({\bf p})$.
\item
Solve the corresponding Sylvester equation and define a matrix
\be
\mathbb{Q}^{(a)}({\bf p})={\rm Syl}[\mathbb{Y}_a^{(a)}({\bf p})]\;.
\ee
\item
Replace the wavevectors in $\mathbb{Q}^{(a)}({\bf p})$ back by the covariant
derivatives, ordering them in a canonical way (see
Sec.~\ref{sec:canonical}). For tensor operators,  
symmetrize 
$\mathbb{Q}_{ij}^{(a)\,kl}$ in the indices $(ij)$ and $(kl)$.
\item
Construct the combination
$\mathbb{Q}^{(0)}\mathbb{Q}^{(a)}+\mathbb{Q}^{(a)}\mathbb{Q}^{(0)}$
and bring it to the canonical form. By construction, this combination
coincides with $\mathbb{Y}_a^{(a)}$, up to terms of order higher than
$a$. Subtract it from $\mathbb{Y}^{(a)}$ to
define a new operator ${\mathbb{Z}}^{(a+1)}$.
\item
Construct other products $\mathbb{Q}^{(b)}\mathbb{Q}^{(c)}$ with
$b,c<a$, $b+c=a+1$, bring them to the canonical form and subtract from
${\mathbb{Z}}^{(a+1)}$. This determines $\mathbb{Y}^{(a+1)}$, according to
Eq.~(\ref{newY}).
\end{enumerate}
In this way we arrive at an iterative procedure for a consecutive
determination of $\mathbb{Q}^{(a)}$. According to (\ref{newY}) the
right hand side of (\ref{Qaeq}) at different steps is given by, 
\bseq
\label{Ys}
\begin{align}
\label{Y2}
&\mathbb{Y}^{(2)}={\mathbb{F}}-(\mathbb{Q}^{(0)})^2\;,\\
\label{Y3}
&\mathbb{Y}^{(3)}=\mathbb{Y}^{(2)}-\mathbb{Q}^{(0)}\mathbb{Q}^{(2)}
-\mathbb{Q}^{(2)}\mathbb{Q}^{(0)}\;,\\
\label{Y4}
&\mathbb{Y}^{(4)}=\mathbb{Y}^{(3)}-\mathbb{Q}^{(0)}\mathbb{Q}^{(3)}
-\mathbb{Q}^{(3)}\mathbb{Q}^{(0)}-(\mathbb{Q}^{(2)})^2\;,\\
\label{Y5}
&\mathbb{Y}^{(5)}=\mathbb{Y}^{(4)}-\mathbb{Q}^{(0)}\mathbb{Q}^{(4)}
-\mathbb{Q}^{(4)}\mathbb{Q}^{(0)}-\mathbb{Q}^{(2)}\mathbb{Q}^{(3)}
-\mathbb{Q}^{(3)}\mathbb{Q}^{(2)}\;,\\
\label{Y6}
&\mathbb{Y}^{(6)}=\mathbb{Y}^{(5)}-\mathbb{Q}^{(0)}\mathbb{Q}^{(5)}
-\mathbb{Q}^{(5)}\mathbb{Q}^{(0)}-\mathbb{Q}^{(2)}\mathbb{Q}^{(4)}
-\mathbb{Q}^{(4)}\mathbb{Q}^{(2)}-(\mathbb{Q}^{(3)})^2\;.
\end{align}
\eseq
We have automated the algorithm described above using the Mathematica
\cite{Mathematica} package {\it
  xAct} \cite{xAct}.

A few comments are in order. First, the coefficient functions of the 5th order
operator $\mathbb{Q}^{(5)}$ contain either
a third derivative of
curvature or a
product of
curvature with its first derivative. None of these combinations can
give rise to a divergent counterterm in the one-loop action. Indeed,
the renormalizability of the theory implies that the counterterms have
the same structure as the terms in the bare action which have order 6
in our power counting, see
\cite{Barvinsky:2015kil,Barvinsky:2017zlx}.
To produce a 6th order contribution the coefficients of
$\mathbb{Q}^{(5)}$ would have to be multiplied by a covariant object
constructed from the metric using a single derivative. But such
objects do not exist. Thus, we conclude that $\mathbb{Q}^{(5)}$ does
not contribute into the beta functions and can be dropped.
Then one can verify that the 5th
order contributions
can be consistently omitted in all $\mathbb{Y}^{(a)}$
at all stages of the
calculation. In particular, instead of solving consecutively for
$\mathbb{Q}^{(5)}$ and then for $\mathbb{Q}^{(6)}$ using the
$\mathbb{Y}$ operators (\ref{Y5}), (\ref{Y6}), we can directly
construct $\mathbb{Q}^{(6)}$ in a single step by solving
Eq.~(\ref{Qaeq}) with the r.h.s.
\be
\label{Y6new}
\mathbb{Y}^{(6)}=\mathbb{Y}^{(4)}-\mathbb{Q}^{(0)}\mathbb{Q}^{(4)}
-\mathbb{Q}^{(4)}\mathbb{Q}^{(0)}
-\mathbb{Q}^{(2)}\mathbb{Q}^{(3)}
-\mathbb{Q}^{(3)}\mathbb{Q}^{(2)}
-\mathbb{Q}^{(2)}\mathbb{Q}^{(4)}
-\mathbb{Q}^{(4)}\mathbb{Q}^{(2)}-(\mathbb{Q}^{(3)})^2\;.
\ee

Second, the most time-consuming part of the calculation are 
steps 5.-6., that involve bringing various operators to the canonical form. 
In detail, this canonicalization proceeds as follows (for
concreteness, we focus on the metric sector):
\begin{itemize}
\item
All (fractional) powers of the Laplacian acting on the metric
fluctuations are commuted through the coefficient functions and
covariant derivatives to the right
  and collapsed to a single fractional Laplacian.
The commutation is performed using the formula
\be
\label{commut}
[A^\al,B]=\sum_{n=1}^\infty C_\al^n\;\underbrace{[A,[A,\ldots,[A}_n,B]]\ldots] A^{\al-n}\;,
\ee
valid for arbitrary operators $A$ and $B$. Here 
\be
\label{binom}
C_\al^n=\frac{\al(\al-1)\ldots(\al-n+1)}{n!}
\ee
are the binomial coefficients. This formula is proved in 
  Appendix~\ref{app:A}.
\item
Next, we bring to the right all covariant derivatives contracted with
the
metric fluctuations. For example, an expression
$\nabla^k\nabla\ldots\nabla (-\D)^{\alpha}h_{kl}$ after bringing it to
the canonical form will read 
$$\nabla\ldots\nabla\nabla^k
(-\D)^{\alpha}h_{kl}+\ldots\;,$$ 
where dots stand for the terms with
curvature that have been generated as the result of commutation.
\item
The remaining derivatives (including possible derivatives acting on
the curvature) are
ordered using the ``SortCovDs'' command of
the {\it xAct} package.
\item
The Riemann tensors appearing from the
commutations are replaced by their expressions in terms of the Ricci
tensor and the
scalar curvature. This step may generate additional Laplacians or
contractions of derivatives with the metric fluctuations,
so the ordering procedure is repeated iteratively
until it converges.
\end{itemize}

When performing the commutation of fractional powers of Laplacian
the formula (\ref{commut})
is truncated at the order relevant for a given step of the
calculation. A single commutator of the Laplacian with a covariant
derivative is proportional to curvature, thus it has 2nd order in
our counting. Every further commutator increases the order at least by
one. In addition, it can be shown that the lowest-order coefficient
function in the nested commutator
\be
\label{Dnabnested}
\underbrace{[\D,[\D,\ldots,[\D}_n,\nabla]]\ldots]
\ee
is a total derivative. Hence, it will not contribute into the
effective action, unless it gets multiplied by another background
tensor. The latter has at least dimension 2, which further limits the
number of nested commutators we need to consider at a given order. A
straightforward analysis of possible cases tells us that in the
commutator   
of the fractional
Laplacian with a covariant derivative we need to go up to: 
\begin{itemize}
\item[-]
 $n=4$ in
the computation of $(\mathbb{Q}^{(0)})^{2}$, 
\item[-]
$n=3$ in the computation of $\mathbb{Q}^{(0)}\mathbb{Q}^{(2)}$
and $\mathbb{Q}^{(2)}\mathbb{Q}^{(0)}$, 
\item[-]
$n=2$ in
$\mathbb{Q}^{(0)}\mathbb{Q}^{(3)}$ and
$\mathbb{Q}^{(3)}\mathbb{Q}^{(0)}$,
\item[-]
$n=1$ in
$\mathbb{Q}^{(0)}\mathbb{Q}^{(4)}$,
$\mathbb{Q}^{(4)}\mathbb{Q}^{(0)}$
and $(\mathbb{Q}^{(2)})^{2}$,
\item[-]
in the 5th order operators $\mathbb{Q}^{(2)}\mathbb{Q}^{(3)}$,
$\mathbb{Q}^{(3)}\mathbb{Q}^{(2)}$ and
6th order operators
$\mathbb{Q}^{(2)}\mathbb{Q}^{(4)}$,
$\mathbb{Q}^{(4)}\mathbb{Q}^{(2)}$,
$(\mathbb{Q}^{(3)})^{2}$ all derivatives can be treated as
commutative. 
\end{itemize}

Similarly, every commutator of the Laplacian with
a coefficient function made of curvature increases the order at
least by one, the lowest order term in
\be
\label{DRnestes}
\underbrace{[\D,[\D,\ldots,[\D}_n,{\cal R}_{(a)}]]\ldots]
\ee
again being a total derivative. By considering possible cases we
conclude that when 
commuting the fractional Laplacian with
the coefficient functions we need to retain up to: 
\begin{itemize}
\item[-]
3 nested commutators in $\mathbb{Q}^{(0)}\mathbb{Q}^{(2)}$,
\item[-]
2 nested commutators in
$\mathbb{Q}^{(0)}\mathbb{Q}^{(3)}$ and  $(\mathbb{Q}^{(2)})^{2}$,
\item[-]
one commutator
in $\mathbb{Q}^{(2)}\mathbb{Q}^{(3)}$,
$\mathbb{Q}^{(3)}\mathbb{Q}^{(2)}$,
\item[-]
in
$\mathbb{Q}^{(0)}\mathbb{Q}^{(4)}$ and 
$\mathbb{Q}^{(4)}\mathbb{Q}^{(2)}$,
$\mathbb{Q}^{(2)}\mathbb{Q}^{(4)}$, $(\mathbb{Q}^{(3)})^{2}$ the
commutator between fractional Laplacians and the coefficient functions
can be omitted altogether.
\end{itemize}

The iterative algorithm of this section provides us with the
square-root operator in the form (\ref{root_structure}) suitable for
further processing with the technique of universal functional
traces. We now describe this technique and draw the list of 
the required UFTs.

\section{Universal functional traces}\label{UFT}

\subsection{Schwinger--DeWitt technique and the method of universal
  functional traces} 
\label{sec:UFTmethod}

The calculation of UFTs of the form (\ref{UFT0}) arising in
(\ref{root_structure}) can be done by means of the heat kernel method
and the Schwinger--DeWitt technique of the proper time expansion on
generic curved spacetime. The heat kernel method allows one to write
down in (\ref{root_structure}) the integral representation for a
generic power of the Laplacian, 
    \be
    \nabla...\nabla\frac{\hat 1}{(-\Delta)^\al}\,\delta(x,y)\,\Big|_{\,y=x}^{\,\rm div} =\frac{1}{\varGamma(\al)}\nabla...\nabla\int_0^\infty ds\,s^{\al-1}\,\e^{s\Delta}\,\hat\delta(x,y)\,
    \Big|_{\,y=x}^{\,\rm div}\;,                    \label{Gammarepr0}
    \ee 
in terms of the kernel of the heat equation $\hat K(\,s\,|\,x,y\,)
=e^{s\Delta}\,\hat\delta(x,y)$ with the ``Hamiltonian''
$-\Delta$. Here $\Delta=g^{ij}\nabla_i\nabla_j$ is a
covariant Laplacian acting on arbitrary set of tensor fields
$\phi^A(x)$ labelled by the index $A$, the hat denoting a matrix in
their vector space, $\hat 1=\delta^A_{~B}$, and the matrix nature of the
delta function $\hat\delta(x,y)\equiv\hat 1\times\delta(x,y)$. 
Note that, in contrast to Eq.~(\ref{tracelog}), the dimension of the
proper-time parameter in this formula is $[s]=-2$ to match the
dimensionality of the Laplacian.

Expansion of $\hat K(\,s\,|\,x,y\,)$ at small values of the proper
time allows one to isolate in the coincidence limit $y=x$ the
integrals diverging at the lower boundary $s=0$, which comprise UV
divergences of the universal functional traces
(\ref{UFT0}).\footnote{For large positive $\alpha$ the operators can
 suffer from infrared (IR) divergences associated with
  the upper integration limit for $s$, but as we will be interested in
  UV divergences, which are clearly separated at one-loop order from
  the IR ones, we will disregard this issue.} This expansion, in its
turn, is based on the Schwinger-DeWitt technique
\cite{DeWitt,PhysRep}. In the most general setting this expansion is
explicitly known for a {\em minimal} second-order operator of the form 
    \begin{eqnarray}
    \hat F(\nabla)=\Box+\hat P-
    \frac{\hat 1}6\,R, \qquad
    \Box=g^{\mu\nu}\nabla_\mu\nabla_\nu,  \label{F}
    \end{eqnarray}
``minimal'' meaning that its second order derivatives form a covariant d'Alembertian determined with respect to the spacetime metric $g_{\mu\nu}$. The operator acts in the representation space of $\phi^A(x)$ in $d$-dimensional spacetime with coordinates $x^\mu$, $\mu=1,...d$ and is
characterized by the set of ``curvatures'' $\Re=(\hat P,\;
\hat{\mbox{\boldmath$R$}}_{\mu\nu},\;R_{\lambda\rho\mu\nu})$ --- the
potential term $\hat 
P\equiv P^A_{~B}$ (the term $-R\hat{1} /6$ 
is singled out from it for convenience), fibre bundle
curvature $\hat{\mbox{\boldmath$R$}}_{\mu\nu}\equiv \mbox{\boldmath$R$}^A_{~B\,\mu\nu}$ ---
commutator of covariant derivatives acting on the vector $\phi^B$ or
a matrix $\hat X\equiv X^B_{~C}$ --- and the Riemann tensor, 
    \begin{eqnarray}
    [\nabla_\mu,\nabla_\nu]\,V^\l=R^\l_{\,\,\rho\mu\nu}V^\rho,\qquad
    [\nabla_\mu,\nabla_\nu]\,\phi=\hat{\mbox{\boldmath$R$}}_{\mu\nu}\phi,\qquad [\nabla_\mu,\nabla_\nu]\,\hat X
    =[\hat{\mbox{\boldmath$R$}}_{\mu\nu},\hat X].      \label{commutators}
    \end{eqnarray}

The heat kernel $\hat K(s|\,x,y)=e^{s\hat F(\nabla)}\delta(x,y)$  for the operator (\ref{F}) has a small (or early) time asymptotic expansion at $s\to 0$,
    \begin{eqnarray} 
    \hat K(s|\,x,y)=\frac{{\cal D}^{1/2}(x,y)}{(4\pi
      s)^{d/2}}\,g^{1/2}(y)\,e^{-\frac{\sigma(x,y)}{2s}} 
    \sum\limits_{n=0}^\infty \,s^n\,
    \hat a_n(x,y),                      \label{heatexpansion}
    \end{eqnarray}
where $\sigma(x,y)$ is the Synge world function --- one half of the
square of geodetic distance between the points $x$ and $y$, and 
\be
{\cal D}(x,y) =g^{-1/2}(x)\,\left|\,{\rm det}\,
\frac{\partial^2\sigma(x,y)}{\partial x^\mu\partial y^\nu} \,\right|\,g^{-1/2}(y)                        \label{PVVleck}
\ee
is the (dedensitized) Pauli--Van Vleck--Morette determinant built of
$\sigma(x,y)$. Both $\hat\delta(x,y)$ and $\hat K(s|\,x,y)$ are
defined above as zero weight tensor densities with respect to $x$ and
tensor densities of weight one with respect to $y$, which explains the
factor $g^{1/2}(y)$ in (\ref{heatexpansion}). 
The two-point matrix quantities $\hat a_n(x,y)$
bear the name of HAMIDEW \cite{Gibbons} or Gilkey-Seely coefficients
praising the efforts of mathematicians and physicists in heat kernel
theory \cite{DeWitt,Gilkey-Seeley} (see review of physics implications
of this theory in \cite{PhysRep,Avramidi,Vassilevich}). 

Substitution of the expansion (\ref{heatexpansion}) in (\ref{Gammarepr0}) 
expresses the UFTs in terms of the
coincidence limits
    \begin{eqnarray}
    &&\nabla_{\mu_1}...\nabla_{\mu_k}\sigma(x,y)\,
    \big|_{\,y=x},
    \qquad \nabla_{\mu_1}...\nabla_{\mu_k}
    {\cal D}^{1/2}(x,y)\,\big|_{\,y=x}, \qquad
    \nabla_{\mu_1}...\nabla_{\mu_k}\hat a_n(x,y)\,
    \big|_{\,y=x}.              \label{limits}
    \end{eqnarray}
Their remarkable property is that they are local functions of the
curvatures and their covariant derivatives. These functions can be
systematically calculated from the equation for the world function
$g^{\mu\nu}\nabla_\mu\sigma\nabla_\nu\sigma=2\sigma$ and the recursive
equations for $\hat a_n(x,y)$, which follow from the heat equation for
$\hat K(\,s\,|\,x,y\,)$. For obvious dimensional reasons the general
structure of these coincidence limits is the sum of various covariant
monomials of curvatures and their covariant derivatives of relevant
powers defined by $k$ and $n$. Since $\sigma(x,y)$ and ${\cal
  D}^{1/2}(x,y)$ are determined solely by the spacetime metric, the
first two sets in (\ref{limits}) are given by the sums of monomials of
the form 
    \begin{eqnarray}
    &&\nabla_1...\nabla_p\,\sigma(x,y)\,\big|_{\,y=x}\propto\;\;
    \stackrel{k}{\overbrace{\nabla\cdots\nabla}}\;
    \stackrel{m}{\overbrace{R
    \cdots R}},\quad 2m+k=p-2,      \label{localterms1}\\
    &&\nabla_1...\nabla_p\,{\cal D}^{1/2}(x,y)\,\big|_{\,y=x}\propto\;\;
    \stackrel{k}{\overbrace{\nabla\cdots\nabla}}\;
    \stackrel{m}{\overbrace{R
    \cdots R}},\quad 2m+k=p.
    \end{eqnarray}
in terms of purely metric curvatures $R$ (we suppress the tensor indices for
clarity), whereas the third set involves all the ``curvatures''
$\Re=(\hat P,\;\hat{\mbox{\boldmath$R$}}_{\mu\nu},\; R_{\mu\nu\alpha\beta})$
pertinent to the operator (\ref{F}) 
    \begin{eqnarray}
    \nabla_1...\nabla_p\,\hat a_n(x,y)\,\big|_{\,y=x}\propto\;\;
    \stackrel{k}{\overbrace{\nabla\cdots\nabla}}\;
    \stackrel{m}{\overbrace{\Re
    \cdots\Re}},\quad 2m+k=p+2n.      \label{localterms}
    \end{eqnarray}
In Appendix~\ref{App:HAMIDEW} we briefly describe the recursive
procedure of calculating all these coincidence limits. 

By adjusting the general technique to our
three-dimensional case, $d=3$, $\mu\mapsto i=1,2,3$, with the operator
$\hat F=\Delta$ ($\hat P=\frac16R\hat 1$) acting for the metric sector
in the space of symmetric covariant tensors $h_{kl}$ ($\hat
1=\delta_{ij}^{\;\;\;kl}$) and in the space of vectors $n^j$ and $c^j$
($\hat 1=\delta^i_j$) for the shift and ghost sectors respectively, we
obtain 
\be
\begin{aligned}
&\nabla_{i_1}\dots\nabla_{i_p}
\frac{\hat{1}}{(-\Delta)^{N+1/2}}\delta(x,y)\,\Big|_{\,y=x}\\
&=\frac{1}{\Gamma(N+1/2)}\frac{1}{8\pi^{3/2}}\int_0^\infty ds\,s^{N-2} \nabla_{i_1}\dots\nabla_{i_p}
 \Big[\,{\cal D}^{1/2}(x,y)
 e^{-\frac{\sigma(x,y)}{2s}}
 \sum_{n=0}^\infty s^n \hat{a}_n(x,y)\, \Big]_{\,y=x}\;.  \label{UFT100}
\end{aligned}
\ee
Here we have used that the UFTs needed for our calculation contain
half-integer powers of the Laplacian, as implied by
Eq.~(\ref{root_structure}). 
These UFTs have UV divergences of degree $p-2N+2$ (recall that the
delta function is three-dimensional), which correspond to the
proper time integrals diverging at $s=0$. In view of the growing power
of $s$ in this expansion, only a few first terms
will
contribute to the UV divergences, which makes the method highly
efficient. Among the divergences we will be interested only in
the logarithmic ones of the form $\int_0^\infty ds/s$ --- the divergent
coefficients of the counterterms of dimensionality 6, which determine
the beta functions of the theory.

\subsection{Types of universal functional traces}
\label{sec:UFTtypes}

Here we consider the types of universal functional traces arising in
the trace of the operator
$\mathbb{Q}$ regarding their number
of derivatives and the powers of the Laplacian acted upon by these
derivatives. We focus first on the tensor sector.
As stated in Sec.~\ref{Sect4}, the curvature expansion of
$\mathbb{Q_{\,D}}$ in
the  canonical form (\ref{root_structure}) reads 
    \begin{eqnarray}
    {\mathbb{Q_{\,D}}}_{ij}^{~~kl}=\Big[\,\sum_{a,p}
    {\cal R}_{(a)}\,\tilde\alpha_{a,p}\nabla_{1}...\nabla_{p}
    \frac1{(-\Delta)^{N+1/2}}\,\Big]^{\;\;\;kl}_{ij},
    \end{eqnarray}
where in each term we redefined the overall negative half-integer
power of the Laplacian as $N+1/2$ and the number of derivatives as $p$. 
Recall that ${\cal R}_{(a)}$ are the
background field tensors built of the curvature and its derivatives of
the dimensionality $a$ in units of inverse length, see
Eq.(\ref{dimR}). For $a=2$ this tensor is just the Ricci curvature
${\cal R}_{(2)}=R^{ij}$, for $a=3$ it is ${\cal R}_{(3)}=\nabla^k
R^{ij}$, etc. Obviously, at any $a$ the tensor quantity has {\em at
  maximum} $a$ indices, ${\cal R}_{(a)}={\cal R}_{(a)}^{i_1...i_r}$,
$r\leq a$. Also, as mentioned in Sec.~\ref{Sect4}, the parameter $N$ is not
independent but follows from the overall dimensionality of the
operator $\mathbb{Q_{\,D}}$ which is 3, so that $a+p-2N-1=3$ or 
    \begin{eqnarray}
    2N=a+p-4,                      \label{2N}
    \end{eqnarray}
whence it follows, in particular, that $a+p$ is always even (denoted
by $2k$ in (\ref{root_structure})). Thus, 
    \begin{eqnarray}
    {\rm Tr}\,\mathbb{Q_{\,D}}=\int d^3x\,\sum_{a,p}
    {\rm tr}\,\Big[\,{\cal R}_{(a)}\,\tilde\alpha_{a,p}\nabla_{1}...\nabla_{p}
    \frac1{(-\Delta)^{N+1/2}}\,
    \delta^{\;\;\;kl}_{ij}(x,y)\,\Big|_{\,y=x}\,\Big]\,,  \label{Trsqrt}
    \end{eqnarray}
where ${\rm tr}$ is the trace which is taken over the multi-indices
$ij$ and $kl$ {\em after} the action of every nonlocal operator on the
tensor delta function has been enforced. 

Another important point, also mentioned in Sec.~\ref{Sect4}, is that
for every $a$ there is an upper bound on the number of derivatives $p$
in these functional traces. Indeed, their $p$ indices can be
contracted at maximum with $r$ indices of ${\cal R}_{(a)}$ and 4
indices of $\delta^{\;\;\;kl}_{ij}(x,y)$.\footnote{If some of these
  $r+4$ indices are contracted with each other, then the possible
  number of derivatives is smaller because their indices cannot be
  contracted with anything else but those of the derivatives
  themselves. These contractions, however, do not count because they
  just shift the power $N$ of the Laplacian.} Therefore 
   $p\leq r+4$,
and in view of $r\leq a$ the upper bound on $p$ is
    \begin{eqnarray}
\label{pbound}
    p\leq a+4\;,
    \end{eqnarray}
which coincides with (\ref{indexbound}) for $K_{\rm free}=4$.
From (\ref{2N}) this leads to the upper bound on $N$,
    \begin{eqnarray}
\label{Nbound}
    N\leq a.
    \end{eqnarray}

In (\ref{Trsqrt}) every UFT with $p$ derivatives, which is conjugated
to the background field tensor ${\cal R}_{(a)}$ of dimensionality $a$, 
    \begin{eqnarray}
    T^{(a)}_p\equiv\nabla_{1}...\nabla_{p}
    \frac1{(-\Delta)^{N+1/2}}
    \delta^{\;\;\;kl}_{ij}(x,y)\,\Big|_{\,y=x}\,,\quad
    N=\frac{a+p}2-2,  \label{T}
    \end{eqnarray}
is divergent when its degree of divergence
$\Omega(T^{(a)}_p)=p-2N+2=6-a$ is positive, or $a\leq 6$. This, of
course, corresponds to the logarithmically divergent counterterms of
dimensionality 6. Therefore, the set of logarithmically divergent
terms in (\ref{Trsqrt}) is given by 
    \begin{eqnarray}
    a=0,2,3,4,6,
    \end{eqnarray}
where the contributions of $a=1$ and $a=5$ are absent
because there are no background field tensors of dimensionality $1$. 
Thus we have the following five sets of universal functional traces
which contribute to logarithmic divergences: 

\subsubsection*{Traces with $a=0$, $p\leq 4$, $p${\rm -even}, $N+\frac12=\frac{p-3}2$:}
    \begin{eqnarray}
    (-\Delta)^{3/2}\hat\delta(x,y)\,\big|_{\,y=x},\quad
    \nabla_{i_1}\nabla_{i_2}
    (-\Delta)^{1/2}
    \hat\delta(x,y)\,\big|_{\,y=x},\quad
    \nabla_{i_1}\nabla_{i_2}\nabla_{i_3}\nabla_{i_4}
    \frac{\hat 1}{(-\Delta)^{1/2}}
    \delta(x,y)\,\big|_{\,y=x}\;. \label{a_zero0}
    \end{eqnarray}
These are the most complicated ones, because they require the
knowledge of coincidence limits up to $\nabla^8\sigma(x,y)\,|_{y=x}$,
$\nabla^6{\cal D}^{1/2}(x,y)\,|_{y=x}$, 
$\nabla^6 \hat a_0(x,y)\,|_{y=x}$,
$\nabla^4\hat a_1(x,y)\,|_{y=x}$,
$\nabla^2\hat a_2(x,y)\,|_{y=x}$, and 
 $\hat a_3(x,x)$.

\subsubsection*{Traces with $a=2$, $p\leq 6$, $p${\rm -even}, $N+\frac12=\frac{p-1}2$:}
    \begin{eqnarray}
    (-\Delta)^{1/2}\hat\delta(x,y)\,\Big|_{\,y=x},\quad
    \nabla^2\frac{\hat 1}{(-\Delta)^{1/2}}
    \delta(x,y)\,\Big|_{\,y=x},\quad
    \nabla^4
    \frac{\hat 1}{(-\Delta)^{3/2}}
    \delta(x,y)\,\Big|_{\,y=x},\quad
    \nabla^6
    \frac{\hat 1}{(-\Delta)^{5/2}}
    \delta(x,y)\,\Big|_{\,y=x}
\label{a_two}
    \end{eqnarray}
Here and in what follows we omit for brevity the indices of derivatives.

\subsubsection*{Traces with $a=3$, $p\leq 7$, $p${\rm -odd}, $N+\frac12=\frac{p}2$:}
    \begin{eqnarray}
    \nabla\frac{\hat 1}{(-\Delta)^{1/2}}\delta(x,y)\,\Big|_{\,y=x},\quad
    \nabla^3\frac{\hat 1}{(-\Delta)^{3/2}}
    \delta(x,y)\,\Big|_{\,y=x},\quad
    \nabla^5
    \frac{\hat 1}{(-\Delta)^{5/2}}
    \delta(x,y)\,\Big|_{\,y=x},\quad
    \nabla^7
    \frac{\hat 1}{(-\Delta)^{7/2}}
    \delta(x,y)\,\Big|_{\,y=x}
\label{a_three}
    \end{eqnarray}

\subsubsection*{Traces with $a=4$, $p\leq 8$, $p${\rm -even},
  $N+\frac12=\frac{p+1}2$:}
    \begin{eqnarray}
    &&\frac{\hat 1}{(-\Delta)^{1/2}}\delta(x,y)\,\Big|_{\,y=x},\quad
    \nabla^2\frac{\hat 1}{(-\Delta)^{3/2}}
    \delta(x,y)\,\Big|_{\,y=x},\quad
    \nabla^4
    \frac{\hat 1}{(-\Delta)^{5/2}}
    \delta(x,y)\,\Big|_{\,y=x},\nonumber\\
    &&\nabla^6
    \frac{\hat 1}{(-\Delta)^{7/2}}
    \delta(x,y)\,\Big|_{\,y=x},\quad
    \nabla^8
    \frac{\hat 1}{(-\Delta)^{9/2}}
    \delta(x,y)\,\Big|_{\,y=x}
\label{a_four}
    \end{eqnarray}

\subsubsection*{Traces with $a=6$, $p\leq 10$, $p${\rm -even},
  $N+\frac12=\frac{p+3}2$:} 
    \begin{align}
    &\frac{\hat 1}{(-\Delta)^{3/2}}\delta(x,y)\,\Big|_{\,y=x},\quad
    \nabla^2\frac{\hat 1}{(-\Delta)^{5/2}}
    \delta(x,y)\,\Big|_{\,y=x},\quad
    \nabla^4\frac{\hat 1}{(-\Delta)^{7/2}}
    \delta(x,y)\,\Big|_{\,y=x}\nonumber\\
    &\nabla^6\frac{\hat 1}{(-\Delta)^{9/2}}
    \delta(x,y)\,\Big|_{\,y=x},\quad
    \nabla^8
    \frac{\hat 1}{(-\Delta)^{11/2}}
    \delta(x,y)\,\Big|_{\,y=x},\quad
    \nabla^{10}
    \frac{\hat 1}{(-\Delta)^{13/2}}
    \delta(x,y)\,\Big|_{\,y=x}
\label{a_six}
    \end{align}
In this fifth group the number of derivatives is high, but these
traces are the simplest ones because they can be calculated in flat
space and reduce to symmetrized products of the metric tensor. 

The classification of the UFTs in the vector sector proceeds
similarly. The only difference is the modification of the bounds
(\ref{pbound}), (\ref{Nbound}) due to the different number of free
indices of $\mathbb{Q_{\,B}}^i_{~j}$. We now have
\be
p\leq a+2~,\qquad N\leq a+1\;,
\ee
which removes the last entry at each order in the list of possible
structures 
(\ref{a_zero0})---(\ref{a_six}). 

We have computed the divergences of all the needed tensor and vector
UFTs using symbolic computer algebra \cite{xAct}. The code is
available at \cite{github}.
The full expressions are very
long, so we do not write them explicitly. 
In Appendix~\ref{App:TUFTs} we present 
the most laborious traces with $a=0$ and $a=2$.
Some relations
between the tensor, vector and scalar functional traces, which can be
obtained by integration by parts, are discussed in Appendix
\ref{UFTcheck}. These relations can be used as a powerful check of the
explicit results for their divergences.

\section{Beta functions}
\label{btfuns}

In the last step of our calculation we combine the operator square root
extracted with the procedure described in Sec.~\ref{sec:pert} with the
results for UFTs enumerated in Sec.~\ref{sec:UFTtypes} to obtain the
divergent part of the one-loop effective action. Namely, we use
Eq.~(\ref{root}), in which we substitute Eq.~(\ref{Trsqrt}) (and a
similar equation for $\Tr\,\mathbb{Q_{\,B}}$). Upon collecting similar
terms, integration by parts and use of Bianchi identities, 
the divergence takes the form,
\begin{equation}
\begin{aligned}\label{ctnu}
\varGamma^{\rm 1-loop}\big|^{\,\rm div} = \ln L^2\int d\tau\,d^3x\,\sqrt{g}\,\bigl( C_{\nu_1} R^3 + C_{\nu_2} R R_{ij}R^{ij} +C_{\nu_3} R^i_j R^j_k R^k_i 
+ C_{\nu_4} \nabla_i R \nabla^i R + C_{\nu_5} \nabla_i R_{jk}
\nabla^i R^{jk} \bigr).
\end{aligned}
\end{equation}
Only the potential part of the action has logarithmic divergences on
our static background. The coefficients $C_{\nu_a}$, which are
functions of the couplings $\l$, $\nu_1,\ldots,\nu_5$, represent the
key result of the calculation. 

The UV divergent factor $\ln L^2$ is related to the integral over the
proper-time parameter,
\be
\label{lnLds/s}
\ln L^2=\int \frac{ds_2}{s_2}\;.
\ee
This integral comes from the heat-kernel representation of the powers
of spatial Laplacian, Eq.~(\ref{Gammarepr0}). Hence, the dimension of
the proper time hear is $[s_2]=-2$,
which we highlighted by the
subscript.\footnote{This is an important
  difference from the diagrammatic calculation of \cite{towards} which
used the proper-time representation for the full field propagators in
$(3+1)$-dimensional spacetime, and hence the proper-time dimension was
$-6$.} This means that the divergent logarithm is related to the
momentum renormalization scale $k_*$ as
\be
\ln L^2 \simeq \ln
\left( \frac{\Lambda^2_{\rm UV}}{k_*^2} \right), \label{ds/s}
\ee
where $\Lambda_{UV}$ is a UV cutoff.

We are now ready to compute the $\beta$-functions of the couplings
$\nu_a$, $a=1,\dots,5$. Comparing Eqs.~(\ref{action31}) and
(\ref{ctnu}), we read off the renormalized combinations of the coupling
constants 
\begin{equation}
\left( \frac{\nu_a}{2G} \right)_{\rm ren}=  \frac{\nu_a}{2G} 
+ C_{\nu_a}\ln L^2,
\end{equation} 
whence
\begin{equation}
\beta_{\nu_a}\equiv \frac{d\nu_{a,\,{\rm ren}}}{d\ln k_*}
= -4 G C_{\nu_a}
+\nu_a\frac{\beta_G}G,\qquad a=1,2,...,5.         \label{beta_nu}
\end{equation}
Therefore, the potential term $\beta$-functions are expressed via
constants $C_{\nu_a}$ and the $\beta$-function of $G$, 
\be
\beta_G \equiv \frac{d G_{\rm ren}}{d\ln k_*},
\ee
which was previously obtained in \cite{towards} (see also
Appendix~\ref{app:betaG}).  

Neither $\beta_G$, not $\beta_{\nu_a}$ are gauge invariant. It is
well-known that a change of gauge adds to the background effective
action a linear combination of the equations of motion
\cite{DeWitt:1967ub,Kallosh:1974yh}. Such contribution vanishes
on-shell, but not for our off-shell background. This leads to gauge
dependence of the one-loop effective action and the renormalized
couplings. 
As shown in Ref.~\cite{towards}, this dependence amounts to a
one-parameter family of transformations which, for an infinitesimal
change of gauge, have the form,
\be
\label{gaugechange}
G\mapsto G-2G^2\epsilon\;,~~~~~\nu_a\mapsto \nu_a-4G\nu_a\epsilon\;,
\ee  
where $\epsilon$ is an infinitesimal parameter. 

We can now construct combinations that are invariant under these
transformations and whose $\beta$-functions, therefore, must be
gauge-invariant. In this way we arrive at the set of essential
couplings (\ref{new_couplings}). Their running is easily obtained from
$\beta_{\nu_a}$, $\beta_{G}$ and $\beta_\l$ (see
Eq.~(\ref{betalambda})):
\bseq
\begin{align}
&\beta_{\cal G} = {\cal G}\left( \frac{\beta_G}G -\frac12
  \frac{\beta_{\nu_5}}{\nu_5} \right),\\
\label{betavs}
&\beta_{v_a} = \frac1{\nu_5}\left( \beta_{\nu_a} -
  \nu_a\frac{\beta_{\nu_5}}{\nu_5} \right), \quad a=1,2,3,\\ 
&\beta_{u_s}=\frac{u_s\beta_\lambda}{(1-\l)(1-3\l)}
+\frac{4(1-\l)\beta_{v_4}}{(1-3\l)u_s}\;,
\end{align}
\eseq
where $v_4=\nu_4/\nu_5$ and its $\beta$-function is defined in the
same way as in (\ref{betavs}) at $a=4$. 
This leads us to our main results, Eqs.~(\ref{betas_new}),
(\ref{Achi}), (\ref{Pol_calG})---(\ref{Pol_v3}). 

We have calculated $\beta_{\cal G}$ in three different gauges {\bf
  a}, {\bf b}, {\bf c} and $\beta_{u_s}$,
$\beta_{v_a},~a=1,2,3$ in four gauges {\bf
  a}, {\bf b}, {\bf c}, {\bf d} from Sec.~\ref{sec:3.1}. 
We have found identical results. All
steps of the calculation were performed by two independent codes ---
one for gauges  {\bf a}, {\bf b} and one for gauges {\bf c}, {\bf
  d}. Notice that though the final results agree, the intermediate
expressions differ significantly in different gauges. In particular,
the coupling-dependent coefficients in
the square-root operator $\mathbb{Q_{\,D}}$ (which contains a few
thousands of distinct tensor structures) are dramatically different in
gauges {\bf a}, {\bf b} and {\bf c}, {\bf d}. In general, they are
rational functions with the denominator being a product of
combinations $(u_\alpha+u_\beta)$, where $u_{\alpha}$,
$\alpha=T,V,S1,S2$, 
are the 
eigenvalues of the principal symbol $\mathbb{Q_{\,D}}({\bf p})$
defined in Eq.~(\ref{allus}). This follows from the formula for the
solution of the
Sylvester equation (\ref{Sylsol}) used at each iteration of the
perturbative procedure to construct $\mathbb{Q_{\,D}}$. In the gauges
{\bf a}, {\bf b} the eigenvalues corresponding to the gauge modes 
coincide with those of the physical modes, so that in the denominator
of the coefficients we get only the powers of $u_s$ and $(1+u_s)$. On
the other hand, in gauges {\bf c} and {\bf d} the gauge eigenvalues
are different and we obtain multiple extra factors $u_V$, $(1+u_V)$,
$u_{S2}$, $(1+u_{S2})$, etc. All these extra factors cancel in the
essential $\beta$-functions, which provides a very powerful check of
the correctness and consistency of our result.  

Finally, let us make the following comment.
Once the gauge invariance of the essential $\beta$-functions has been
explicitly checked, we can invert the logic and derive the coeffcients
$C_{\nu_a}$ in the one-loop effective action for arbitrary values of
the gauge parameters $\sigma,\xi$. Indeed, the gauge invariance
of the $\beta$-functions for the ratios $v_a=\nu_a/\nu_5$, $a=1,2,3,4$,
implies that gauge-dependent
parts of the coefficients $C_{\nu_a}$ in the one-loop effective action
are proportional to
the couplings $\nu_a$ themselves with a common proportionality
factor,
\be
\label{Cgauge}
C_{\nu_a}^{\rm gauge}=\nu_a\,
\Xi(\lambda,\{\nu\},\sigma,\xi)~,~~~~~a=1,\ldots,5\;. 
\ee 
Adding to this the invariance of the $\beta$-function for the
essential coupling ${\cal G}$, one derives the gauge dependent parts
of the $\beta$-functions for $G$ and $\nu_a$,
\be
\label{betagauge}
\beta_{G}^{\rm gauge}= -4G^2\,\Xi~,~~~~~
\beta_{\nu_a}^{\rm gauge}= -8G\nu_a\,\Xi\;.
\ee 
The function $\Xi$ can be fixed by a calculation of the effective
action on a simple special background that can be carried out in a
general $(\sigma,\xi)$-gauge. This task is performed in the next
section and yields a remarkably simple result, see
Eq.~(\ref{gfromsphere}). In the Supplemental
Material we provide a Mathematica file with the coefficients
$C_{\nu_a}$ in arbitrary $(\sigma,\xi)$-gauge, obtained by adding the
gauge-dependent piece (\ref{Cgauge}) to our explicit results in gauges
{\bf a}, {\bf b}, {\bf c}, {\bf d}.

\section{Additional check: Effective action on $R^{1}\times S^3$ }
\label{spherecheck} 

The complexity of our calculation for the full set of beta functions
imposes the necessity of its efficient verification. 
It is based on the UFT method of \cite{PhysRep,twoloop,Scholarpedia}
which, despite its power, is not commonly used in the literature and
therefore requires detailed validation and caution. While the gauge
independence of the essential $\beta$-functions discussed above
already provides a strong argument in favor of the validity of our
approach, we perform one more check using an alternative calculational
scheme. Namely, we compute the divergence of the one-loop effective
action of projectable HG on a static background with spherical
spatial slices using spectral decomposition for the differential
operators $\mathbb{D}$ and $\mathbb{B}$ entering the potential part of
the action. The traces of their square roots are found by means of
spectral summation, for which we use two different regularizations ---     
dimensional and
$\zeta$-functional one. Due to the simplicity of the background, this
calculation can be carried out in an arbitrary $(\sigma,\xi)$-gauge
introduced in Sec.~\ref{gaugefix}. As a byproduct, it fixes 
the function
$\Xi$ from Eq.~(\ref{Cgauge}), and hence allows us to 
completely determine the
dependence of the coefficients $C_{\nu_a}$ in Eq.~(\ref{ctnu}) on the
gauge choice.
As another byproduct, we derive
the logarithmic dependence of the renormalized partition function of
HG on $R^1\times S^3$ on the radius of the sphere.

Consider a static spacetime with spherical 3-dimensional slices of
inverse square radius radius $\kappa$. We have
    \begin{eqnarray}
    \quad R_{ij}=2\kappa g_{ij},\quad R=6\kappa.        \label{S-metric}
    \end{eqnarray} 
On this background, the general expression (\ref{ctnu}) for the
divergent part of the effective action reduces to 
    \begin{eqnarray}\label{GammaS3}
    \varGamma^{\rm 1-loop}\,\Big|^{\,\rm div}_{\,R^1\times S^3} 
=\ln L^2\sum\limits_{a=1}^3 C_{\nu_a}\int d\tau\, I_a,
    \end{eqnarray}
where $I_a$ are the following three nonvanishing invariants 
\bseq
\label{Is}
    \begin{eqnarray}
    &&I_1=\int d^3x\,\sqrt{g}\,R^3\,\Big|_{\,S^3}=9\times
    48\pi^2\kappa^{3/2},\label{I1}\\ 
    &&I_2=\int d^3x\,\sqrt{g}\,RR_{ij}R^{ij}\,\Big|_{\,S^3}=3\times
    48\pi^2\kappa^{3/2},\label{I2}\\ 
    &&I_3=\int d^3x\,\sqrt{g}\,R^i_jR^j_kR^k_i\,\Big|_{\,S^3}
    =48\pi^2\kappa^{3/2}.
\label{I3}
    \end{eqnarray}
\eseq
Therefore, an independent calculation of this divergent part of
$\varGamma^{\rm 1-loop}$ on $R^1\times S^3$ provides a check of the
linear combination $9C_{\nu_1}+3C_{\nu_2}+C_{\nu_3}$. Notice that this
combination is gauge-dependent, so the comparison between the general
result and the calculation on the sphere must be performed in the same
gauge.

\subsection{Tensor and vector operators on $S^3$}

Our starting point is the formula (\ref{root}) for the one-loop
effective action.
On the homogeneous space --- a sphere $S^3$ --- the tensor and
vector operators can be explicitly diagonalized, and the
functional traces of their square roots can be represented as spectral
sums of square roots of their eigenvalues. Then the UV divergences can
be obtained under appropriate (dimensional or $\zeta$-functional)
regularization of these spectral sums.

Diagonalization of the tensor operator takes place in the complete
orthonormal basis of tensor harmonics $H^{A\,(n)}_{ij}$ which we
present, for the sake of dimensional regularization, on the
$d$-dimensional sphere $S^d$. Here $A=t,\,v,\,s1,\,s2$ is the helicity
index running over tensor, vector and two scalar polarizations
contained in the metric, whereas $(n)$ enumerates all other quantum
numbers at the level $n$.
In the basis of these harmonics,
    \begin{eqnarray}
    &&h_{ij}(x) = \sum_{A,(n)}h_{A\,(n)}
    H_{ij}^{A\,(n)}(x),   \label{2.3}
    \end{eqnarray}
the operator ${\mathbb{D}}$, takes the following block-diagonal form
    \begin{equation}
    {\mathbb{D}}\,\Big|_{\,S^3} =
    \begin{bmatrix}
    \;\mathbb{D}_t & 0 & 0 & 0\; \\
    \;0 & \mathbb{D}_v & 0 & 0\; \\
    \;0 & 0 & \mathbb{A}_{11} & \mathbb{A}_{12}\; \\
    \;0 & 0 & \mathbb{A}_{21} & \mathbb{A}_{22}\;
    \end{bmatrix}\equiv {\rm diag}\;[\;\mathbb{D}_t,\mathbb{D}_v,\mathbb{D}_s\,].     \label{2.14}
    \end{equation}
The harmonics which provide this property can in their turn be
expressed in terms of complete and orthonormal sets of
transverse-traceless tensor $h^{TT\,(n)}_{ij}(x)$, transverse vector
$\xi^{(n)}_i(x)$ and scalar $\phi^{(n)}(x)$ eigenfunctions of the
covariant Laplacian $\Delta=g^{ij}\nabla_i\nabla_j$ (see
Appendix~\ref{app:modes} for details),
\bseq
\label{allHs}
   \begin{align}
    &H^{t\,(n)}_{ij}(x)
    =h^{TT\,(n)}_{ij}(x),\qquad &n\geq 2\;,\\
&H_{ij}^{v\,(n)}(x)
    = 2\nabla_{(i}
    \frac1{\sqrt{2\left(-\Delta-\frac1d R\right)}}\,
    \xi^{(n)}_{j)}(x), \qquad& n\geq
    2\;,                                   \label{2.4}\\
    &H_{ij}^{s1\,(n)}(x)= \Big(\nabla_i\nabla_j - \frac1d\,g_{ij}\Delta\Big)\,
    \frac{1}{\sqrt{\frac{d-1}{d}(-\Delta)^2-\frac1d R(-\Delta)}}\,\phi^{(n)}(x),\qquad &n\geq 2\;,\label{Hs1}\\
    &H_{ij}^{s2\,(n)}(x)=
    \frac1{\sqrt d}\, g_{ij}\,\phi^{(n)}(x),\qquad &n\geq 0\;.              \label{2.5}
    \end{align}
\eseq
We will denote their relevant eigenvalues as $\Delta_n^t$,
$\Delta_n^v$ and $\Delta_n^s$ and write their orthonormality
conditions in the form
\bseq
\label{allmodes}
    \begin{align}
    &\Delta h^{TT\,(n)}_{ij}(x)=\Delta_n^t\,h^{TT\,(n)}_{ij}(x),\quad
    &&\int d^dx\,\sqrt g\,h^{TT\,(n)}_{ij}(x)\,h_{TT\,(m)}^{ij}(x)
    =\delta^{(n)}_{(m)},                        \label{tensor_modes}\\
    &\Delta \xi^{(n)}_i(x)=\Delta_n^v\,
    \xi^{(n)}_i(x), \quad
    &&\int d^dx\,\sqrt g\,\xi^{(n)}_{i}(x)\,\xi_{(m)}^{i}(x) =\delta^{(n)}_{(m)},                     \label{vector_modes}\\
    &\Delta \phi^{(n)}(x)=\Delta_n^s\,\phi^{(n)}(x), \quad
    &&\int d^dx\,\sqrt g\,\phi^{(n)}(x)\,
    \phi_{(m)}(x)=\delta^{(n)}_{(m)}           \label{scalar_modes}
    \end{align}
\eseq
Integer quantum numbers $(n)$ enumerating these eigenfunction are of
course different for tensor, vector and scalar modes, but we will not
introduce for them different notations, for in what follows we will
need for each $(n)$ only the eigenvalue $\Delta_n$ and its degeneracy
$D_n$ -- the dimensionality of eigenvalue subspace. In generic
dimension $d$, which we need for the sake of dimensional
regularization, they read on the sphere of unit radius 
\cite{GibbonsPerry,Duff,Tseytlin,Litim}
\bseq
    \begin{align}
    &-\Delta_n^s=n(n+d-1),\quad  D_n^s=\frac{(2n+d-1)(n+d-2)!}{n!(d-1)!},\quad &n\geq 0\,,   \label{2.13a}\\
    &-\Delta_n^v=n(n+d-1)-1,\quad  D_n^v=\frac{n(n+d-1)(2n+d-1)(n+d-3)!}{(d-2)!(n+1)!},\quad &n\geq 1\,,   \label{2.13b}\\
    &-\Delta_n^t=n(n+d-1)-2,\;
    D_n^t=\frac{(d+1)(d-2)(n+d)(n-1)(2n+d-1)(n+d-3)!}{2(d-1)!(n+1)!},\;& n\geq 2\,.   \label{2.13c}
   \end{align}
\eseq
In three dimensions the above complicated expressions for degeneracies
simplify to
    \begin{eqnarray}
    &&D_n^s=(n+1)^2,  \quad D_n^v=2n(n+2),\quad D_n^t=
    2(n-1)(n+3). \label{d3dims}
    \end{eqnarray}

The blocks of the matrix (\ref{2.14}) for the $n$-th level have the form of
the functions of $\Delta_n$ times the relevant $D_n\times D_n$
unit matrices $\delta_{(n)}^{(m)}$,
\bseq
\label{matrixel}
    \begin{eqnarray}
    &&\mathbb{D}_t(n)\,\delta_{(n)}^{(m)}=\int d^dx\,\sqrt{g}\, H^{ij}_{t\,(n)}(x)\, \mathbb{D}_{ij}^{\:\:\:\:kl}H_{kl}^{t\,(m)}(x),\\
    &&\mathbb{D}_v(n)\,\delta_{(n)}^{(m)}=\int d^dx\,\sqrt{g}\, H^{ij}_{v\,(n)}(x)\,
    \mathbb{D}_{ij}^{\:\:\:\:kl}
    H_{kl}^{v\,(m)}(x),                              \label{2.15}\\
    &&\mathbb{A}_{ab}(n)\,\delta_{(n)}^{(m)}=\int d^dx\,\sqrt{g}\, H^{ij}_{sa,\,(n)}(x)\, \mathbb{D}_{ij}^{\:\:\:\:kl}
    H_{kl}^{sb,\,(m)}(x), \quad a=1,2,\quad b=1,2.   \label{2.16}
    \end{eqnarray}
\eseq
They are calculated using the mode normalization  and
relations (\ref{2.11})---(\ref{2.12}) of Appendix~\ref{app:modes}.
Their expressions, which are too lengthy to be presented
explicitly, schematically read 
\bseq
\label{Dns}
    \begin{eqnarray}
    &&\mathbb{D}_t(n)=\kappa^3\,T_{(3)}(-\Delta_n^t),
              \label{TensorLap}\\
    &&\mathbb{D}_v(n)=\kappa^3\,V_{(3)}(-\Delta_n^v)\,,
                 \label{2.18}\\
    &&\mathbb{A}_{11}(n)=\kappa^3
    \frac{S_{(5)}(-\Delta_n^s)}{(\Delta_n^s)^2
    +d\,\Delta_n^s} 
    ,\,\,
    \mathbb{A}_{12}(n)=\mathbb{A}_{21}(n)=\kappa^3
    \frac{S_{(4)}(-\Delta_n^s)}{\sqrt{(\Delta_n^s)^2
    +d\,\Delta_n^s}},\,\,
    \mathbb{A}_{22}(n)=\kappa^3
   S_{(3)}(-\Delta_n^s),
                \label{2.19}
    \end{eqnarray}
\eseq
where $T_{(q)}$, $V_{(q)}$, $S_{(q)}$, $q=3,4,5$, are polynomials of
$q$-th order in their argument, and the denominators in
$\mathbb{A}_{11}(n)$, $\mathbb{A}_{12}(n)$ and $\mathbb{A}_{21}(n)$
follow from the normalization factor in (\ref{Hs1}). Cancellation of
similar denominators in (\ref{2.18}) occurs due to equations
(\ref{diagvec}) and (\ref{2.11}) of Appendix~\ref{app:modes}. 

The total $2\times 2$ scalar block of the operator (\ref{2.14}),
$\mathbb{D}_s=\mathbb{D}_{s,ab}$, is still not diagonal, but in each
$n$-th eigenvalue subspace it can be diagonalized in the basis of
finite-dimensional matrix eigenvectors
$\Upsilon(n)=\Upsilon_{a(b)}(n)$ and $\Upsilon^\dagger(n)=
\Upsilon^\dagger_{(b)a}(n)$, $(b)=\pm$, $a=1,2$,
    \begin{eqnarray}
    &&\mathbb{A}(n)=\Upsilon(n)\,
    \begin{bmatrix}
    \;\varLambda_+(n) & 0\; \\
    \;0 & \varLambda_-(n)\;
    \end{bmatrix}\,\Upsilon^\dagger(n),\quad
    \Upsilon^\dagger(n)\Upsilon(n)\equiv
    \sum_c\Upsilon^\dagger_{(a)c}(n)\Upsilon_{c(b)}(n)=\delta_{(a)(b)},\\
    &&\Lambda_\pm(n) = \frac12\left(\mathbb{A}_{11}(n)+\mathbb{A}_{22}(n)\pm\sqrt{\mathbb{A}_{11}^2(n)
    +\mathbb{A}_{22}^2(n)-2\mathbb{A}_{11}(n)\mathbb{A}_{22}(n)
    +4\mathbb{A}_{12}(n)\mathbb{A}_{21}(n)}\right).
    \end{eqnarray}
As the result, the operator $\mathbb{D}$ becomes diagonal in all of
its sectors, and the unregulated spectral sum representation of the
functional trace of its square root, $\mathbb{Q}=\mathbb{D}^{1/2}$,
takes the form
    \begin{equation}\label{decouple}
    \begin{aligned}
    {\rm Tr}\,\mathbb{Q_{\,D}}\,\Big|_{\,S^3} &=
\int d^dx\, \sqrt{g}\,\sum\limits_{A,(n)} H^{kl}_{A\, (n)}(x)
\big(\sqrt{{\mathbb{D}}}\big)_{\,kl}^{\,\:\:\:\:ij}\,
    H_{ij}^{A\,(n)}(x)\\
    &=\sum_{n={2}}^\infty D_n^t \sqrt{\mathbb{D}_t(n)}
    +\sum_{n=2}^\infty D_n^v \sqrt{\mathbb{D}_v(n)}
    +\sum_{n=2}^\infty D_n^s \sqrt{\varLambda_+(n)}+
     \sum_{n=0}^\infty D_n^s \sqrt{\varLambda_-(n)}.
    \end{aligned}
    \end{equation}
Summation in the tensor, vector and scalar ``$+$'' sectors starts with
$n=2$, whereas in the scalar ``$-$'' sector it starts from $n=0$, in
accordance with the restrictions on $n$ in (\ref{allHs}).

The calculation of ${\rm Tr}\,\mathbb{Q_{\,B}}$ on $S^3$ in the ghost
and shift sectors proceeds along the same lines, except that we can,
in view of simplicity of these sectors, explicitly present the
expressions for the corresponding operators. In particular, the operator
$\mathbb{B}$ defined by Eq.~(\ref{B}), when converted to the canonical
form,  reads on $S^3$ as
\begin{equation}\label{BonS3}
\begin{aligned}
\mathbb{B}^i_{~j}\,\Big|_{\,S^3} =
\frac{1}{2\sigma} \Bigl[&\delta^i_j (-\Delta)^3 +
\big(1-2(1-\l)(1+\xi)\big)\nabla^i\nabla_j(-\Delta)^2 - 
2\kappa\delta^i_j(-\Delta)^2 \\
&-4(1-2\lambda)\xi\kappa\,\nabla^i\nabla_j(-\Delta)
+8\l\xi\kappa^2\,\nabla^i\nabla_j\Bigr]\;.
\end{aligned}
\end{equation}
In the basis of transverse and longitudinal vector modes,
\begin{equation}\label{decomp1}
c^j(x) = \sum_{(n)}c_{(n)}^T\xi^j_{(n)}(x)
+ \sum_{(n)}c_{(n)}^L\nabla^j
\frac{1}{\sqrt{-\Delta}} \phi^{(n)}(x),
\end{equation}
where $\xi^j_{(n)}\equiv g^{ji}\xi_i^{(n)}$ and $\phi^{(n)}$ are
orthonormal sets of transverse vector and scalar Laplacian
eigenfunctions (\ref{vector_modes}) and (\ref{scalar_modes})
introduced above, this operator similarly to (\ref{2.14}) becomes
diagonal $\mathbb{B} ={\rm diag}\, [\,\mathbb{B}_T,
\,\mathbb{B}_L\,]$. Here, as it follows from (\ref{BonS3}),
\bseq
\begin{align}
&\mathbb{B}_T = \frac{\kappa^3}{2\sigma} \left[(-\Delta^v_n)^3 -
  2(-\Delta^v_n)^2\right]\,\delta^{(n)}_{(m)}
\equiv\mathbb{B}_T(n)\,\delta^{(n)}_{(m)}\;,\\
&\mathbb{B}_L = \frac{\kappa^3}{2\sigma} \Big[2(1-\l)(1+\xi)(-\Delta^s_n)^3 - 4(\xi-2\l+3)(-\Delta^s_n)^2 + 8(3-\lambda)(-\Delta^s_n)-16\Big]
\equiv\mathbb{B}_L(n)\,\delta^{(n)}_{(m)}\;,
\end{align}
\eseq
and
\begin{equation}
{\rm Tr}\,\mathbb{Q_{\,B}}\,\Big|_{\,S^3} = \sum_{n=1}^\infty D_n^v
\sqrt{\mathbb{B}_v(n)}
+ \sum_{n=1}^\infty D_n^s \sqrt{\mathbb{B}_s(n)}.
\end{equation}
The vector modes sum starts with $n=1$ because $D_0^v=0$, whereas the
scalar modes sum begins with $n=1$ because the expansion
(\ref{decomp1}) does not include the zero mode of the scalar
Laplacian.

\subsection{Dimensional and $\zeta$-functional regularization of spectral sums}
Regularization and extraction of divergences by dimensional
regularization consists in extension of these sums to space
dimensionality $d=3-\varepsilon$ with $\varepsilon\to 0$. In the
$\zeta$-functional regularization the square root power of the
operator $\mathbb{F}=(\mathbb{D}, \mathbb{B})$ is analytically
continued to $1/2-\varepsilon'$.
\begin{equation}\label{zetapower}
    {\rm Tr}\,\sqrt{{\mathbb{F}}}\,\Big|_{\,\zeta-{\rm
        reg.}}={\rm Tr}\,{\mathbb
      F}^{\frac12-\varepsilon'},
    \end{equation}
which is provided by the replacement of all square roots in the above
formula (\ref{decouple}) by this power while keeping $d=3$.
We identify
\be
\label{dimzeta}
\varepsilon'=\varepsilon/6\;,
\ee
as implied by the dimensionality of
the operator.\footnote{
\label{foot:epsilons}
Indeed, both in dimensional and $\zeta$-regularizations
one has to introduce a dimensionful parameter $k_*$ to keep the
dimension of the operator trace equal to 3.
Equating the powers of this
parameter in the two cases, we obtain the relation
(\ref{dimzeta}).}
One can
show then that the resulting divergent parts of the trace --- the pole terms in
$\varepsilon$ --- coincide in both regularizations. Below we
demonstrate this on the example of the tensor sector.

For the dimensionality $d=3-\varepsilon$ the multiplicity of the
$n$-th eigenvalue in the transverse-traceless tensor sector has for
large $n\gg 1$ the following form
    \begin{eqnarray}
    D_n^t=
    \frac{(d+1)(d-2)(n+d)(n-1)(2n+d-1)(n+d-3)!}{2(d-1)!(n+1)!}
=2\,n^{2-\varepsilon}\left(1+\frac2n-\frac3{n^2}+O(\varepsilon)\right),
    \end{eqnarray}
where we used that
    \begin{equation}
    \frac{\Gamma(n+1-\varepsilon)}{\Gamma(n+1)}
    = n^{-\varepsilon}\left(1 + O(\varepsilon) \right),\quad n\to\infty\;.
    \end{equation}
This yields,
    \begin{equation}
    \begin{aligned}
    {\rm Tr}\,\sqrt{\mathbb{D}_t}\,\Big|_{\,\rm dim. reg.}
    =\kappa^{3/2}\sum_{n=2}^\infty D^t_n\,\Big\{&\,T_{(3)}\big(n(n+d-1)-2\big)\Big\}^{1/2}
    =\sum_{n=2}^\infty n^{5-\varepsilon}
    G_t\Big(\frac1n,\varepsilon\Big).    \label{4.5}
    \end{aligned}
    \end{equation}
Here we have explicitly disentangled the fractional power of the
growing factor $n^{5-\varepsilon}$ as the coefficient of the function
$G_t\Big(\frac1n,\varepsilon\Big)$ which is regular at $n\to\infty$,
\begin{equation}
\begin{aligned}\label{Treg}
G_t\Big(\frac1n,\varepsilon\Big)
=\,2\kappa^{3/2}\bigg(1+\frac2n-\frac3{n^2}\bigg)\Bigg\{&\,\frac1{n^6}\,T_{(3)} (n^2+2n-2)\,\Bigg\}^{1/2}+O(\varepsilon)
\end{aligned}
\end{equation}
(remember that $T_{(3)} (n^2+2n-2)$ is six order polynomial in $n$).

The divergent pole in $\varepsilon$ of this series can be extracted
using the Abel--Plana formula which expresses a discrete series $\sum_nf(n)$
in terms of the sum of the integral along the real axes of $n$ and the
integral of $f(z)$ on the imaginary axis,
    \begin{equation}
    \begin{aligned}
    \sum_{n=0}^\infty f(n) = \int_0^\infty
dz\,f(z)+\frac12\,f(0)+i\int_0^\infty dz\,
    \frac{f(iz)-f(-iz)}{\exp(2\pi z)-1}.
    \end{aligned}
    \end{equation}
With $f(n)=(n+2)^{5-\varepsilon} G_t(1/(n+2),\varepsilon)$ the latter integral
is convergent being exponentially damped at infinity, while the divergence of
the integral over the real axes at $n\to\infty$ can be isolated by changing
the integration variable, $n+2=1/y$, and integrating the needed number of
times by parts.
In the domain of convergence $\varepsilon>6$ integration by
parts does not give extra terms at $y=0$, so that the analytic
continuation to $\varepsilon=0$ yields UV divergences as a pole term
    \begin{equation}
    \begin{aligned}
    {\rm Tr}\,\sqrt{\mathbb{D}_t}\,\Big|^{\,\rm div}_{\rm dim.reg.}&
= \int_2^{\infty} dn\, n^{5-\varepsilon} G_t\Big(\frac1n,\varepsilon \Big)\,\Big|^{\,\rm div}
    = \int_0^{1/2} \frac{dy}{y^{7-\varepsilon}}
    G_t(y,\varepsilon)\,\Big|^{\,\rm div}\\
    &=\frac{1}{\varepsilon-6}\frac{1}{\varepsilon-5}
    \dots\frac{1}{\varepsilon-1}\frac1\varepsilon\int_0^{1/2} dy\,
    y^\varepsilon \frac{d^6 G_t(y,\varepsilon)}{dy^6}\,\Big|^{\,\rm div}= \frac1{6!\,\varepsilon}
    \frac{d^6 G_t(y,0)}{dy^6}\,\Big|_{\,y=0}. \label{dimreg}
    \end{aligned}
    \end{equation}

This result agrees with the zeta-function
regularization. Indeed, we have
\be
\begin{aligned}
{\rm Tr}\sqrt{\mathbb{D}_t}\Big|_{\,\zeta-{\rm reg.}}
=\kappa^{3/2-\varepsilon}\sum_{n=2}^\infty
2(n-1)(n+3)\Big\{&T_{(3)}\big(n(n+2)-2\big)\Big\}^{\frac12-\frac\varepsilon6}
= \sum_{n=2}^\infty n^{5-\varepsilon}\, F_t\Big(\frac1n,\varepsilon\Big),
\end{aligned}
\ee
where $F_t\Big(\frac1n,\varepsilon\Big)$ is a function different from
$G_t\Big(\frac1n,\varepsilon\Big)$, but coinciding with it at
$\varepsilon=0$,
$F_t\Big(\frac1n,0\Big)=G_t\Big(\frac1n,0\Big)$. Then, by expanding
this function in Taylor series in $1/n$, one acquires a series of
Riemannian zeta functions
    \begin{equation}
    \begin{aligned}
    {\rm Tr}\,\sqrt{\mathbb{D}_t}\,\Big|^{\,\rm div}_{\,\zeta-{\rm reg.}}
    &=\sum_{n=2}^\infty  n^{5-\varepsilon}\sum_{m=0}^\infty \frac1{m!}F_t^{(m)}(0,\varepsilon)\frac1{n^m}\,\Big|^{\,\rm div}\\
    &=\sum_{m=0}^\infty \frac1{m!}F_t^{(m)}(0,\varepsilon)\, \zeta_R(m+\varepsilon-5)\,\Big|^{\,\rm div}
    =\frac1{6!\,\varepsilon}
    \frac{d^6 F_t\left(y,0\right)}{dy^6}\,\Big|_{\,y=0},
    \end{aligned}
    \end{equation}
which coincides with (\ref{dimreg}). Here we used the fact that
Riemann zeta-function $\zeta_R(z)$ has a simple pole only at $z=1$
with unit residue.

Analogously to the formula (\ref{4.5}) we regularize the vector and scalar traces
    \begin{equation}
    {\rm Tr}\,\sqrt{\mathbb{D}_v} = \sum_{n=2}^\infty n^{5-\varepsilon} G_v\Big(\frac1{n},\varepsilon\Big), \qquad
    {\rm Tr}\,\sqrt{\mathbb{D}_s} = \sum_{n=2}^\infty n^{5-\varepsilon} G_+\Big(\frac1{n},\varepsilon\Big)+
   \sum_{n=0}^\infty n^{5-\varepsilon} G_-\Big(\frac1{n},\varepsilon\Big)\,.
    \end{equation}
Then the total divergence reads
    \begin{equation}\label{tensorS3}
    {\rm Tr}\,\mathbb{Q}\,\Big|^{\,\rm div}_{\,S^3} = \frac1{6!\,\varepsilon}\frac{d^6 }{dy^6}\,\Big[\,G_t\left(y,0\right)
    +G_v\left(y,0\right)+G_+\left(y,0\right)+G_-\left(y,0\right)\,\Big]\,\Big|_{\,y=0}.
    \end{equation}
Full expressions for functions $G_t(y,0)$, $G_v(y,0)$, $G_\pm(y,0)$
can be obtained for arbitrary gauge in the two-parameter family of
Sec.~\ref{gaugefix}, but are too lengthy to be
presented here. The Mathematica code to calculate them can be found 
at \cite{github}.

The same procedure applies to the regularization of the
trace of the vector operator square roots. The simplicity
of this sector allows us to present the final result, 
\be
\label{trBsph}
{\rm Tr}\,\mathbb{Q_{\,B}}\,\Big|^{\,\rm div}_{\,S^3} \!=\!
\frac{\kappa^{3/2}}\varepsilon
\sqrt{\nu_5} \left[- 4u_V\! -\!
  \frac{(3\!-\!\lambda)^2(1\!+\!\lambda) u_{S2}}{16(1-\lambda)^3} 
\!+\! \frac{(5\!-\!6\lambda\!+\!5\lambda^2)u_V^2}{2(1-\lambda)u_{S2}}
\!+\! \frac{4(1\!-\!\lambda)(5-3\lambda)u_V^4}{u_{S2}^3}
\!-\!\frac{32(1\!-\!\lambda)^3u_V^6}{u_{S2}^5} \right]\;,
\ee
where $u_V$ and $u_{S2}$ have been defined in Eq.~(\ref{allus}).
Combining this result with the tensor contribution,
we finally find the divergent part of the one-loop effective action on
the static spacetime with spherical 3-space in an arbitrary
$(\sigma,\xi)$-gauge,
\bseq
\label{divonS3}
\begin{align}
\varGamma^{\rm 1-loop}\,\Big|^{\,\rm div}_{\,R^1\times S^3} 
&=\int d\tau\,\frac{\kappa^{3/2}}\varepsilon\,P(\l,u_s,v_1,v_2,v_3,\sigma,\xi),\\
P(\l,u_s,v_1,v_2,v_3,\sigma,\xi)&
=\frac{\sqrt{\nu_5}}{32\,(1-\lambda)^3 (1-3\lambda)^3 {u_s^5}}
\Big\{\,(1 - \lambda)^6 (144 v_1 + 50 v_2 + 18 v_3 -1)^3 \notag\\
 &+ 2 u_s^2 (1 - \lambda)^4 (1 - 3\lambda) (144 v_1 + 50 v_2 + 18 v_3 -1) \Big[72 v_1 (\lambda + 1)\notag\\
 &\qquad+ 2 v_2 (17 \lambda + 8)+ 18 v_3 \lambda  - 5 \lambda  + 4 \Big]\notag\\
 &+ 4 u_s^4 (1 - \lambda)^2 (1 - 3\lambda)^2 \Big[\,72 v_1  ( 16 \lambda^2 - 9\lambda -3 )+ 2 v_2 (200  \lambda^2  - 120  \lambda - 33) \notag\\
&\qquad+ 6 v_3 (24 \lambda^2 - 16 \lambda - 3)  - 8 \lambda^2 + 12 \lambda - 3\,  \Big]\notag\\
&+ 6 u_s^5 (1 - \lambda)^3 (1 - 3\lambda)^3 \Big[ \,12 (v_2 + v_3) \left(v_2 ( 6 v_2+ 25) + 3 v_3 ( 4 v_2 + 3)  + 6 v_3^2\right)\notag\\
&\qquad+432 v_1 ( 2 v_2 + 2 v_3 + 3) + 430 v_2 + 142 v_3 - 11\,\Big]
+ 8 u_s^6 (1 - 3\lambda)^3 \lambda (4\lambda^2 - 8\lambda + 3) \Big\}\notag\\ 
&+ 3 \nu_5 (9 v_1 +3 v_2 +v_3)\left\{4
  \sqrt{2\sigma} 
+ \frac{1}{(1-\l)}\sqrt{\frac{\sigma}{(1-\l)(1+\xi)}}  \right\}.
\label{Psphere}
\end{align}
\eseq
Note that the $R^1\times S^3$ background
   for a generic radius
  $\kappa^{-1/2}$ of the 3-sphere is not a solution of equations of motion,
  so that this divergent part of the effective action is off-shell
  and, therefore, is gauge dependent. Equations of motion on static
  $R^1\times S^3$ imply that the derivative of the action with respect
  to $\kappa$ should vanish
  and hold only at flat space geometry, $\kappa=0$, of infinitely
  large 3-sphere.
The gauge dependence is described by the last term in (\ref{Psphere}) and
is remarkably simple. 

In comparing this expression to our previous results, we need
the relation between $1/\varepsilon$ and the divergent
logarithm $\ln L^2$, Eq.~(\ref{lnLds/s}). In dimensional
regularization the latter is 
regularized as
\be
\int \frac{ds_2}{s_2}\;\;\mapsto\;\; k_*^{\varepsilon}\int
\frac{ds_2}{s_2^{(1-\varepsilon/2)}}\simeq  
\frac{2}{\varepsilon}\;,
\ee
where $k_*$ is a parameter with units of momentum to keep the
expression dimensionless (cf. footnote \ref{foot:epsilons}).
This gives  
$\ln L^2=2/\varepsilon$.  Comparing with 
Eqs.~(\ref{GammaS3}), (\ref{Is}), we find
\be
96\pi^2\kappa^{3/2}(9C_{\nu_1}+3C_{\nu_2}+C_{\nu_3}) 
=\kappa^{3/2}P(\l,u_s,v_1,v_2,v_3,\sigma,\xi)\;.
\ee
We have checked that this equality is indeed satisfied in the four
gauges {\bf a}, {\bf b}, {\bf c}, {\bf d}. 
 This accomplishes the verification of our results
on the static homogeneous spacetime. 

As a corollary we obtain the expression for the 
function $\Xi$ introduced in Eq.~(\ref{Cgauge}) which parameterizes the
gauge dependence of the divergent coefficients in the effective action, 
\be
\label{gfromsphere}
\Xi=\frac{1}{32\pi^2}\bigg\{4\sqrt{2\sigma}
+\frac{1}{(1-\l)}\sqrt{\frac{\sigma}{(1-\l)(1+\xi)}} \bigg\}\;.
\ee
Its knowledge allows us to generalize our expressions 
for $C_{\nu_a}$,
$a=1,\ldots,5$, to arbitrary $(\sigma,\xi)$-gauge. The result is
contained in the form of Mathematica file in Supplemental Material.

The final remark here is that the knowledge of the logarithmic divergences
(\ref{divonS3}) allows one to extract the logarithmic dependence of
the finite part of the effective action on the radius if the sphere 
$\kappa^{-1/2}$. This is easily seen within the $\zeta$-functional
regularization in which the overall scale of operators
$\mathbb{F}=(\mathbb{D}, \mathbb{B})\propto\kappa^3$ is raised to the
fractional power in (\ref{zetapower}) and gives  
\be
\frac1\varepsilon\,\kappa^{\frac32-\frac\varepsilon2}=
\kappa^{\frac32}\left(\frac1\varepsilon-\frac12\ln\frac\kappa{k_*^2}\right),
\ee 
where $k_*$ 
is a normalization scale of the $\zeta$-function
regularization. This leads to the expression for the full
effective action as a function of $\kappa$, 
\be
\begin{aligned}
\varGamma^{\rm 1-loop}\,\Big|_{\,R^1\times S^3} 
=\int d\tau\,\left(\frac{\kappa^{3/2}}\varepsilon-\frac{\kappa^{3/2}}2\,
\ln\frac{\kappa}{k_*^2}\right)\,P(\l,u_s,v_1,v_2,v_3,\sigma,\xi)
+\int d\tau\,\kappa^{3/2}\,Q(\l,u_s,v_1,v_2,v_3,\sigma,\xi),
\end{aligned}
\ee
where the logarithmic term plays the role of the Coleman--Weinberg
effective potential on the metric background of size
$\kappa^{-1/2}$. 
In contrast to the logarithmic contribution, the second term is not
controlled by the UV divergent coefficient and, contrary to the case
of single-charge models cannot be absorbed into the redefinition of
the normalization $k_*$, because it carries a nontrivial dependence on
multiple couplings.

\section{Discussion}
\label{concl}

In this paper
we have obtained the full set of one-loop $\beta$-functions for marginal
essential coupling constants in projectable HG. The
results underwent a number of very powerful checks that confirm gauge
independence of these beta functions in a wide set of gauge conditions
--- the corner stone of the physically invariant content of quantum
gauge theories. These checks also provide a very deep verification and
show high efficiency of
the method of universal functional traces which replaces within the
background field approach the standard Feynman diagrammatic technique. 
This method implicitly performs summation
of a humongous number of Feynman graphs and leads to a final result
hardly achievable by standard momentum space methods in flat
spacetime. 
As a byproduct of our calculation we derived an expression for the
divergence of the one-loop effective action of HG on static background
in a two-parameter family of gauges. 

The complexity of the 
expressions (\ref{betas_new}) for the $\beta$-functions with the
polynomials ${\cal P}^{\chi}_n$ collected in
Eqs.~(\ref{Pol_calG})---(\ref{Pol_v3}) is high and we postpone a
comprehensive analysis of the resulting RG flow for future. At this
point, we content ourselves with a few preliminary observations.  

Clearly, the $\beta$-functions (\ref{betas_new}) are in general
singular at $\l\to 1/3$, $\l\to 1$ or $u_s\to 0$.\footnote{We do not
  consider the singularity at $u_s=-1$, because $u_s$ is assumed to be
  positive by construction, see Eq.~(\ref{new_couplings}).}
This is not surprising, since the two first limits correspond to the
boundaries of the unitarity domain (\ref{lambdaunitary}), whereas in
the last limit the dispersion relation of the scalar mode becomes
degenerate, see Eq.~(\ref{disp2}).
Remarkably, however, for a special choice of the values
\be
\{v^*\}: v_1 = 1/2,\quad v_2 = -5/2, \quad v_3 = 3 
 \label{crit_point}
\ee
the limit $u_s\to 0$ of the $\beta$-functions (\ref{betas_new})
becomes regular for any $\l$ in the unitary domain:
\bseq
\label{sing_in_us} 
\begin{align}
&\beta_{v_a}\Big|_{\{v^*\},\,u_s\to 0}=0\;,~~~~~a=1,2,3\;,\\
&\beta_{u_s}\Big|_{\{v^*\},\,u_s\to 0}=\frac{ 1893 \lambda^2 - 6720 \lambda + 4576 }{6720 \pi^2 (1-\lambda) (1-3\lambda) }\,{\cal G}\;,\\
&\beta_{\cal G}\Big|_{\{v^*\},\,u_s\to 0}=-\frac{159}{80\pi^2}\,{\cal G}^2\;.
\end{align}
\eseq
The point $\{v^*\},\;u_s\to 0$ is special since it corresponds to the
version of HG, in which the potential term is a square of the
Cotton tensor $C_{ij}$, 
\begin{align}
S&= \frac{1}{2G} \int d\tau\, d^3x \,
\sqrt{\gamma}\,(K_{ij}K^{ij}
-\lambda K^2 + \nu_5\,C^{ij}C_{ij})\notag\\
&=\frac{2}{G} \int d\tau\, d^3x \,\sqrt{\gamma}\, (K_{ij}+\sqrt{\nu_5}C_{ij})
\,\mathbb{G}^{ij,kl}\,(K_{kl}+\sqrt{\nu_5}C_{kl}),           \label{actionCS}\\
C^{ij}&=\varepsilon^{ikl}\nabla_k\Big(R^j_l-\frac14\,R\,\delta^j_l\Big)
=\varepsilon^{kl(i}\nabla_k R^{j)}_{\,l},
\end{align}
where $\varepsilon^{ikl}=\epsilon^{ikl}/\sqrt{g}$,
$\epsilon^{123}=1$. In the second equality in (\ref{actionCS}) we used
the tracelessness of the Cotton tensor and integration by parts. 

This version of HG was originally suggested in \cite{Horava:2009uw}
and its quantum properties were studied in \cite{Orlando:2009en}. It is
known as HG with detailed balance and is interesting because
the Cotton tensor can we rewritten as a variational derivative of the
3-dimensional gravitational Chern--Simons theory, 
\begin{eqnarray}
&&C^{ij}=-\frac1{\sqrt{g}}\,\frac{\delta W_{\rm CS}[\,g\,]}{\delta g_{ij}(x)},\\
&&W_{\rm CS}[\,g\,]=\frac{1}{2}\int
d^3x\,\epsilon^{ijk}\Big(\,\Gamma^m_{il}\partial_j\Gamma^l_{km}+ 
\frac23\Gamma^n_{il}\Gamma^l_{jm}\Gamma^m_{kn}\Big),
\end{eqnarray}
defined in terms of the metric Christoffel symbol as a functional of
$g_{ij}$. Further, there exists a deformation of the action
(\ref{actionCS}) by relevant operators which preserves the detailed balance
structure and is related to the topological massive gravity
\cite{Deser:1982vy,Deser:1981wh,Deser:1990bj}. The detailed balance
relation between $d$ and $(d+1)$-dimensional theories appears in the
context of stochastic quantization \cite{Parisi:1980ys,Damgaard:1987rr} 
and establishes nontrivial
connection between the renormalization properties of the two theories
\cite{Zinn-Justin:1986nph}. In our case this suggests an intriguing
connection between the $(3+1)$-dimensional projectable HG and the
$3$-dimensional gravitational Chern--Simons / topological massive
gravity \cite{Orlando:2009en}. 
 
It is important to emphasize, however, that the point $\{v^*\},\,u_s\to
0$ is {\em not} a fully regular point of the RG flow in HG, because
the $\beta$-function (\ref{betalambda})
of the remaining essential coupling $\l$
diverges in this limit,
\be
\label{bl_us0}
\beta_{\l}\,\Big|_{\{v^*\},\,u_s\to 0}\sim
\frac{9}{40\pi^2}\frac{1-\l}{u_s}\,{\cal G}\;.
\ee
Thus, the physical significance of the result (\ref{sing_in_us}) is
unclear at the moment. It will be interesting to understand if the
inclusion of fermionic degrees of freedom appearing in the stochastic
quantization framework \cite{Orlando:2009en} can change the picture. 

An important question is the existence and nature of fixed points of
the RG flow. As already observed, the dependence of the
$\beta$-functions on the coupling ${\cal G}$ factorizes. This coupling
determines the overall strength of interactions in HG and must be
small for the validity of the perturbative expansion. Its UV behavior
determines whether the model is asymptotically free (${\cal G}\to 0$)
or has a Landau pole (${\cal G}\to\infty$). On the other hand, the
rest of the couplings $\l,u_s, v_a$ are ratios of the coefficients in
the action and need not be small. The search for fixed points of the
RG flow thus splits into two steps. One first identifies the fixed
points of the flow in the subspace of the couplings $\l,u_s, v_a$ by
solving the system,
\bseq
\label{fpeq}
\begin{align}
\label{fpeq1}
&\beta_\l/{\cal G}=0\;,\\
\label{fpeq2}
&\beta_\chi/{\cal G}=0\;,~~~~~\chi=u_s,v_1,v_2,v_3\;.
\end{align}
\eseq
In the full parameter space, these solutions correspond to flow lines
along the ${\cal G}$-direction. One then evaluates $\beta_{\cal G}$ at
a given solution, whose sign determines whether the flow line goes to a
Gaussian fixed point or a Landau pole.

Omitting the denominators in the expressions (\ref{betalambda}),
(\ref{beta_chi}), the system (\ref{fpeq}) becomes a system of 5
polynomial equations for 5 unknowns $\l,u_s,v_a$, $a=1,2,3$.  
We have studied it numerically with the following results:
\begin{itemize}
\item[i)]
We have found no solutions in the right
part of the unitary domain, $\l>1$. In this respect
$(3+1)$-dimensional HG appears to be different from its
$(2+1)$-dimensional counterpart, which possesses an asymptotically
free fixed point at $\l=15/14$ \cite{Barvinsky:2017kob}.

\item[ii)]
In the left part of the unitary domain, $\l<1/3$, we found 4 solutions
summarized in Table~\ref{tabFP1}. All these fixed points turn out to be
asymptotically free. Note that the two last points correspond to very
large values of $v_1$ and their validity requires further
investigation. As discussed in \cite{towards}, the fixed points at
$\l<1/3$ are UV repulsive along the $\l$-direction. We indicate this
in the last column of Table~\ref{tabFP1}. We do not know if these
fixed points are attractive or not along the other directions.  
\end{itemize}

\begin{table}[h] 
\begin{center}
\begin{tabular}{| c | c | c | c | c | c |c |c|}
 \hline
$\lambda$  &$u_s$ & $v_1$ & $v_2$ & $v_3$ &  $\beta_{\cal G}/{\cal
  G}^2$ & \makecell{asymptotically\\ free?}
 &\makecell{UV attractive \\along $\l$?}\\ [0.5ex] 
 \hline\hline
0.1787&  60.57 &-928.4 & -6.206 & -1.711 &  -0.1416 & yes& no \\ [0.5ex]
  \cline{1-8}
 0.2773 & 390.6  &-19.88& -12.45 & 2.341 &  -0.2180  & yes & no\\ [0.5ex]
  \cline{1-8}
0.3288 & 54533 & 3.798$\times 10^8$ & -48.66 & 4.736 &  -0.8484 & yes&
no\\ [0.5ex]
  \cline{1-8}
 0.3289 & 57317 &-4.125$\times 10^8$ & -49.17 & 4.734 & -0.8784 &yes &
 no\\ [0.5ex]
 \hline   
  \end{tabular}
  \caption{\label{tabFP1}
Solutions of the system (\ref{fpeq}). 
The sixth column gives the value of the
    $\beta$-function for ${\cal G}$ at the respective solution and the
    seventh column indicates whether it corresponds to an
    asymptotically free fixed point. The eighth column tells if the
    fixed point is UV attractive along the $\lambda$-direction.}
\end{center}
\end{table}

It is worth stressing that the results above should be taken with a grain
of salt. Currently we do not have a firm proof of the absence of fixed
points at $\l>1$, nor do we claim that the list of fixed points at
$\l<1/3$ in Table~\ref{tabFP1} is exhaustive.    

It was conjectured in \cite{Gumrukcuoglu:2011xg} that the UV fixed
points of HG can lie at infinite $\l$ and that the limit $\l\to\infty$
is well-defined. We find that all $\beta$-functions (\ref{betas_new})
are finite at $\l\to\infty$, whereas $\beta_\l$ is proportional to
$\l$,\footnote{The directionality of the limit is not important: the
  resulting expressions for the $\beta$-functions are the same at
  $\l=\pm\infty$.}
\be
\label{betallimit}
\beta_\l=-\frac{3(3-2u_s)}{40\pi^2 u_s}\,\l\,{\cal G}\;,~~~~~\l\to\infty\;.
\ee
This behavior is compatible with the conjecture of
\cite{Gumrukcuoglu:2011xg}. Notice that the point $\l=\infty$ is UV
attractive (repulsive) for $u_s>3/2$, ($u_s<3/2$). To identify fixed
points of the RG flow at $\l=\infty$, we looked for solutions of the
system
\be
\label{betalimiteq}
\beta_\chi/{\cal G}\Big|_{\l=\infty}=0\;,~~~~~\chi=u_s,v_1,v_2,v_3\;.
\ee
We have found eight solutions listed in Table~\ref{tabFP2}. Three among
them are UV attractive along the $\l$-direction and correspond to
asymptotically free fixed points. Clearly, the structure of the RG
flow around these points deserves further investigation. More
generally, this strongly motivates a detailed study of the
$\l\to\infty$ limit of HG. 
It is worth mentioning that similar limit naturally arises in
connection of non-relativistic gravity to Perelman--Ricci 
flows~\cite{Frenkel:2020dic}.

\begin{table}[h] 
\begin{center}
\begin{tabular}{| c | c | c | c | c | c |c |}
 \hline
$u_s$  & $v_1$ & $v_2$ & $v_3$ &   $\beta_{\cal G}/{\cal G}^2$&  \makecell{asymptotically\\ free?}
 &\makecell{UV attractive \\along $\l$?}\\ [0.5ex]
\hline\hline
0.01950& 0.4994 & -2.498 & 2.999 &  -0.2004 &yes&no\\ [0.5ex]
\cline{1-7}
 0.04180 & -0.01237 & -0.4204 & 1.321 &  -1.144&yes&no \\ [0.5ex]  
  \cline{1-7}
0.05530 & -0.2266 & 0.4136 & 0.7177 &  -1.079 &yes&no\\ [0.5ex]
 \cline{1-7}
12.28 & -215.1 & -6.007 & -2.210 &   -0.1267 &yes&yes\\ [0.5ex]
 \cline{1-7}
21.60 & -17.22 & -11.43 & 1.855 &   -0.1936 &yes&yes\\ [0.5ex]
 \cline{1-7}
440.4 & -13566 & -2.467 & 2.967 &   0.05822 &no&yes\\ [0.5ex]
 \cline{1-7}
571.9 & -9.401 & 13.50 & -18.25 &  -0.07454 &yes&yes\\ [0.5ex]
 \cline{1-7}
950.6 & -61.35 & 11.86 & 3.064 &   0.4237 &no&yes\\ [0.5ex]
 \hline
  \end{tabular}
  \caption{ \label{tabFP2} Solutions of the system (\ref{betalimiteq})
  corresponding to fixed points of Ho\v rava gravity at $\l=\infty$. The
  fourth column lists the value of the $\beta$-function for the
  coupling ${\cal G}$ at each solution, whose sign determines whether the
  flow is asymptotically free or runs into strong coupling, as
  indicated in the fifth column. The last column tells if the point is UV
attractive along the $\lambda$-direction in the space of all couplings.}
\end{center}
\end{table}

Let us stress again that presently we do not know if
Table~\ref{tabFP2} is exhaustive. We plan to return to a systematic
classification of fixed points of HG and its RG flow in our future
work.

\section*{Acknowledgments}

We thank Diego Blas, Mario Herrero-Valea, Benjamin Knorr, Vladimir
Nechitailo  
and Christian Steinwachs
for invaluable
discussions. The work was partially supported by the Russian
Foundation for Basic Research grant 20-02-00297. 
The work of A.B. and A.K. is 
supported by the Foundation for Theoretical Physics Development ``Basis''.
The work of S.S. is supported by
the 
Natural Sciences and Engineering Research Council (NSERC) of Canada.
Research at Perimeter Institute is supported in part by the Government
of Canada through the Department of Innovation, Science and Economic
Development Canada and by the Province of Ontario through the Ministry
of Colleges and Universities.

\appendix

\renewcommand{\thesection}{\Alph{section}}
\renewcommand{\thesubsection}{\Alph{section}.\arabic{subsection}}
\renewcommand{\theequation}{\Alph{section}.\arabic{equation}}

\section{Explicit expressions}
\subsection{Beta-function of $G$}
\label{app:betaG}
The coupling $G$ of HG is not {\it essential}, i.e. it is not defined using
the on-shell quantities. Hence its $\beta$-function depends on the
gauge choice. Ref.~\cite{towards} obtained this $\beta$-function for the
$(3+1)$-dimensional projectable model in
a
subset of the two-parameter family of regular gauges described in
Sec.~\ref{gaugefix}. The results are: 
\begin{itemize}
\item[i)]
$\sigma$-arbitrary, $\xi=-\frac{1-2\l}{2(1-\l)}$
\begin{eqnarray}
\beta_G=\sqrt{\nu_5}\,\frac{{\cal G}^2 }{40 \pi ^2
(1-\lambda ) (1-3 \lambda)\,\left(1+u_s\right)u_s}
\Big[&&\!\!-27+74\lambda-57\lambda^2
-u_s \big(5(1-3
\lambda) (5-4 \lambda) \sqrt{2\sigma  \nu_5}\nonumber\\
+53
-142 \lambda+99 \lambda^2\big)
&&\!\!- u_s^2 (1-3 \lambda ) \big(5 (5-4 \lambda)
\sqrt{2\sigma  \nu_5}+18-14 \lambda\big)
 \Big]\;,
\label{betaGa}
\end{eqnarray}

\item[ii)]
$\sigma=\frac{1}{2\nu_5}$, $\xi=\frac{\nu_s}{2\nu_5(1-\l)}-1$
\begin{eqnarray}
    \beta_G=-\sqrt{\nu_5}\frac{{\cal G}^2}
    {40\pi^2(1-\lambda)
    (1-3\lambda)\,(1+u_s)\,u_s}\Big[&&32-89\lambda+57\lambda^2
    +3u_s(26-79\lambda+53\lambda^2)\nonumber\\
    &&+2u_s^2(19-74\lambda+51\lambda^2)\,\Big]\;,    \label{betaGc}
    \end{eqnarray}

\item[iii)]
$\sigma=\frac{1}{2\nu_s}$, $\xi=\frac{\nu_5}{2\nu_s(1-\l)}-1$
\begin{eqnarray}
    \beta_G=-\sqrt{\nu_5}\frac{{\cal G}^2}
    {40\pi^2(1-\lambda)
    (1-3\lambda)\,(1+u_s)\,u_s}\Big[&&47-154\lambda+117\lambda^2
    +3u_s(26-79\lambda+53\lambda^2)\nonumber\\
    &&+u_s^2(23-83\lambda+42\lambda^2)\,\Big]\;.    \label{betaGiii}
    \end{eqnarray}
\end{itemize}
This set of gauges overlaps with the gauges used in the present 
work. Thus, the gauge ii) coincides with the gauge {\bf (a)},
Eq.~(\ref{cparam}), whereas the gauge i) reduces to the gauges {\bf
  (b)} and {\bf (d)} for the appropriate choices of $\sigma$, see
Eqs.~(\ref{bparam}), (\ref{dparam}). The expressions (\ref{betaGa}),
(\ref{betaGc}) are used in Sec.~\ref{btfuns} to derive the
$\beta$-function of the essential coupling ${\cal G}$.

\subsection{Polynomials in the $\beta$-functions of essential
  couplings}
\label{betaexpl}

In this Appendix we collect the expressions for the polynomials appearing in
Eqs.~(\ref{betas_new}). For the $\beta$-function of the 
coupling ${\cal G}$ the polynomials read,
\bseq
\label{Pol_calG}
\begin{align}
{\cal P}^{\cal G}_0 =&(1 - \lambda)^4\big(1809 v_3^2 +  832 v_2^2+ 16 v_2 ( 159 v_3 - 217) - 4494 v_3  + 2401  \big),\\
{\cal P}^{\cal G}_1 =&3 {\cal P}^{\cal G}_0,\\
{\cal P}^{\cal G}_2 =&-(1 - \lambda)^2\big[ 3 (15779 \lambda^2  -
20362 \lambda + 3967)   + 64 v_2^2 (81 \lambda^2   - 82 \lambda+ 1) 
\notag \\
& +27 v_3^2 ( 279 \lambda^2   - 238 \lambda - 41) - 6 v_3 (8823 \lambda^2   - 10620 \lambda + 1561)  \notag\\
&+16 v_2 \big(    v_3 (675 \lambda^2 - 582 \lambda -93) - 2307 \lambda^2  + 2732 \lambda -365 \big ) \big], \\
{\cal P}^{\cal G}_3 =&(1 - \lambda)^2\big[27 v_3^2 (961 \lambda^2 - 1434 \lambda + 401) +    64 v_2^2 (1717\lambda^2 - 2298 \lambda + 581) \notag\\
&  + 16 v_2 \big(19409 \lambda^2 - 26004 \lambda + 6415  +  3 v_3 (2741 \lambda^2   - 3690 \lambda + 949)\big)  \notag  \\
& +6 v_3 (39331 \lambda^2 - 58728 \lambda + 14873)  - 345977 \lambda^2   + 276750 \lambda  -52741\big], \\
{\cal P}^{\cal G}_4 =&
2 (1 - 3 \lambda)\big\{138545 \lambda^3 - 328263 \lambda^2 - 5888  (1 - \lambda)^3 v_2^2\notag\\
&-(1 - \lambda)^2  \big[16 v_2\big(3119 \lambda  + 840 v_3 (1 - \lambda)
  - 2396 \big)  \notag\\
&-  3 v_3 \big(9 v_3 (353 \lambda - 299 ) - 5012 \lambda + 8210\big)\big] +239597 \lambda  -49947     \big\},\\
{\cal P}^{\cal G}_5 =&2 (1 - 3 \lambda)\big\{159709 \lambda^3 -  378471 \lambda^2  +    (1 - \lambda)^2 \big[16 v_2 (1243 \lambda-412  ) \notag\\
&-    3 v_3 \big(243 v_3 (1 - 3 \lambda)   - 13280 \lambda + 4366\big)\big]    + 273933 \lambda    -55375   \big\}, \\
{\cal P}^{\cal G}_6 =&-6 (1 - 3 \lambda)^2 \big(8465 \lambda^2 - 16310 \lambda + 3(1 - \lambda)^2 v_3  (254 + 27 v_3)   + 7811 \big),\\
{\cal P}^{\cal G}_7 =& 4  (1 - 3 \lambda)^2 (48 \lambda^2  - 38 \lambda + 7)\;.
\end{align}
\eseq
Polynomials in the $\beta$-function of $u_s$ are:
\bseq
\label{Pol_us}
\begin{align}
{\cal P}^{u_s}_0 =& -3 (1 - \lambda)^5 \big[  537600 v_1^2 + 78992 v_2^2  + 14205 v_3^2 +   2688 v_1 (154 v_2 + 67 v_3 - 16)  \notag\\
&+    16 v_2 (4236 v_3 - 959) - 5838 v_3 + 329 \big],\\
{\cal P}^{u_s}_1 =& 3 {\cal P}^{u_s}_0,\\
{\cal P}^{u_s}_2 =& -2 (1 - \lambda)^3 \big[ 2419200 v_1^2 (1 - \lambda)^2 +  8 v_2^2 (42645 \lambda^2 - 86482 \lambda + 43837) \notag\\
&+ v_3^2(58698 - 106947 \lambda + 48249 \lambda^2)  +  4032 v_1 \big(462 v_2 (1 - \lambda)^2 +  201 v_3 (1 - \lambda)^2+ 30 \lambda^2 - 44 \lambda  - 10\big)\notag\\
&+ 8 v_2 (6252 \lambda^2  - 9188
\lambda -1468)+  8 v_2  v_3 (34335 \lambda^2  - 71196 \lambda + 36861)\notag\\ 
& + v_3 (20556  \lambda^2  - 30792 \lambda - 3696)
+ 4533 \lambda^2  - 3881 \lambda + 1448\big],\\
{\cal P}^{u_s}_3 =&-2 (1 - \lambda)^3 \big[806400v_1^2 (1 -
\lambda)^2+  8v_2^2 (20709 \lambda^2 - 32026 \lambda + 14957) +
v_3^2(61686\l^2 \notag\\ 
&  - 52875 \lambda + 20241) +    4032v_1 \big(98 + 154v_2 (1 - \lambda)^2 +       67v_3 (1 - \lambda)^2+ 218 \lambda^2 - 388 \lambda \big)\notag \\
& +    8v_2 \big(3v_3 (7833 \lambda^2 - 9656 \lambda + 4231) +  4 (8658 \lambda^2 - 16817 \lambda + 4324)\big)\notag\\
&+    v_3 (81594\lambda^2 -    189660 \lambda  + 50262)    - 2970 \lambda^2 - 1529 + 6235 \lambda \big],\\
{\cal P}^{u_s}_4 =& (1 - \lambda)(1 - 3\lambda)
\big[32 v_2^2 (1 - \lambda)^2 (4081 \lambda - 1191)
+ v_3^2 (133083 \lambda^3- 303453 \lambda^2+ 207657 \lambda - 37287) \notag\\
& +  48384 v_1 (1 - \lambda)^2 (13 \lambda - 19) - 16 v_2 (1 - \lambda) \big(7873 \lambda^2 - 25922 \lambda +  18109 \notag\\
& +     3 v_3 (5419 \lambda^2  - 6970 \lambda
+ 1551)\big) -  v_3 (31938 \lambda^3 + 15042 \lambda^2- 127314 \lambda + 80334) \notag \\
&+ 10415 \lambda^3+ 11815 \lambda^2- 30239
\lambda+10017\big],\\ 
{\cal P}^{u_s}_5 =& (1 - \lambda)(1 - 3\lambda)\big[ 
32 v_2^2 (1 - \lambda)^2 ( 661 \lambda - 203 ) 
+  v_3^2(104787 \lambda^3 - 240381 \lambda^2+ 168345 \lambda- 32751)   \notag\\
&  +  16128 v_1 (1 - \lambda)^2 ( 13 \lambda -
19 ) + 16 v_2 (1 - \lambda) \big(12761 \lambda^2- 14690 \lambda  + 69  \notag\\ 
& -    3 v_3 (2677 \lambda^2  - 3534 \lambda +
857)\big)- v_3(178962  \lambda^3 - 468990  \lambda^2 +  347070  \lambda  -  57042)\notag\\
& +  379967 \lambda^3 - 512385 \lambda^2 + 126609 \lambda    + 1081\big],\\
{\cal P}^{u_s}_6 =&- 4(1 - 3\lambda)^2 \big[  
6584 v_2^2 (1 - \lambda)^3 - 27 v_3^2 (1 - \lambda)^2 (311 \lambda -284 )+ 24 v_2 (1 - \lambda)^2 \big(405 \lambda- 584  \notag\\
& + 581 v_3 (1 - \lambda) \big) 
-  3 v_3 (1 - \lambda)^2 (2507 \lambda + 2452)  - 92671 \lambda^3 +
205653 \lambda^2- 130039 \lambda  +  17539\big],\\
{\cal P}^{u_s}_7 =&- 2(1 - 3\lambda)^2 \big[  (1 - \lambda)^2\big(
729v_3^2 (1 - 3 \lambda) - 16 v_2  (3133 \lambda-1042 )- 6v_3  (11680
\lambda - 3863) \big) 
\notag \\
& - 212947 \lambda^3 + 494301 \lambda^2 - 341005 \lambda + 61647\big],\\
{\cal P}^{u_s}_8 =&- 2(1 - 3\lambda)^3 \big((1 - \lambda)^2 ( 243 v_3^2 + 3360 v_2  +    5646 v_3  ) + 31443 \lambda^2  -  61026 \lambda  +29033 \big),\\
{\cal P}^{u_s}_9 =& 4(1 - 3\lambda)^3 \left(48 \lambda^2 -  38 \lambda + 7\right).
\end{align}
\eseq
Polynomials in the $\beta$-function of $v_1$ are:
\bseq
\label{Pol_v1}
\begin{align}
{\cal P}^{v_1}_0 =& - (1 - \lambda)^6 \big[ 11612160  v_1^3 + 472088
v_2^3 + 241920  v_1^2 (50  v_2+18  v_3-1)\notag\\ 
&+12  v_2^2 (40758  v_3-427) + 1008  v_1 \bigl(4124  v_2^2+4  v_2 (726  v_3-23)+6  v_3 (81  v_3+4)-31\bigr)\notag\\
&+78  v_2 \bigl(6  v_3 (345  v_3+28)-119\bigr) +18  v_3 \bigl(3  v_3 (318  v_3+77)-119\bigl) -385\big],\\
{\cal P}^{v_1}_1 =& 3{\cal P}^{v_1}_0,\\
{\cal P}^{v_1}_2 =&-(1 - \lambda)^4 \Bigl\{ 34836480 v_1^3 (1-\lambda)^2 +24 v_2^3 (54595 \lambda ^2-112134 \lambda +57539 ) \notag\\
&+ 108 v_3^3 (213 \lambda ^2-602 \lambda +389) +161280 v_1^2  \bigl[ 225 v_2 (1-\lambda)^2+81v_3 (1-\lambda)^2-6 \lambda ^2+8 \lambda-4 \bigr] \notag\\
& +4 v_2^2  \bigl[ 6 v_3 (52401 \lambda ^2-110626 \lambda +58225 )  -72285 \lambda ^2+86204 \lambda- 22411 \bigr]\notag\\
& -36 v_3^2 (1947 \lambda ^2-2236 \lambda +375 )  +2 v_1  \bigl[32
v_2^2 (190749 \lambda ^2-384238 \lambda +193489 )\notag\\
& -224  v_2 (2613 \lambda ^2-3196 \lambda +1051 )+528 v_2 v_3 (7935 \lambda ^2-16124 \lambda +8189 )\notag\\
&+243  v_3^2  (2703 \lambda ^2-5620 \lambda +2917 )-42 v_3 (8535
\lambda^2-9652 \lambda +1885 )  +7  (22587 \lambda^2 -26516 \lambda
+3353 ) \bigr]\notag\\ 
& +2 v_2  \bigl[ 18 v_3^2 (9687 \lambda^2-21886 \lambda +12199 ) -24
v_3  (6447 \lambda ^2-7402 \lambda +1387 ) +52401 \lambda ^2-62686 \lambda+8885 \bigr] \notag\\
&+18 v_3 (1245 \lambda ^2-1506 \lambda +205)
+ 14805 \lambda ^2-18928 \lambda+4151\Bigr\},\\
{\cal P}^{v_1}_3 =&-(1 - \lambda)^4 \Bigl\{ 11612160 v_1^3 (1-\lambda)^2 +8 v_2^3  (19267 \lambda ^2-65030 \lambda +45763 )\notag\\
&-324  v_3^3  (211 \lambda ^2-246 \lambda +35 ) +483840 v_1^2  \bigl[25 v_2 (1-\lambda)^2 +9 v_3 (1-\lambda)^2 -2 \lambda ^2 \bigr]\notag\\
&-12 v_2^2  \bigl[ 6  v_3 (1943 \lambda ^2+1938 \lambda -3881 )+68869 \lambda ^2-79372 \lambda +18995 \bigr]\notag\\
&-108   v_3^2 (2255 \lambda^2-2852 \lambda +683 ) +6 v_1  \bigl[ 32 v_2^2 (17541 \lambda^2-37822 \lambda +20281 )\notag\\
&- 224  v_2 (2061 \lambda^2-2092 \lambda +499 )+528 v_2 v_3 (543 \lambda^2-1340 \lambda +797 )\notag\\
&+243  v_3^2 (15 \lambda^2-244 \lambda +229) - 42  v_3 (9303 \lambda^2-11188 \lambda +2653 )\notag\\
&+7  (28539 \lambda^2-38420 \lambda +9305 )\bigr] -6 v_2  \bigl[ 18  v_3^2 (2273 \lambda ^2-2034 \lambda -239 )\notag\\
&+24 v_3 (7175 \lambda ^2-8858 \lambda +2115 )  -64777 \lambda ^2+87438 \lambda-21261 \bigr]\notag\\
&+18  v_3 (4687 \lambda ^2-6422 \lambda +1567 )  + 7  (6785 \lambda ^2-8992 \lambda +2219 )\Bigr\},\\
{\cal P}^{v_1}_4 =&-2(1 - \lambda)^2(1-3\lambda) \Bigl\{ 1024 v_2^3 (1-\lambda)^2 (45 \lambda -38) +1728  v_3^3 (1-\lambda )^2 (7 \lambda -6)\notag\\
&+120960 v_1^2 (1-\lambda )^2 (\lambda +1) -4 v_2^2 (1-\lambda)   \bigl[ 384  (59 \lambda ^2-109 \lambda +50 ) v_3\notag\\
&+29133 \lambda ^2-55225 \lambda +25452 \bigr] -9 v_3^2 (877 \lambda ^3+871 \lambda ^2-4213 \lambda +2465 ) \notag\\
&+v_1  \bigl[64 v_2^2 (1-\lambda )^2 (1263 \lambda -1343)  -16 v_2 (1-\lambda)   \bigl( 3 v_3 (2463 \lambda ^2-5182 \lambda +2719 ) \notag\\
&+33534 \lambda^2-52670 \lambda +19076 \bigr)+3  \bigl( 9 v_3^2 (1-\lambda)^2 (1513 \lambda -1833) \notag\\
&+2 v_3  (82239 \lambda ^3-226251 \lambda ^2+205549 \lambda -61537 )  -64219 \lambda ^3+203973 \lambda ^2\notag\\
&-210641 \lambda +71335 \bigr) \bigr] +4 v_2  \bigl[ 144 v_3^2 (1-\lambda)^2 (101 \lambda -86) +12  v_3  (1961 \lambda ^3-6699 \lambda ^2\notag\\
&+7435 \lambda -2697 ) +8085 \lambda ^3-7434 \lambda ^2-9300 \lambda+8755 \bigr] +6   v_3 (4487 \lambda ^3\notag\\
&-9281 \lambda ^2+4807 \lambda +7 ) + 55452 \lambda ^3-123853 \lambda ^2+81624 \lambda-13195\Bigr\},\\
{\cal P}^{v_1}_5 =&-2(1 - \lambda)^2(1-3\lambda) \Bigl\{ 168 v_2^3 (51 \lambda ^3-149 \lambda ^2+125 \lambda -27 )  -108 v_3^3 (9 \lambda ^3+9 \lambda^2\notag\\
&-25 \lambda +7 )   -4 v_2^2  (1-\lambda )   \bigl[ 18 v_3 (117 \lambda ^2-366 \lambda +109 )   -284 \lambda ^2-7265 \lambda+5425 \bigr]\notag\\
&+40320 v_1^2 (1-\lambda )^2 (\lambda +1) -9 v_3^2 (3467 \lambda ^3-8839 \lambda ^2+6237 \lambda -865 ) \notag\\
&+v_1  \bigl[ 64  v_2^2(1-\lambda)^2 (1717 \lambda -581) - 16 v_2 (1-\lambda ) \bigl( 3 v_3  (2741 \lambda ^2-3690 \lambda +949 )\notag\\
&+25940 \lambda ^2-40662 \lambda+12022 \bigr) +27 v_3^2 (961 \lambda ^3-2395 \lambda ^2+1835 \lambda -401 )\notag\\
&+6  v_3 (52267 \lambda ^3-148963 \lambda ^2+129881 \lambda -33185 ) -288353 \lambda ^3+542255 \lambda ^2\notag\\
&-333355 \lambda+83485 \bigr] -2  v_2  \bigl[ 162 v_3^2 (3 \lambda ^3+35 \lambda ^2-51 \lambda +13 )  + 24  v_3  (1265 \lambda ^3\notag\\
&-2191 \lambda ^2+691 \lambda +235 ) +30971 \lambda ^3-40323 \lambda ^2+13167 \lambda-4451 \bigr]-12  v_3  (6551 \lambda ^3\notag\\
&-11593 \lambda ^2+6124 \lambda -1112 ) +109519 \lambda ^3-252396 \lambda ^2+177357 \lambda -34396\Bigr\},\\
{\cal P}^{v_1}_6 =&2(1-3\lambda)^2 \Bigl\{ 56 v_2^3  (1-\lambda )^3  (103 \lambda -13)  + 108 v_3^3 (1-\lambda)^3 (41 \lambda -11) \notag\\
&+4 v_2^2 (1-\lambda)^3  \bigl(2315 \lambda +54 (89 \lambda -19) v_3+807\bigr) -36 v_3^2 (1-\lambda )^3 (657 \lambda -239)\notag \\
&-2  v_1 (1-\lambda)  \bigl[ 5888 v_2^2 (1-\lambda)^3 +16 v_2 (1-\lambda )^2  \bigl(  840 v_3 (1-\lambda ) + 284 \lambda -1451\bigr)\notag\\
&-27  v_3^2 (1-\lambda )^2 (353 \lambda -299) - 6 v_3 (1-\lambda)^2  (5054 \lambda +1585)  - 146609 \lambda ^3\notag\\
&+330783 \lambda ^2-220781 \lambda +36675 \bigr] - 2 v_2 (1-\lambda)^2   \bigl[  54  v_3^2  (169 \lambda ^2-212 \lambda +43 )\notag\\
&+ 96 v_3  (\lambda ^2+29 \lambda -30 ) +49685 \lambda ^2-66892 \lambda +16249 \bigr]  -6 v_3 (1-\lambda)^2  (7601 \lambda^2\notag\\
&-11994 \lambda +4203 ) -15115 \lambda ^4+38758 \lambda ^3-23950 \lambda ^2-8038 \lambda +8337\Bigr\},\\
{\cal P}^{v_1}_7 =&  -2(1-3\lambda)^2  \Bigl\{ 420 v_2^2 (6  v_2+18  v_3+17) (1-\lambda)^3 (1-3 \lambda)\notag\\
&+108 v_3^2 (15 v_3 - 67  ) (1-\lambda)^3 (1-3 \lambda ) -2  v_1 (1 - \lambda )   \bigl[ 16 v_2 (1-\lambda)^2 (4078 \lambda -1357)\notag\\
&  - 729 v_3^2 (1-\lambda )^2 (1-3 \lambda ) + 6 v_3 (1-\lambda)^2 (14200 \lambda -4703)   +193645 \lambda ^3-448191 \lambda ^2\notag\\
&+313917 \lambda-59575 \bigr] + 2 v_2 (1-\lambda )^2    \bigl[  3402 v_3^2 (3 \lambda ^2-4 \lambda +1 )  +2016  v_3 (3 \lambda ^2-4 \lambda +1 )\notag\\
&+39661 \lambda ^2-53228 \lambda +13045 \bigr] +6  v_3 (1-\lambda)^2  (1021 \lambda ^2-1958 \lambda +799 ) \notag \\
&+25751 \lambda ^4-95078 \lambda ^3+122898 \lambda ^2-63194
\lambda+9647 \Bigr\},\\ 
{\cal P}^{v_1}_8 =& -  2(1-3\lambda)^3  \Bigl\{  4 (1-\lambda)^3  \bigl[210 v_2^3 + 9 v_3^2 ( 15 v_3 - 67) + 35 v_2^2 ( 18 v_3+17)\bigr] \notag  \\
&+6 v_1 (1-\lambda)    \bigl[ 1680 v_2 (1-\lambda)^2  +81 v_3^2 (1-\lambda )^2  +2442 v_3 (1-\lambda )^2   +10201 \lambda ^2\notag\\
&-19558 \lambda +9323 \bigr] + 14 v_2 (1-\lambda)^2    \bigl[  v_3^2 162 (1-\lambda)  + 96 v_3 (1-\lambda ) -951 \lambda + 935  \bigr]\notag\\
&-6  v_3 (1-\lambda)^2(515 \lambda -499)  +3349 \lambda ^3-5135 \lambda ^2-105 \lambda+1879\Bigr\},\\
{\cal P}^{v_1}_9 =&  -4(1-3\lambda)^3  \big[ 2  v_1 (48 \lambda ^3-86 \lambda ^2+45 \lambda -7 ) +163 \lambda ^3-537 \lambda ^2+477 \lambda -105\big].
\end{align}
\eseq
Polynomials in the $\beta$-function of 
$v_2$ are:
\bseq
\label{Pol_v2}
\begin{align}
{\cal P}^{v_2}_0 =& -3  (1 - \lambda)^6 (8 v_2 + 9 v_3 - 7)^2 (30v_3 +106v_2 +336v_1 + 7)  ,\\
{\cal P}^{v_2}_1 =& 3 {\cal P}^{v_2}_0,\\
{\cal P}^{v_2}_2 =&   (1 - \lambda)^4 \Big\{-192 v_2^3 (1197 \lambda^2
- 1808 \lambda + 611)  - 16 v_2^2 \big[ 9 v_3 (3429 \lambda^2  - 5288
\lambda + 1859) \notag\\ 
&+ 4 ( 3921 \lambda^2 - 4322 \lambda + 291)\big]   - 162 v_3^3 (447
\lambda^2 - 686 \lambda + 239) -     18 v_3^2 (4119 \lambda^2 - 4628
\lambda + 371)  \notag\\
&+ 2 v_3 (6957 \lambda^2 - 8772 \lambda + 975  ) - 336   v_1  \big[192 v_2^2 (9 \lambda ^2-14 \lambda +5  )\notag\\
&+ 8 v_2  \big(18 v_3 (21 \lambda ^2-34 \lambda +13  )+ 273 \lambda ^2-292 \lambda +11 \big) + 27 v_3^2 (51 \lambda ^2-86 \lambda +35  )\notag\\
&+18  v_3 (117 \lambda ^2-120 \lambda -1  ) -1371 \lambda ^2+1618
\lambda -191\big] - v_2  \big[ 9 v_3^2  (37983 \lambda ^2-59248
\lambda +21265 )  \notag\\
&+ 6 v_3 (54417 \lambda ^2-59380 \lambda +3275  )  - 143457
\lambda ^2 + 173720 \lambda - 24327  \big]
+14  (2481 \lambda ^2-3182 \lambda +687) \Big\} ,\\
{\cal P}^{v_2}_3 =&  -3 (1 - \lambda)^4\Bigl\{ 336 v_1 \bigl[ 64 v_2^2
(19 \lambda ^2-26 \lambda +7) +8 v_2  \bigl(18 v_3 (13 \lambda ^2-18
\lambda +5 )+385 \lambda ^2 - 516 \lambda  +123 \bigr)\notag \\ 
&+27 v_3^2 (27 \lambda ^2-38 \lambda +11 )
+18 v_3 (173 \lambda ^2-232 \lambda +55 )  - 1763 \lambda ^2 +2402
\lambda - 583 \bigr] \notag\\
&   + 64  v_2^3  (2743 \lambda ^2-3728
\lambda +985 ) +16 v_2^2  \bigl(3 v_3 (7423 \lambda ^2-10136 \lambda
+2713 )
 \notag\\
&+4  (5349 \lambda ^2-7178 \lambda +1719 )  \bigr) + v_2  \bigl( 9 v_3^2 (26511 \lambda ^2-36304 \lambda +9793 ) \notag\\
&+6 v_3 (75361 \lambda ^2-101268 \lambda +24219 ) -178737 \lambda
^2+244280 \lambda-59607 \bigr) \notag\\
&+162 v_3^3 (327 \lambda ^2-446 \lambda +119 ) +18 v_3^2  (5547
\lambda ^2-7484 \lambda +1799 )\notag \\
& - 6 v_3 (3103 \lambda ^2-4492 \lambda +1109 )  -14  (2677 \lambda ^2-3574 \lambda +883 )\Bigr\},\\
{\cal P}^{v_2}_4 = & (1 - \lambda)^2(1 - 3\lambda)\Bigl\{64 v_2^3
(1-\lambda)^2 (2369 \lambda -1953)+1728  v_3^3 (1-\lambda)^2 (21
\lambda -20) \notag\\
&-16 (1-\lambda )  v_2^2  \bigl[ 3 v_3  (6145 \lambda ^2-11298 \lambda
+5153 )+2  (5124 \lambda ^2-14591 \lambda +9193 )   \bigr] \notag\\
&- 9 v_3^2 (16909 \lambda ^3-30841 \lambda ^2+11563 \lambda +2369 )
+1344  v_1  \bigl[ -192  v_2^2 (1-\lambda)^3 \notag\\
&+6 (1-\lambda )^2  v_2 \bigl(149 \lambda -48 v_3 (1-\lambda)
-137\bigr)-108 v_3^2 (1-\lambda )^3  \notag\\
&+9 v_3(1-\lambda)^2(98 \lambda -93) - 372 \lambda ^3+1225 \lambda
^2-1317 \lambda+462 \bigr] \notag\\
&+ v_2  \bigl[ 9 v_3^2(1-\lambda)^2 (20741 \lambda -17925) - 6 v_3
(7139 \lambda ^3+39053 \lambda ^2-97327 \lambda +51135 ) \notag\\
&+293769 \lambda ^3 - 562239 \lambda^2+237523 \lambda+29971 \bigr]+6
v_3  (29825 \lambda ^3-68713 \lambda ^2 +46923 \lambda -8099 )\notag\\
&  +245651 \lambda ^3-551007 \lambda ^2+363249
\lambda-57837\Bigr\}, \\
{\cal P}^{v_2}_5 =&  (1 - \lambda)^2(1 - 3\lambda)\Bigl\{64 v_2^3
(\lambda -1)^2 (329-961 \lambda )  +324 v_3^3 (39 \lambda ^3-73
\lambda ^2+41 \lambda -7 ) \notag \\
&-16 v_2^2 (\lambda -1)   \bigl[ 3 v_3 (1085 \lambda ^2-1482 \lambda
+397 ) +22990 \lambda ^2-27850 \lambda+3776 \bigr] \notag\\
&-27  v_3^2 (7625 \lambda ^3-19677 \lambda ^2+14703 \lambda -2651 )+
1344 v_1  \bigl[ 2 v_2 (1-\lambda)^2 (181 \lambda -137) \notag\\
&+9 v_3(1-\lambda)^2 (38 \lambda -31) +140 \lambda ^3-53 \lambda
^2-215 \lambda+122 \bigr] +v_2  \bigl[ 27 v_3^2  (683 \lambda ^3 
-1273 \lambda ^2+681 \lambda -91 )
\notag\\
& - 6 v_3 (102523 \lambda ^3-238935
\lambda ^2+159777 \lambda -23365 ) 
+193117 \lambda ^3-427691 \lambda ^2+258807 \lambda-27161 \bigr]
\notag\\
& -6 v_3 (31359 \lambda ^3-58643 \lambda ^2+33565 \lambda
-6089 )  +499453 \lambda ^3-1131897 \lambda ^2+782671 \lambda-150059\Bigr\},\\
{\cal P}^{v_2}_6 =& -2  (1 - 3\lambda)^2 \Bigl\{ 8576 (\lambda -1)^4
v_2^3 - 54  v_3^3 (1-\lambda)^3 (85 \lambda -31) 
- 18  v_3^2 (1-\lambda)^3 (3397 \lambda -1203)
\notag \\
& +16 v_2^2 (1-\lambda
)^3  \bigl( 1116 v_3 (1-\lambda ) +634 \lambda -709\bigr)
+2016  v_1  (\lambda -1)^2 (35 \lambda ^2-48 \lambda +12 )
 \notag\\
&  - v_2
(1-\lambda) \bigl[ 9 v_3^2 (1-\lambda )^2 (1669 \lambda -1255)
+6  v_3 (1-\lambda)^2 (8860 \lambda -1737) \notag\\
&+14503 \lambda ^3-16053
\lambda ^2-9653 \lambda+11135 \bigr]
-6 v_3 (1-\lambda)^2  (2180 \lambda ^2-7623 \lambda +5187 ) \notag \\
&-29405 \lambda ^4+75256 \lambda ^3-46026 \lambda ^2-16152 \lambda
+16351\Bigr\},\\ 
{\cal P}^{v_2}_7 =& -2  (1 - 3\lambda)^2 \Bigl\{  16 v_2^2
(1-\lambda)^3 (727-2188 \lambda )  +1458  v_3^3 (1-\lambda)^3 (1-3
\lambda) \notag\\
&+19602 v_3^2 (1-\lambda)^3 (1-3 \lambda) + 672  v_1 (1-\lambda)^2 (35
\lambda ^2-48 \lambda +12 ) \notag\\
&+  v_2 (1-\lambda)  \bigl[ 1863 v_3^2 (1-\lambda)^2 (1-3 \lambda ) -
78v_3 (1-\lambda )^2 (1102 \lambda -365) -111931 \lambda ^3 \notag\\
&+ 255965 \lambda ^2 - 176459 \lambda + 32629 \bigr] +6
v_3(1-\lambda)^2  (14814 \lambda ^2-18371 \lambda +3829 ) \notag \\
&- 47811 \lambda ^4+180304 \lambda ^3-236654 \lambda ^2+123472
\lambda-19239\Bigr\},\\ 
{\cal P}^{v_2}_8 =& -2  (1 - 3\lambda)^3 \Bigl\{ 6 \bigl(280  v_2^2+9
v_3^2 (9  v_3+121) \bigr)  (1-\lambda)^3 - 6 v_3 (1-\lambda)^2 (707
\lambda -715) \notag\\
&  +v_2 (1-\lambda)    \bigl[ 621  v_3^2 (1-\lambda)^2 +7410  v_3
(1-\lambda)^2  +10597 \lambda ^2-18998 \lambda +8299 \bigr] \notag\\
& -9739 \lambda ^3+18073 \lambda ^2-6487 \lambda-1883 \Bigr\},\\
{\cal P}^{v_2}_9 =& -2  (1 - 3\lambda)^3 \big[ 2 v_2 (48 \lambda ^3-86 \lambda ^2+45 \lambda -7)  -295 \lambda ^3+1253 \lambda ^2-1271 \lambda +301\big]\;.
\end{align}
\eseq
Polynomials in the $\beta$-function of $v_3$ are:
\bseq
\label{Pol_v3}
\begin{align}
{\cal P}^{v_3}_0 = &-4  (1 - \lambda)^6 (8 v_2 + 9 v_3 - 7)^3,\\
{\cal P}^{v_3}_1 = &3 {\cal P}^{v_3}_0,\\
{\cal P}^{v_3}_2 = & -(1 - \lambda)^4 (8 v_2 + 9 v_3 - 7)\Big\{ 768
v_2^2 (1 - \lambda) (5 \lambda-1)  +  3 v_3^2 (1 - \lambda) (741
\lambda-31)  \notag\\
&+  8 v_2 \left( v_3 (1 - \lambda) ( 687 \lambda - 85) - 306 \lambda^2
  + 496 \lambda  -  206 \right) + v_3 (651 \lambda^2  - 76 \lambda -
719) \notag\\
& - 5286 \lambda^2  + 6824 \lambda -1426 \Big\},\\
{\cal P}^{v_3}_3 =&    -(1 - \lambda)^4 (8 v_2 + 9 v_3 - 7)\Big\{
256 v_2^2 (1 - \lambda) (53 \lambda - 17) +  9 v_3^2 (1 - \lambda) (
1029 \lambda - 319)  \notag\\
&+  8 v_2 \big(3 v_3 (1 - \lambda) ( 879 \lambda - 277) - 470
  \lambda^2 + 592 \lambda  - 170 \big) \notag\\
&+ 3 v_3 (1995 \lambda^2 - 2764 \lambda + 625) - 17426 \lambda^2   + 23608 \lambda  - 5846 \Big\},\\
{\cal P}^{v_3}_4 =& -(1 - \lambda)^2 (1 -  3 \lambda) \Big\{ -64 v_2^2
(1 - \lambda) \big[384 v_2 (1 - \lambda)^2  +  v_3 (495 \lambda^2   -
1358 \lambda + 863) \notag\\ 
&- 6 (697 \lambda^2  - 942 \lambda + 261)\big]  +   16 v_2 \big[ 3
  v_3^2 (1 - \lambda)^2 ( 63 \lambda - 895 ) \notag\\ 
&+   v_3 (1 - \lambda)(38823 \lambda^2
 - 53248 \lambda + 15061)+  31674 \lambda^3  - 76106 \lambda^2  +
 57078 \lambda  - 12630  \big] \notag\\
&-  27 v_3^3 (1 - \lambda)^2 ( 279 \lambda + 425)-  12 v_3^2 ( 24879
\lambda^3 - 59880 \lambda^2 + 45529 \lambda -10528 ) \notag\\
&+  3 v_3 ( 87697 \lambda^3  - 220167 \lambda^2 + 178243 \lambda -
45677 ) + 240090 \lambda^3 - 542894 \lambda^2 + 357966 \lambda -55386
\Big\},\\
{\cal P}^{v_3}_5 =&-(1-\lambda)^2 (1- 3\lambda) \Big\{-64 v_2^2 (1 -
\lambda) \big[ v_3 (1069 \lambda^2 - 1434 \lambda + 365) - 6026
\lambda^2+ 7932 \lambda -2002 \big]   \notag\\ 
&  + 16 v_2 \big[3 v_3^2 (1 - \lambda)^2
  (1445 \lambda - 517) -  v_3 ( 38317 \lambda^3 - 88093 \lambda^2 +
  62567 \lambda -12791 ) \notag\\
& +   2 ( 2335 \lambda^3 - 9175 \lambda^2 +
9161 \lambda   -2297) \big]    +  27 v_3^3 ( 169 \lambda^3 - 499
\lambda^2 + 451 \lambda -121   )   \notag   \\
&-  12 v_3^2 ( 15961 \lambda^3
- 36412 \lambda^2 + 26015 \lambda  -5564  )  
- v_3 ( 216917 \lambda^3 - 386819 \lambda^2 + 188983 \lambda - 19945 )\notag \\
& +  2 ( 264149 \lambda^3 - 575591 \lambda^2+ 382375 \lambda -71269
) \Big\},\\ 
{\cal P}^{v_3}_6 =&2 (1- 3\lambda)^2 \Big\{ (1 - \lambda)^3 \big[ 3
v_3^2 \big(45 v_3 ( 73 \lambda -67) - 2 (  6835 \lambda - 5644)\big)
- 256 v_2^2 ( 32 v_3 - 75 ) (1 - \lambda) \big] 
\notag\\ 
&   - 16 v_2 (1 -
\lambda)^2 \big[1056 v_3^2 (1 - \lambda)^2 -     v_3 ( 3911 \lambda^2
  - 7243 \lambda 
+3332)+  6 (213 \lambda^2 - 457 \lambda + 226)\big]
\notag\\
& - v_3 (4877
\lambda^4 -  40820 \lambda^3 + 91880 \lambda^2 - 80876 \lambda
+ 24939  )\notag\\ 
& -  4 (  3330 \lambda^4 - 7283 \lambda^3 + 1162 \lambda^2 +
6195 \lambda  - 3396  ) \Big\},\\ 
{\cal P}^{v_3}_7 =& -2 (1- 3\lambda)^2 \Big\{  3 v_3^2 (1 - \lambda)^3 
\big[135 v_3 (1 - 3 \lambda)  -  14 ( 289 \lambda - 92)\big]  \notag\\
&+16 v_2 (1 - \lambda)^2 \big[  v_3 (1 - \lambda) (412 - 1243 \lambda)
- 5542 \lambda^2 + 7230 \lambda - 1788\big] \notag \\
&+  v_3 (1 - \lambda) (93119 \lambda^3 -  211785 \lambda^2 + 147207
\lambda - 28337) \notag\\
&+  4 ( 4889 \lambda^4 - 22219 \lambda^3 + 32834 \lambda^2 - 18597
\lambda + 3117 )\Big\}, \\ 
{\cal P}^{v_3}_8 =& -6 (1- 3\lambda)^3 \Big\{  3  v_3^2 (15 v_3 - 88 )
(1 - \lambda)^3  - v_3 (1289 \lambda^3 - 3119 \lambda^2 + 2337 \lambda
- 507 )  \notag\\ 
&+  4 ( 1297 \lambda^3 - 2877 \lambda^2 + 1858 \lambda - 282 )\Big\},\\
{\cal P}^{v_3}_9 =&-4 (1- 3\lambda)^3 \big[  v_3 ( 48 \lambda^3 - 86
\lambda^2 + 45 \lambda - 7 )  - 151 \lambda^3 + 257 \lambda^2 - 135
\lambda + 21 \big]\; . 
\end{align}
\eseq

\section{Commutator of an operator in a fractional power}
\label{app:A}

Here we derive the asymptotic series (\ref{commut}) for the
commutator of an operator in a fractional power. 
The derivation starts with the representation
\be
\label{Gammarepr}
A^\al=\frac{1}{\Gamma(-\al)}\int_0^\infty ds\,s^{-\al-1}\e^{-sA}\;.
\ee
Next, for $X(s)\equiv [\e^{-sA},B]$  we have an equation,
\be
\frac{dX}{ds}=-AX-[A,B]\e^{-sA}\;,
\ee
with initial condition $X(0)=0$. It is straightforward to verify that
the
solution has the form,
\be
\label{Xs}
X(s)=-\int_0^s ds_1 \e^{-(s-s_1)A}[A,B]\e^{-s_1A}\;.
\ee
Using Eqs.~(\ref{Gammarepr}) and (\ref{Xs}) we obtain,
\be
\label{step1}
\begin{split}
[A^\al,B]&=\frac{-1}{\Gamma(-\al)}\int_0^\infty ds s^{-\al-1}\int_0^s ds_1
\e^{-(s-s_1)A}[A,B]\e^{-s_1 A}\\
&=\frac{-1}{\Gamma(-\al)}\int_0^\infty ds s^{-\al-1}
\bigg(s[A,B]\e^{-sA}+\int_0^s
ds_1[\e^{-(s-s_1)A},[A,B]]\e^{-s_1A}\bigg)\\
&=-\frac{\Gamma(-\al+1)}{\Gamma(-\al)}[A,B]A^{\al-1}-
\frac{1}{\Gamma(-\al)}\int_0^\infty ds s^{-\al-1}\int_0^s
ds_1[\e^{-(s-s_1)A},[A,B]]\e^{-s_1A}\;.
\end{split}
\ee
In the last term we can again use the formula (\ref{Xs}) that will
generate the second term in the expansion (\ref{commut}).
Proceeding by induction we obtain the
representation
\be
\begin{split}
[A^\al,B]=&\sum_{n=1}^N C_\al^n\;\underbrace{[A,[A,\ldots[A}_n,B]\ldots]
A^{\al-n}\\
&+\frac{(-1)^N}{\Gamma(-\al)}\int_0^\infty ds\,s^{-\al-1}
\int_0^s ds_1\frac{s_1^{N-1}}{(N-1)!}
[\e^{-(s-s_1)A},\underbrace{[A,[A,\ldots[A}_N,B]\ldots]\e^{-s_1A}\;,
\end{split}
\ee
which is valid for arbitrary $N$. This yields the
formula (\ref{commut}).

\section{More on universal functional traces}
\label{AppTTraces}

\subsection{Coincidence limits of the heat kernel coefficients}
\label{App:HAMIDEW}
The heat kernel $\hat K(s|\,x,y)=e^{s\hat F(\nabla)}\hat\delta(x,y)$ for
the operator (\ref{F}) satisfies the initial value problem
for its heat equation 
\bseq
\begin{align}
&\partial_s\hat K(s|\,x,y)=\hat F(\nabla)\,K(s|\,x,y),   \label{heat_equation}\\
&K(0|\,x,y)=\hat\delta(x,y).  \label{K_initial_condition}
\end{align}
\eseq
The expansion (\ref{heatexpansion}) when substituted into
Eq.~(\ref{heat_equation})  
leads to the sequence of equations for the HAMIDEW
coefficients $\hat a_n(x,y)$, 
\bseq
\label{HAMeqs}
\begin{align}
&\s^\mu\n_\mu \hat{a}_0=0,\\
&(n+1)\,\hat{a}_{n+1} + \s^\mu\n_{\mu} \hat{a}_{n+1} =
{\cal D}^{-1/2}\,{\Box}\Big({\cal D}^{1/2} \hat{a}_n \Big)+\Big(\hat P-\frac16R\hat 1\Big)\, \hat{a}_n,\quad n\ge 0,
\end{align}
\eseq
where $\sigma^\mu=\nabla^\mu\sigma(x,y)$ is the vector at the point
$x$ tangential to the geodetic curve connecting $x$ and $y$. In
deriving these recurrence equations we took into account that in the
lowest two orders of the $s$-expansion the heat equation is satisfied
in virtue of the equation for the Synge world function $\s(x,y)$ 
\bseq
\label{Syngeeq}
\be
\s = \frac12 \nabla^\mu\s\,\nabla_\mu\s
\ee
and its corollary, the equation for the Van Vleck--Morette
determinant (\ref{PVVleck}) 
\be
{\cal D}^{-1}\n_\mu \left( {\cal D} \s^\mu \right) = d.
\ee
\eseq
These equations in their turn should be amended by the initial
conditions in the coordinate spacetime, 
\be
\s\,|_{\,y=x} = 0,\quad \n_\mu\s\, |_{\,y=x}=0,
\quad \hat{a}_0|_{\,y=x} = \hat{1},
\ee
the latter following from the initial condition (\ref{K_initial_condition}).

By consecutively differentiating Eqs.~(\ref{HAMeqs}), (\ref{Syngeeq})
and taking their coincidence limits one can
systematically derive in the closed form the expressions for the
needed coincidence limits (\ref{limits}). The result of these
calculations which basically reduce to the commutator algebra of
covariant derivatives (\ref{commutators}) begins with the following
list \cite{PhysRep},\footnote{The first and last indices in round
  brackets are symmetrized with the factor $1/2$.} 
\bseq
\label{firstUFTs}
    \begin{align}
    &\sigma\,\big|_{\,y=x}=0,\quad\nabla_{\mu}\nabla_{\nu}\sigma\,
    \big|_{\,y=x}=g_{\mu\nu},\quad
    \nabla_{\mu}\nabla_{\nu}\nabla_{\alpha}\sigma\,
    \big|_{\,y=x}=0,\quad
\nabla_{\mu}\nabla_{\nu}\nabla_{\alpha}\nabla_{\beta}\sigma\
    \big|_{\,y=x}=-\frac23\,R_{\mu(\alpha\nu\beta)},\\
&
\nabla_\lambda\nabla_\mu\nabla_\nu\nabla_\al\nabla_\beta\sigma\big|_{\,y=x}=-\frac{1}{2}
\big(\nabla_\lambda R_{(\al\mu\beta)\nu}+\nabla_\mu
R_{(\al\nu\beta)\lambda}
+\nabla_\nu R_{(\al\lambda\beta)\mu}\big)\;,
\\
    &{\cal D}^{1/2}\,
    \big|_{\,y=x}=1,\quad
    \nabla_{\mu}{\cal D}^{1/2}\,\big|_{\,y=x}=0,\quad
    \nabla_{\mu}\nabla_{\nu}{\cal D}^{1/2}\,\big|_{\,y=x}
    =\frac16\,R_{\mu\nu},\\
&\nabla_\mu\nabla_\nu\nabla_\al {\cal D}^{1/2}\big|_{\,y=x}
=\frac{1}{12} \big(\nabla_\mu R_{\nu\al}+\nabla_\nu
R_{\al\mu}+\nabla_\al R_{\mu\nu} \big)\;,
    \end{align}
\eseq
and
\bseq
\label{firstUFTs1}
	\begin{eqnarray}
    &&\hat a_0\,\big|_{\,y=x}=\hat 1,\quad \nabla_\mu\hat a_0\,\big|_{\,y=x}=0,
    \quad \nabla_\mu\nabla_\nu\hat a_0\,\big|_{\,y=x}
    =\frac12\,\hat{\mbox{\boldmath$R$}}_{\mu\nu},\quad
    \nabla_\mu\nabla_\nu\nabla_\alpha\hat a_0\,\big|_{\,y=x}=
    \frac23\nabla_{(\mu}\hat{\mbox{\boldmath$R$}}_{\nu)\alpha},\\
    &&\hat a_1\,\big|_{\,y=x}=\hat P, \quad
    \nabla_\mu\hat a_1\,\big|_{\,y=x}=\frac12\nabla_\mu\hat P-\frac16\nabla^\nu\hat{\mbox{\boldmath$R$}}_{\nu\mu},\\
    &&\nabla_\mu\nabla_\nu\hat a_1\,\big|_{\,y=x}=\frac1{180}(2R^{\al\bt}R_{\al\m\bt\nu}
    +2R_{\al\bt\l\m}R^{\al\bt\l}_{\;\;\;\;\;\;\;\nu}
    -4R_{\mu\al}R^\al_\nu+3\Box R_{\mu\nu}-\nabla_\mu\nabla_\nu R)\,\hat 1
    \notag\\
    &&\qquad\qquad\qquad\quad-\frac16\hat{\mbox{\boldmath$R$}}_{\al(\mu}
\hat{\mbox{\boldmath$R$}}_{\nu)}^{\;\;\;\al} 
    +\frac16\nabla_{(\mu}\nabla^\al\hat{\mbox{\boldmath$R$}}_{\nu)\al}+\frac12\hat
    P\,\hat{\mbox{\boldmath$R$}}_{\mu\nu}+\frac13[\hat{\mbox{\boldmath$R$}}_{\mu\nu},\hat   P]+ 
    \frac13\nabla_\mu\nabla_\nu\hat P,\\
	&&\hat{a}_2\,\big|_{\,y=x}=
	\frac1{180}\,(R_{\alpha\beta\gamma\delta}
	R^{\alpha\beta\gamma\delta}
	-R_{\mu\nu} R^{\mu\nu}+\Box R)\,\hat{1}
	+\frac1{12}\hat{\mbox{\boldmath$R$}}_{\mu\nu}
\hat{\mbox{\boldmath$R$}}^{\mu\nu}  
	+\frac12\hat{P}^2+\frac16\Box\hat{P}\;.
	\end{eqnarray}
\eseq
Higher order coincidence limits, which we need up to
$\nabla^8\sigma\,|_{\,y=x}$, $\nabla^6{\cal D}^{1/2}\,|_{\,y=x}$,
$\nabla^6\hat{a}_0\,|_{\,y=x}$,
$\nabla^4\hat{a}_1\,|_{\,y=x}$,
$\nabla^2\hat a_2\,|_{\,y=x}$ and $\hat a_3\,|_{\,y=x}$, take pages, 
so we do not present them here. 

To obtain the needed list of traces (\ref{a_zero0})---(\ref{a_six}),
we apply this method in three spatial dimensions
$d=3$, $\mu\mapsto i=1,2,3$ using the symbolic tensor algebra package
{\it xAct} \cite{xAct} for Mathematica \cite{Mathematica}. 
For the tensor sector of the theory we have
$\hat 1= \delta_{ij}^{\;\;kl}\equiv\delta_{(i}^k\delta^l_{j)}$, 
$({\mbox{\boldmath$R$}}_{mn})_{ij}^{\;\;kl}
=-2\delta^{(k}_{(i}R^{l)}_{~~j)mn}$, whereas for the vector sector
$\hat 1=\delta^i_j$, $({\mbox{\boldmath$R$}}_{mn})^i_{~j}=R^i_{~jmn}$. 

\subsection{Divergences of tensor and vector traces}
\label{App:TUFTs}
To evaluate the one-loop effective action of HG, we compute the UFTs
in the vector and tensor sectors using the method described in
Sec.~\ref{sec:UFTmethod}.  
We keep only the logarithmically divergent part proportional to
the infinite integral over the proper-time parameter,
\be
\ln L^2\equiv\int\frac{ds}s\;.  \label{C15}
\ee
The UFTs are classified in Sec.~\ref{sec:UFTtypes} by the
dimensionality $a$ of the coefficient function that they multiply in
the action; the dimension of the UFT itself is then $6-a$. Within each
group we order the UFTs by the number of derivatives acting on the
powers of Laplacian. 

Let us start with the simplest traces of zero dimension ($a=6$). Their
divergences cannot depend on curvature and 
are expressed purely in terms of the
metric. They have the universal form,
\begin{align}
&\nabla_{i_1}\dots\nabla_{i_{2N-2}}
\frac{\hat{1}}{(-\Delta)^{N+1/2}}\delta(x,y)\,\Big|_{\,y=x}^{\,\rm div}
=\frac{\ln L^2}{4\pi^{2}}\frac{(-1)^{N-1}}{(2N-1)!!}\;\sqrt{g}\;  
 g^{(N-1)}_{i_1\dots i_{2N-2}} \hat{1},
\end{align}
which is valid both for the tensor and vector sectors. Here
$g^{(N-1)}_{i_1\dots i_{2N-2}}$ is the completely symmetric tensor
built of the metric by the recurrence relations
\bseq
\begin{align}
&g^{(1)}_{ij} = g_{ij},
\quad g^{(2)}_{ijkl} =  g_{ij} g_{kl} + g_{ik} g_{jl} + g_{il} g_{jk},\\
&g^{(N)}_{i_1\dots i_{2N}} = \sum_{k=2}^{2N} g_{i_1 i_k}
g^{(N-1)}_{i_2\dots i_{k-1}i_{k+1}\dots i_{2N}}\;.
\end{align}
\eseq

For other traces the full expressions are very lengthy because of the
large number of free indices that they carry in general. To reduce 
the length, we can contract the indices in various combinations which
appear in the effective action. Still, for the traces with $a=4$ and
$3$ (linear in curvature and in derivatives of curvature,
respectively) the number of possible contraction is too high to be
listed explicitly. On the other hand, these traces are obtained by a
straightforward algebra from the lower heat-kernel coefficients listed
in (\ref{firstUFTs}), (\ref{firstUFTs1}). Thus, we focus below on the
most laborious traces with $a=2$ and $a=0$, which are quadratic and
cubic in curvature. Full expressions for uncontracted UFTs for all
$a=0,2,3,4,6$
can be
obtained with the Mathematica code available at~\cite{github}.

\subsubsection*{Traces with $a=2$, quadratic in curvature}
These UFTs enter the divergent part of the action with the 
coefficient functions linear in curvature (to form the logarithmic
divergences of the overall cubic power in the curvature), so that they
represent local quantities quadratic in the curvature and having two
free indices. 

\vspace{1.5mm} 

\noindent
{\bf Tensor traces} can have an even number of derivatives running from
zero to six. 
For the trace without derivatives
there are three possibilities of index contractions:
\bseq
\begin{align}
&g^{ij}(-\Delta)^{1/2}\delta_{ij}^{~\,\,kl}(x,y)|_{y=x}^{\rm div}
= -\frac{\ln L^2}{16\pi^2} \,\sqrt g\,
g^{kl}\frac{1}{30}\left(\frac12R^{mn}R_{mn} + \frac14 R^2 + \Delta
  R\right).\\ 
&g_{kl}(-\Delta)^{1/2}\delta_{ij}^{~\,\,kl}(x,y)|_{y=x}^{\rm div}
= -\frac{\ln L^2}{16\pi^2} \,\sqrt g\, g_{ij}\frac{1}{30}\left(\frac12R^{mn}R_{mn} + \frac14 R^2 + \Delta R\right).\\
&\delta_k^i(-\Delta)^{1/2}\delta_{ij}^{~\,\,kl}(x,y)|_{y=x}^{\rm div}
= -\frac{1}{16\pi^2} \ln L^2\notag\\ 
&\qquad\qquad\times\sqrt g\,\left(-\frac{43}{60}\,\delta_j^l R^{mn}R_{mn}
 + \frac7{12}R_{mj}R^{ml}-\frac7{12}RR_j^l + \frac7{20}\delta_j^l R^2 
+ \frac1{15}\delta_j^l \Delta R\right).
\end{align}
\eseq
For the two-derivative case there are nine contractions,
\bseq
\begin{align}
&\delta_{mn}^{~\,\,kl}\nabla_i\nabla_j
\frac{1}{(-\Delta)^{1/2}}\delta_{kl}^{~\,\,mn}(x,y)\,\big|_{\,y=x}^{\,\rm
  div}
=\frac{\ln L^2}{8\pi^2} \,\notag\\
&\quad\times\sqrt g\,
\left(\frac{19}{15}R_{ik}R^k_j-\frac{41}{60}g_{ij}R_{mn}R^{mn}-\frac{19}{15}RR_{ij}+\frac12g_{ij}R^2 + \frac{3}{10}\nabla_i\nabla_j R + \frac1{10}\Delta R_{ij} - \frac1{10}g_{ij} \Delta R\right),\\
&g_{mn}g^{kl}\nabla_i\nabla_j\frac{1}{(-\Delta)^{1/2}}\delta_{kl}^{~\,\,mn}(x,y)\,\big|_{\,y=x}^{\,\rm div}
=\frac1{8\pi^2}\ln L^2\,\notag\\
&\quad\times\sqrt g\,
\left(-\frac{1}{5}R_{ik}R^k_j+\frac{3}{40}g_{ij}R_{mn}R^{mn}
+\frac{1}{5}RR_{ij}-\frac1{16}g_{ij}R^2 + \frac{3}{20}\nabla_i\nabla_j
R + \frac1{20}\Delta R_{ij} - \frac1{20}g_{ij} \Delta R\right),\\
&g_{mn}\nabla^i\nabla^k\frac{1}{(-\Delta)^{1/2}}
\delta_{kj}^{~\,\,mn}(x,y)\,\Big|_{\,y=x}^{\,\rm div} 
=\frac1{8\pi^2}\ln L^2\,\notag\\
&\quad\times\sqrt g\,\left(
-\frac{1}{15}R^{ik}R_{jk}
+\frac{1}{40}\delta^i_jR_{mn}R^{mn}+\frac{1}{15}RR^{i}_{j}-\frac{1}{48}\delta^{i}_{j}R^2 
+ \frac{1}{20}\nabla^i\nabla_j R + \frac{1}{60}\Delta R^{i}_{j} -
\frac1{60}\delta^{i}_{j} \Delta R
\right),\\
&\delta_n^l\nabla^i\nabla^k\frac{1}{(-\Delta)^{1/2}}\delta_{kl}^{~\,\,jn}(x,y)\,\Big|_{\,y=x}^{\,\rm div}
=\frac1{8\pi^2}\ln L^2\,\notag\\
&\quad\times\sqrt g\, \left(\frac{149}{120}R^i_{k}R^{kj}
-\frac{8}{15}g^{ij}R_{mn}R^{mn}-\frac{119}{120}RR^{ij}+\frac{13}{48}g^{ij}R^2
+ \frac{37}{120}\nabla^i\nabla^j R - \frac{7}{40}\Delta R^{ij} -
\frac1{30}g^{ij} \Delta R\right),\\
&g^{kl}\nabla^i\nabla_m\frac{1}{(-\Delta)^{1/2}}
\delta_{kl}^{~\,\,mj}(x,y)\,\Big|_{\,y=x}^{\,\rm div}
=\frac1{8\pi^2}\ln L^2\,\notag\\
&\quad\times\sqrt g\, \left(
-\frac{1}{15}R^{ik}R^j_{k}
+\frac{1}{40}g^{ij}R_{mn}R^{mn}
+\frac{1}{15}RR^{ij}-\frac{1}{48}g^{ij}R^2 
+ \frac{1}{20}\nabla^i\nabla^j R + \frac{1}{60}\Delta R^{ij} -
\frac1{60} g^{ij} \Delta R
\right),\\
&\delta^l_n\nabla^i\nabla_m\frac{1}{(-\Delta)^{1/2}}
\delta_{jl}^{~\,\,mn}(x,y)\,\Big|_{\,y=x}^{\,\rm div}
=\frac1{8\pi^2}\ln L^2\,\notag\\
&\quad\times\sqrt g\, \left(
-\frac{1}{120}R^{ik}R_{jk}
-\frac{7}{60}\delta^i_jR_{mn}R^{mn}+\frac{1}{20}RR^{i}_{j}
+\frac{1}{16}\delta^{i}_{j}R^2 
- \frac{13}{120}\nabla^i\nabla_j R + \frac{29}{120}\Delta R^{i}_{j} -
\frac{1}{30}\delta^{i}_{j} \Delta R
\right),\\
&\nabla^k\nabla^l\frac{1}{(-\Delta)^{1/2}}\delta_{kl}^{~\,\,ij}(x,y)\,\big|_{\,y=x}^{\,\rm
  div}=\frac1{8\pi^2}\ln L^2\,\notag\\
&\quad\times\sqrt g\,
\left(\frac{14}{15}R^i_{k}R^{kj}-\frac{37}{120}g^{ij}R_{mn}R^{mn}\right.
\left.-\frac{13}{30}RR^{ij}+\frac7{48}g^{ij}R^2 + \frac{13}{60}\nabla^i\nabla^j R - \frac3{20}\Delta R^{ij} - \frac1{60}g^{ij} \Delta R\right),\\
&\nabla_m\nabla^k\frac{1}{(-\Delta)^{1/2}}
\delta_{kl}^{~\,\,mn}(x,y)\,\Big|_{\,y=x}^{\,\rm div}
=\frac1{8\pi^2}\ln L^2\,\notag\\
&\quad\times\sqrt g\, \left(\frac{11}{120}R_{kl}R^{kn}-\frac{3}{20}\delta_l^nR_{mn}R^{mn}-\frac{13}{60}RR_l^n+\frac1{80}\delta_l^nR^2 
+ \frac{1}{40}\nabla_l\nabla^n R + \frac1{120}\Delta R_l^n \right),\\
&\nabla_m\nabla_n\frac{1}{(-\Delta)^{1/2}}
\delta_{ij}^{~\,\,mn}(x,y)\,\Big|_{\,y=x}^{\,\rm div}
=\frac1{8\pi^2}\ln L^2\,\notag\\ 
&\quad\times\sqrt g\, \left(
-\frac{1}{15}R_{i}^{k}R_{jk}
+\frac{1}{40}g_{ij}R_{mn}R^{mn}
+\frac{1}{15}RR_{ij}-\frac{1}{48}g_{ij}R^2 
- \frac{7}{60}\nabla_i\nabla_j R + \frac{11}{60}\Delta R_{ij} -
\frac1{60} g_{ij} \Delta R
\right)\;.
\end{align}
\eseq
For the four-derivative trace there are 5 ways to contract indices:
\bseq
\begin{align}
&g_{mn}\nabla_{i}\nabla_{j}\nabla^{k}\nabla^{l}
\frac{1}{(-\Delta)^{3/2}}\delta_{kl}^{~\,\,mn}(x,y)\,
\Big|_{\,y=x}^{\,\rm div}=\frac{1}{4\pi^2}\ln L^2  \notag \\
&\quad\times\sqrt g\,\left(\frac{1}{30}R_{ik}R^k_j-\frac{1}{80}g_{ij}R_{mn}R^{mn}
-\frac{1}{30}RR_{ij}+\frac{1}{96}g_{ij}R^2 - \frac{1}{40}\nabla_i\nabla_j R - \frac{1}{120}\Delta R_{ij} + \frac1{120}g_{ij} \Delta R \right),\\
& \delta_n^m \nabla_{i}\nabla_{j}\nabla_{k}\nabla^{l}
\frac{1}{(-\Delta)^{3/2}}\delta_{lm}^{~\,\,kn}(x,y)\,\Big|_{\,y=x}^{\,\rm div}=\frac{1}{4\pi^2}\ln L^2   \notag\\
&\quad\times\sqrt g\, \left(-\frac{247}{120}R_{ik}R^k_j+\frac{179}{240}g_{ij}R_{mn}R^{mn}
+\frac{41}{60}RR_{ij}-\frac{19}{96}g_{ij}R^2 -
\frac{13}{10}\nabla_i\nabla_j R - \frac{1}{60}\Delta R_{ij} +
\frac1{60}g_{ij} \Delta R \right),\\
&g^{mn}\nabla_{i}\nabla_{j}\nabla_{k}\nabla_{l}
\frac{1}{(-\Delta)^{3/2}}\delta_{mn}^{~\,\,kl}(x,y)\,\Big|_{\,y=x}^{\,\rm
  div}=\frac{1}{4\pi^2}\ln L^2   \notag\\
&\quad\times\sqrt g\, \left(
\frac{1}{30}R_{ik}R^k_j-\frac{1}{80}g_{ij}R_{mn}R^{mn}
-\frac{1}{30}RR_{ij}+\frac{1}{96}g_{ij}R^2 -
\frac{1}{40}\nabla_i\nabla_j R - \frac{1}{120}\Delta R_{ij} +
\frac{1}{120}g_{ij} \Delta R
\right)\;,\\
& \nabla^{i}\nabla_m\nabla^{k}\nabla^{l}
\frac{1}{(-\Delta)^{3/2}}\delta_{kl}^{~\,\,mj}(x,y)\,\Big|_{\,y=x}^{\,\rm div}=\frac{1}{4\pi^2}\ln L^2  \notag \\
&\quad\times\sqrt g\, \left(-\frac{77}{40}R^{ik}R_k^j
+\frac{307}{240}g^{ij}R_{mn}R^{mn}
+\frac{113}{60}RR^{ij}-\frac{55}{96}g^{ij}R^2 
- \frac{17}{30}\nabla^i\nabla^j R + \frac{1}{30}\Delta R^{ij} 
+ \frac{1}{120}g^{ij} \Delta R \right)\;,\\
& \nabla_{i}\nabla_k\nabla_{l}\nabla^{m}
\frac{1}{(-\Delta)^{3/2}}\delta_{mj}^{~\,\,kl}(x,y)\,\Big|_{\,y=x}^{\,\rm div}=\frac{1}{4\pi^2}\ln L^2  \notag \\
&\quad\times\sqrt g\, \left(
-\frac{17}{40}R_{ik}R^k_j-\frac{13}{240}g_{ij}R_{mn}R^{mn}
-\frac{9}{20}RR_{ij}+\frac{3}{32}g_{ij}R^2 -
\frac{29}{60}\nabla_i\nabla_j R - \frac{1}{20}\Delta R_{ij} +
\frac{1}{120}g_{ij} \Delta R
\right)\;.
\end{align}
\eseq
There is only one way to contract indices for the trace with 6 derivatives
\be
\begin{aligned}
& \nabla_{i}\nabla_{j}\nabla_{k}\nabla_{l}\nabla^{m}\nabla^{n}
\frac{1}{(-\Delta)^{3/2}}\delta_{mn}^{~\,\,kl}(x,y)\,\Big|_{\,y=x}^{\,\rm div}=\frac{1}{6\pi^2}\ln L^2   \\
&\quad\times\sqrt g\,\left(\frac{57}{10}R_{ik}R^k_j-\frac{457}{160}g_{ij}R_{mn}R^{mn}
-\frac{49}{20}RR_{ij}+\frac{47}{64}g_{ij}R^2 + \frac{283}{80}\nabla_i\nabla_j R + \frac{1}{80}\Delta R_{ij} + \frac{19}{80}g_{ij} \Delta R \right).
\end{aligned}
\ee

\vspace{2mm}

\noindent
{\bf Vector traces} In this groups there is one trace 
without derivatives:
\be
\begin{aligned}
&(-\Delta)^{1/2}\delta^i_{\,\,j}(x,y)\Big|_{y=x}^{\rm div}= 
-\frac{\ln L^2}{16\pi^2} \sqrt{g}
\left(-\frac{3}{20}\delta^i_j \, R^{kl}R_{kl} +\frac16 R^{ik}R_{kj}-\frac16 R^i_j R + \frac{11}{120}\delta^i_j \, R^2 +\frac{1}{30}\delta^i_j \Delta R\right)\;,
\end{aligned}
\ee
three traces with two derivatives:
\bseq
\begin{align}
&\delta^l_k\nabla_i\nabla_j\frac{1}{(-\Delta)^{1/2}}\delta^l_{~k}(x,y)
\Big|_{y=x}^{\rm div} 
=\frac{\ln L^2}{8\pi^2}\notag\\ 
&\quad\times\sqrt{g}\left(\frac{2}{15}R_{ik}R^k_j-\frac{11}{120}g_{ij}R_{kl}R^{kl}
-\frac{2}{15}RR_{ij}+\frac{1}{16}g_{ij}R^2 +
\frac{3}{20}\nabla_i\nabla_j R + \frac1{20}\Delta R_{ij} -
\frac1{20}g_{ij} \Delta R\right),\\ 
&\nabla_i\nabla_k
\frac{1}{(-\Delta)^{1/2}}\delta^k_{\,\,j}(x,y)\,\Big|_{\,y=x}^{\,\rm div}
=\frac{\ln L^2}{8\pi^2}\notag\\
&\quad\times\sqrt{g}
\left(\frac{7}{20}R_{ik}R^k_j-\frac{17}{120}g_{ij}R_{kl}R^{kl}
-\frac{4}{15}RR_{ij}+\frac1{16}g_{ij}R^2 +
\frac{2}{15}\nabla_i\nabla_j R - \frac1{15}\Delta R_{ij} -
\frac1{60}g_{ij} \Delta R\right)\;,\\
&\nabla_i\nabla^j\frac{1}{(-\Delta)^{1/2}}
\delta^k_{\,\,j}(x,y)\,\Big|_{\,y=x}^{\,\rm div} =\frac{\ln
  L^2}{8\pi^2}\notag\\
&\quad\times\sqrt{g}\left(-\frac{3}{20}R_{ij}R^{jk}
+\frac{1}{40} \delta_i^k R_{jl}R^{jl} +\frac{3}{20}R R_i^k
-\frac{1}{48} \delta_i^k R^2-\frac{1}{30}\nabla_i\nabla^k R
+\frac{1}{10} \Delta R_i^k
-\frac{1}{60} \delta_i^k \Delta R 
\right)
\end{align}
\eseq
and a single trace with four derivatives:
\be
\begin{aligned}
&\nabla_{i}\nabla_{j}\nabla^l\nabla_{k}
\frac{1}{(-\Delta)^{3/2}}\delta^k_{\,\,l}(x,y)\,\Big|_{\,y=x}^{\,\rm
  div}=\frac{\ln L^2}{4\pi^2} \\
&\quad\times\sqrt{g}
\left(-\frac{11}{20}R_{ik}R^k_j+\frac{47}{240}g_{ij}R_{kl}R^{kl}
+\frac{2}{15}RR_{ij}-\frac{1}{32}g_{ij}R^2 - \frac{21}{40}\nabla_i\nabla_j R - \frac{1}{120}\Delta R_{ij} + \frac1{120}g_{ij} \Delta R \right).
\end{aligned}
\ee

\subsubsection*{Traces with $a=0$, cubic in curvature}

These traces have dimension $6$ and thus enter the divergent part of
the action without extra curvature coefficients. Hence, for our
purposes, it is sufficient to calculate them with all indices
contracted and integrated over the whole 3-dimensional space. In the
integrals 
we freely integrate by parts in order to convert
them to the sum of basic curvature invariants of
Eq.~(\ref{action31}). 

\vspace{1.5mm}

\noindent
{\bf Tensor traces} can have no derivatives, 
two derivatives or four derivatives. There are two possible index
contractions in the traces without 
derivatives, which are expressed directly in terms of the coincidence
limit of the third Schwinger-DeWitt coefficient
${a_3}_{ij}^{~~kl}(x,x)$ 
\bseq
\label{TTcub0div}
\begin{align}
\label{TTr3ordKr}
\int d^3x&\,
\delta_{kl}^{\,\,~ij}(-\Delta)^{3/2}\delta_{ij}^{\,\,~kl}(x,y)\,\Big|_{\,y=x}^{\,\rm 
  div} 
 = \frac{3\ln L^2}{32\pi^2} \int d^3x\,\sqrt g\,
 \delta_{kl}^{\,\,~ij}\, {a_3}_{ij}^{~\,\,kl}(x,x)\notag\\ 
 &=\frac{3\ln L^2}{32\pi^2}\int d^3x\,\sqrt g\,
 \left(\frac{31}{45}R^i_jR^j_kR^k_i - \frac{233}{210}R_{ij}R^{ij}R 
 + \frac{673}{2520}R^3 + \frac{5}{84} R\Delta R  -
 \frac{67}{420}R_{ij}\Delta R^{ij} \right), \\
\label{TTr3ord2m}
\int d^3x&\, g_{kl} g^{ij}(-\Delta)^{3/2}\delta_{ij}^{~\,\,kl}(x,y)\,\Big|_{\,y=x}^{\,\rm div}
 = \frac{3\ln L^2}{32\pi^2} \int d^3x\,\sqrt g\, g_{kl} g^{ij}\,
 {a_3}_{ij}^{~\,\,kl}(x,x)\notag\\ 
 &=\frac{3\ln L^2}{32\pi^2}\int d^3x\,\sqrt g\, \left(-\frac{1}{60}R^i_jR^j_kR^k_i + \frac{1}{35}R_{ij}R^{ij}R
 - \frac{3}{560}R^3 + \frac{1}{112} R\Delta R  + \frac{1}{280}R_{ij}\Delta R^{ij} \right).
\end{align}
\eseq
The traces with two derivatives admit three possible
contractions of indices, 
\bseq
\begin{align}\label{3o3d}
\int d^3x&\, g_{ij}\nabla^{k}\nabla^{l}(-\Delta)^{1/2}\delta_{kl}^{~~ij}(x,y)\,\Big|_{\,y=x}^{\rm div}=
\int d^3x\, g_{kl}\nabla^{i}\nabla^{j}(-\Delta)^{1/2}\delta_{ij}^{~~kl}(x,y)\,\Big|_{\,y=x}^{\rm div}\notag\\
&=-\frac{\ln L^2}{16\pi^2} \int d^3x\,\sqrt g\,  \left(-\frac{1}{120}R^i_jR^j_kR^k_i + \frac{1}{70}R_{ij}R^{ij}R
 - \frac{3}{1120}R^3 + \frac{1}{224} R\Delta R  + \frac{1}{560}R_{ij}\Delta R^{ij} \right),\\
\int d^3x&\, \delta^j_l\nabla_{k}\nabla^{i}(-\Delta)^{1/2}\delta_{ij}^{~\,\,kl}(x,y)\,\Big|_{\,y=x}^{\rm div}\notag\\
&=-\frac{\ln L^2}{16\pi^2} \int d^3x\,\sqrt g\, \left(-\frac{23}{80}R^i_jR^j_kR^k_i + \frac{753}{1120}R_{ij}R^{ij}R
 - \frac{22}{105}R^3 - \frac{1}{84} R\Delta R  -
 \frac{61}{560}R_{ij}\Delta R^{ij} \right).
\label{twodivUFT}
\end{align}
\eseq
The trace with four derivatives admits only one type of index
contractions and reads 
\be
\label{fourdivUFT}
\begin{aligned}
& \int d^3x\,\nabla_{k}\nabla_{l}\nabla^{i}\nabla^{j}
\frac{1}{(-\Delta)^{1/2}}\delta_{ij}^{~\,\,kl}(x,y)\, \Big|_{\,y=x}^{\,\rm div}\\
&=\frac{\ln L^2}{8\pi^2}\int d^3x\,\sqrt g\, \left(-\frac{211}{240}R^i_jR^j_kR^k_i + \frac{2987}{1680}R_{ij}R^{ij}R
 - \frac{703}{2240}R^3 + \frac{31}{1344} R\Delta R  + \frac{141}{1120}R_{ij}\Delta R^{ij} \right).
\end{aligned}
\ee

\vspace{2mm}

\noindent
{\bf Vector traces:} There are two traces in this group, 
\bseq
\begin{align}
\int& d^3x\, \delta^j_i(-\Delta)^{3/2}\delta^i_{~j}(x,y)\,\Big|_{\,y=x}^{\,\rm div}
 \notag\\
 &=\frac{3\ln L^2}{32\pi^2} \int d^3x\,\sqrt{g}\left(\frac{23}{180}R^i_jR^j_kR^k_i - \frac{43}{210}R_{ij}R^{ij}R
 + \frac{253}{5040}R^3 + \frac{29}{1680} R\Delta R  - \frac{5}{168}R_{ij}\Delta R^{ij} \right),\\
\label{3o3dv}
\int& d^3x\, \nabla^j\nabla_i(-\Delta)^{1/2}\delta^i_{\,\,j}(x,y)\,\Big|_{\,y=x}^{\,\rm div}\notag\\
&=-\frac{ \ln L^2}{16\pi^2}\int d^3x\,\sqrt{g}  \left(\frac{1}{20}R^i_jR^i_kR^k_i - \frac{67}{1680}R_{ij}R^{ij}R
 - \frac{3}{1120}R^3 -\frac{9}{1120} R\Delta R  - \frac{13}{560}R_{ij}\Delta R^{ij} \right).
\end{align}
\eseq

\subsection{Relations between universal functional traces of different
  tensor ranks}\label{UFTcheck} 

Tensor traces with $a=0$, $p\leq 4$, $N+1/2=(p-3)/2$ listed in
Eq.~(\ref{TTcub0div})---(\ref{fourdivUFT})   
are the most complicated ones, because they require the knowledge of
$\nabla^8\sigma(x,y)\,|_{y=x}$,
$\nabla^6{\cal D}^{1/2}(x,y)\,|_{y=x}$, 
$\nabla^6\hat a_0(x,y)\,|_{y=x}$,
$\nabla^4\hat a_1(x,y)\,|_{y=x}$,
$\nabla^2 \hat a_2(x,y)\,|_{y=x}$ and
$\hat a_3(x,x)$. 
When used in the calculation of the effective
action, the indices of derivatives in them are necessarily
contracted with indices of the tensor delta function. Such traces can
be simplified to the case of lower tensor rank --- to vector and scalar
ones.  

Consider first the 4-derivative trace. Since this trace is already
cubic in curvature it is not multiplied by any dimensionful quantity
and enters as an integral (\ref{fourdivUFT}). We use the general
formula valid for an arbitrary kernel $\hat G(x,y)$,
\be
\nabla^x_k \hat G(x,y)\big|_{y=x}+\nabla^y_k \hat G(x,y)\big|_{y=x}=
\nabla^x_k \hat G(x,x)
\ee 
and denote
\begin{eqnarray}
    \hat G(x,y)\overleftarrow{\nabla}_k\equiv
    \nabla^y_k\hat G(x,y),\quad
    \hat G(x,y)\overleftarrow{\nabla_k\nabla_l}\equiv
    \nabla^y_l\nabla^y_k\hat G(x,y),
    \end{eqnarray}
Note the order of derivatives is important and gets reversed
with this notation. Neglecting the total-derivative terms, we obtain,
    \begin{eqnarray}
    \int d^3x\nabla_k\nabla_l\nabla^i\nabla^j
    \frac1{(-\Delta)^{1/2}}
    \delta_{ij}^{\;\;\; kl}(x,y)\,\big|_{\,y=x}=
    \int d^3x\nabla^i\nabla^j
    \frac1{(-\Delta)^{1/2}}
    \delta_{ij}^{\;\;\; kl}(x,y)
    \overleftarrow{\nabla_k\nabla_l}\,\big|_{\,y=x}\,\label{by_parts}
    \end{eqnarray}
By the definition of the tensor delta function we have
    \begin{eqnarray}
    \delta_{ij}^{\;\;\; kl}(x,y)\overleftarrow{\nabla_k\nabla_l}=
    \nabla_i\nabla_j\delta_{\rm scalar}(x,y),
    \end{eqnarray}
because $\int
d^3y\,\delta_{ij}^{\;\;\;
  kl}(x,y)\overleftarrow{\nabla_k\nabla_l}\varphi(y)\equiv\int
d^3y\,\delta_{ij}^{\;\;\; 
  kl}(x,y)\nabla_k\nabla_l\varphi(y)=\nabla_i\nabla_j\varphi(x)$
for any scalar test function $\varphi(x)$, and $\delta_{\rm
  scalar}(x,y)$ means the scalar delta function which is a
scalar with respect to the argument $x$ and the scalar {\em density}
with respect to the second argument $y$. Thus (\ref{by_parts}) can be
rewritten as 
    \begin{align}
    \int d^3x&\,\nabla_k\nabla_l\nabla^i\nabla^j
    \frac1{(-\Delta)^{1/2}}
    \delta_{ij}^{\;\;\; kl}(x,y)\,\Big|_{\,y=x}\nonumber\\
    &=
    \int d^3x\,\left((\nabla_i\nabla_j)^2\frac1{(-\Delta_{\rm
          scalar})^{1/2}}
+\nabla_i\nabla_j\big[\,(-\Delta)^{-1/2},\nabla^i\nabla^j\,\big]
    \right)
    \delta_{\rm scalar}(x,y)\,\Big|_{\,y=x}\nonumber\\
    &=
    \int d^3x\,\left( (-\Delta)^{3/2}
+\nabla^i R_{ij}\nabla^j (-\Delta)^{-1/2}
+\nabla_i\nabla_j \big[\,(-\Delta)^{-1/2},\nabla^i\nabla^j\,\big]
  \right)
    \delta_{\rm scalar}(x,y)\,\Big|_{\,y=x}\,.
    \end{align}
The commutator in the last term can be calculated
using Eq.~(\ref{commut}).
We have used this identity as a check of the symbolic computation
result and verified that an  
independent evaluation of the left hand side (purely tensorial)
coincides with the scalar type functional trace on the
right hand side. 

The two-derivative traces contain three possible
index contractions. Two of them, Eq.~(\ref{3o3d}), trivially reduce
to the purely 
scalar case,
    \begin{align}
    \int d^3x\,g^{ij}\nabla_k\nabla_l
    (-\Delta)^{1/2}
    \delta_{ij}^{\;\;\; kl}(x,y)\,\big|_{\,y=x}&= \int
    d^3x\,\nabla_k\nabla_l(-\Delta)^{1/2}\, g^{kl}
\delta_{\rm scalar}(x,y)\,\big|_{\,y=x}
\notag\\
    &=-\int d^3x\,
    (-\Delta)^{3/2}\delta_{\rm scalar}(x,y)\,\big|_{\,y=x}\,.
    \end{align}
For the third trace (\ref{twodivUFT}) we have
    \begin{align}
    \int d^3x\,\delta^i_k\nabla_l\nabla^j
    (-\Delta)^{1/2}
    \delta_{ij}^{\;\;\; kl}(x,y)\,\big|_{\,y=x}&=
    -\int d^3x\,\delta^i_k\nabla^j
    (-\Delta)^{1/2}\delta_{ij}^{\;\;\; kl}(x,y)\overleftarrow{\nabla}_l\,\big|_{\,y=x}\nonumber\\
   &=\int d^3x\,\delta^i_k\nabla^j
    (-\Delta)^{1/2}\nabla_{(i}\delta_{j)}^{\;\;\; k}(x,y)\,\big|_{\,y=x}\,,
    \end{align}
where we took into account that $\delta_{ij}^{\;\;\;
  kl}(x,y)\overleftarrow{\nabla}_l=-\nabla_{(i}\delta_{j)}^{\;\;\;
  k}(x,y)$ from the definition of the vector delta
function\footnote{Note, that unlike in the rest of the text,
  $\delta_{j}^{\; k}(x,y)$ here acts on vectors with {\em lower}
indices.}
$\delta_j^{\;\;\; k}(x,y)$. 
Commuting the Laplacian to the right, we arrive at
    \begin{align}
    \int d^3x\,\delta^i_k\nabla^j
    (-\Delta)^{1/2}\nabla_{(i}\delta_{j)}^{\;\;\; k}(x,y)\, \big|_{\,y=x}
    =&\frac{1}{2}\int
    d^3x\,\Big(\,\nabla^j\nabla_k(-\Delta)^{1/2}\delta_j^{~k}(x,y)
-\delta^i_k(-\Delta)^{3/2}\delta_i^{~k}(x,y)\notag\\
&+\nabla^j\big[\,(-\Delta)^{1/2},\nabla_k\,\big]\delta_{j}^{~k}(x,y)
+\delta^i_k\nabla^j\big[\,(-\Delta)^{1/2},\nabla_j\,\big]\delta_{i}^{~k}(x,y)
   \Big)\, \Big|_{\,y=x}\,.
    \end{align}
The first term can be further reduced to scalar traces, but we do not
do it here.
This relation has also been used as a check by independent calculation of the
left and right hand sides.

\section{Metric decomposition on a sphere $S^d$}
\label{app:modes}
Orthonormal basis of harmonics (\ref{allHs}) on the d-dimensional
sphere, $H^{A\,(n)}_{ij}$, is motivated by the tensor
decomposition into transverse-traceless tensor, vector and scalar
parts  
\bseq
    \begin{eqnarray}\label{decom}
    &&h_{ij} = h^{TT}_{ij} + \nabla_i\xi_j+ \nabla_j\xi_i
    + \Big(\nabla_i\nabla_j - \frac1d\,g_{ij}\Delta\Big)E + \frac1d g_{ij}\phi,\\
    &&\label{constaints}
    g^{ij}h^{TT}_{ij}=0,\qquad\nabla^i h^{TT}_{ij}=0,\qquad \nabla_i \xi^i=0,
    \end{eqnarray}
\eseq
in which their ingredients $h^{TT}_{ij}(x)$, $\xi_i(x)$, $E(x)$ and
$\phi(x)$ are in their turn expanded in the complete basis of
irreducible tensor, vector and scalar eigenmodes of the covariant
Laplacian (\ref{allmodes}). The square-root
normalization factors of the latter easily follow from the relations
which are valid for generic transverse vectors $\xi_{i}$, $\zeta_{i}$
and scalars $\varPhi$, $\varPsi$ on $S^d$, 
    \begin{gather}
    4\int d^dx\,\sqrt{g}\,\nabla_{(i}\,\xi_{j)}
    \nabla^{(i}\zeta^{j)}=  2\int d^dx\,\sqrt{g}\,
    \xi_{i}\Big(-\Delta -\frac1d R\Big)\zeta^i,        \label{2.11}\\
    \int d^dx\,\sqrt{g}\,\Big(\nabla^i\nabla^j - \frac1d\,
    g^{ij}\Delta\Big)\varPhi(x)\, \Big(\nabla_i\nabla_j -\frac1d\,
    g_{ij}\Delta\Big)\,\varPsi(x)
    =\int d^dx\,\sqrt{g}\, \varPhi(x)\Big(\frac{d-1}{d}\Delta^2+\frac1d R\,\Delta\Big)\,\varPsi(x).    \label{2.12}
    \end{gather}
This provides orthonormality and completeness relations for $H_{ij}^{A\,(m)}(x)$
    \begin{eqnarray}
    &&\int d^dx \sqrt{g}\, H_{ij}^{A\,(m)}(x)
    H^{ij}_{B\,(n)}(x) =\delta_{(n)}^{(m)}\,\delta^A_B,  \label{norm}\\
    &&
    \sum_{A,(n)} H_{ij}^{A\,(n)}(x)H^{kl}_{A\,(n)}(y) =
    \frac{1}{\sqrt{g}}\delta_{ij}^{~~kl}(x,y)\;.   \label{compl} 
    \end{eqnarray}

On $S^d$
the operator $\mathbb{D}_{ij}^{\:\:\:\:kl}$ is composed only of
covariant derivatives and covariantly conserved metric. In the
calculation of the matrix elements (\ref{matrixel}) all covariant derivatives,
including those in the definition of the spherical harmonics
(\ref{allHs}), can be grouped into powers of the Laplacian
$\Delta$. Then the replacement of $\Delta$ by its eigenvalues tells us
that the matrix elements in each sector are functions of the
eigenvalues of the Laplacian in this sector. This leads to
Eqs.~(\ref{Dns}). 

Clearly, in the tensor sector the function $\mathbb{D}_t(n)$ is
polynomial in the Laplacian eigenvalues 
and has cubic order, according to the dimensionality of $\mathbb{D}$. 
In the vector and scalar sectors this need not be the
case a priori, because of the normalization factors in (\ref{2.4}),
(\ref{Hs1}) with the inverse powers of the Laplacian. In the vector
sector, however, the normalization factors cancel, 
 due to the identity
    \begin{equation}
\label{diagvec}
    \mathbb{D}_{ij}^{\:\:\:\:kl}\nabla_{(k}\xi_{l)}
    = \nabla_{(i}\mathbb{D}_v \xi_{j)},
    \end{equation}
valid for any transverse vector $\xi_i$. Equation (\ref{diagvec}) is
just the statement that $\mathbb{D}$ transforms a vector polarization
of the metric into a vector polarization.

\bibliographystyle{JHEP}
\bibliography{lambda31nus}

\end{document}